# Molecular design for cardiac cell differentiation using a small dataset and decorated shape features

Fatemeh Etezadi[1], Shunichi Ito[1,2], Kosuke Yasui[3], Rodi Kado Abdalkader[4], Itsunari Minami[5], Motonari Uesugi[1,6], Ganesh Pandian Namasivayam[1], Haruko Nakano[7], Atsushi Nakano[7,8], Daniel M. Packwood[1*]

[1] Institute for Integrated Cell-Material Sciences (iCeMS), Kyoto University, Kyoto 606-8501, Japan
[2] Faculty of Pharmaceutical Sciences, Kyoto University, Kyoto, 606-8501, Japan
[3] Department of Applied Chemistry, Graduate School of Engineering, Osaka University, Osaka, 565-0871, Japan
[4] Ritsumeikan Global Innovation Research Organization (R-GIRO), Ritsumeikan University, Shiga, 525-8577, Japan
[5] Myoridge Co. Ltd. Kyoto, 606-8305, Japan
[6] Institute for Chemical Research, Kyoto University, 611-0011, Kyoto, Japan
[7] Department of Molecular Cell and Developmental Biology, University of California Los Angeles, Los Angeles, 90095, USA
[8] Department of Cell Physiology, School of Medicine, Jikei University, Tokyo, 105-8461, Japan
* Corresponding author. dpackwood@icems.kyoto-u.ac.jp

## Abstract

The discovery of small organic compounds for inducing stem cell differentiation is a time- and resource-intensive process. While data science could, in principle, facilitate the discovery of these compounds, novel approaches are required due to the difficulty of acquiring training data from large numbers of example compounds. In this paper, we demonstrate the design of a new compound for inducing cardiomyocyte differentiation using simple regression models trained with a data set containing only 80 examples. We introduce decorated shape descriptors, an information-rich molecular feature representation that integrates both molecular shape and hydrophilicity information. These models demonstrate improved performance compared to ones using standard molecular descriptors based on shape alone. Model overtraining is diagnosed using a new type of sensitivity analysis. Our new compound is designed using a conservative molecular design strategy, and its effectiveness is confirmed through expression profiles of cardiomyocyte-related marker genes using real-time polymerase chain reaction experiments on human iPS cell lines. This work demonstrates a viable data-driven strategy for designing new compounds for stem cell differentiation protocols and will be useful in situations where training data is limited.

## 1. Introduction

Breakthroughs in stem cell science will accelerate research in developmental biology and regenerative medicine. Pluripotent stem cells have the potential to differentiate, in principle, into tissue of all kinds under particular treatment protocols [1]. Following the discovery of induced pluripotent stem (iPS) cells in 2006 [2], a considerable number of differentiation protocols have been developed, and numerous cell types can now be generated [3]. Until now, these protocols have been established entirely through experimental efforts. In the age of advancing digitization, it is natural to ask whether data science can assist further developments.

Figure 1 shows a widely used protocol for generating cardiomyocytes, the major cells that comprise cardiac tissue [4 - 8]. In this protocol, undifferentiated iPS cells undergo step-wise differentiation into mesoderm, cardiac mesoderm, and cardiac progenitor cells before reaching the final cardiomyocyte state. Two of the steps – the Wnt activation step and the Wnt inhibition step – are of particular importance: they involve the addition of specific chemical compounds to modulate the Wnt signaling pathway, a group of cellular processes that transmit information into the cell through receptor proteins at the cell surface [6]. The mechanism by which these compounds affect Wnt signaling appears to vary. For example, XAV939, a compound widely used during the Wnt inhibition step, is known to interfere directly with the processes involved in Wnt signaling [9]. On the other hand, another compound, KY0211, is known to indirectly modulate Wnt signaling to promote cardiomyogenesis [10]. While these lines research continue to develop, there are currently no guidelines or principles for tailoring these compounds in order to fine-tune cardiomyocyte quality or maturity.

Data science has made a considerable impact on chemistry and the life sciences; however its uptake in the stem cell field has been comparatively slow. Most efforts in this area have aimed to develop machine-learned models to discriminate between cell types observed from optical microscopy images taken at various stages of differentiation (see [11 - 14] for reviews). For the case of iPS-cell-derived cardiomyocytes, several works have developed models using calcium transient signals, tissue contraction profiles, and microscopy images to classify cell maturity states, predict tissue quality, and predict the effects of drugs (see [15 - 20] and references therein). However, to the best of our knowledge, there are no works which have attempted to apply these methods to design chemical compounds for stem cell differentiation protocols. At least two barriers to progress in this direction can be identified: (i) the restriction to small data sets and (ii) the difficulty of identifying effective input features (‘descriptors’) for regression modeling.

To illustrate how the problem (i) arises, consider the task of discovering new compounds for the protocol shown in Figure 1 by screening a chemical library. While this task is simpler than developing an entire protocol from scratch, it still requires a massive investment of time and resources. The maintenance of undifferentiated iPS cell lines requires precise incubation conditions and careful monitoring. The differentiation protocol requires over two weeks to complete, during which the cells need the application of chemical compounds, changes of the medium, and the maintenance of optimal incubation conditions. Once the protocol is complete, optical microscopy imaging and biological assays are required to confirm the successful differentiation of cells. Due to unavoidable biological variation between cells, the protocol also needs to be repeated several times for each candidate compound to establish statistically reliable results. When the differentiation efficiency is low, further work is required to identify the origin of the problem. These labor-intensive processes are compounded by the high price of the chemical reagents involved. These severe demands on time and resources mean that many chemical screenings are often limited to around 100 compounds at most. While emerging developments in automated screening and cell line maintenance may improve matters [21, 22], the limitation of small data sets is unavoidable in many areas of stem cell science at present.

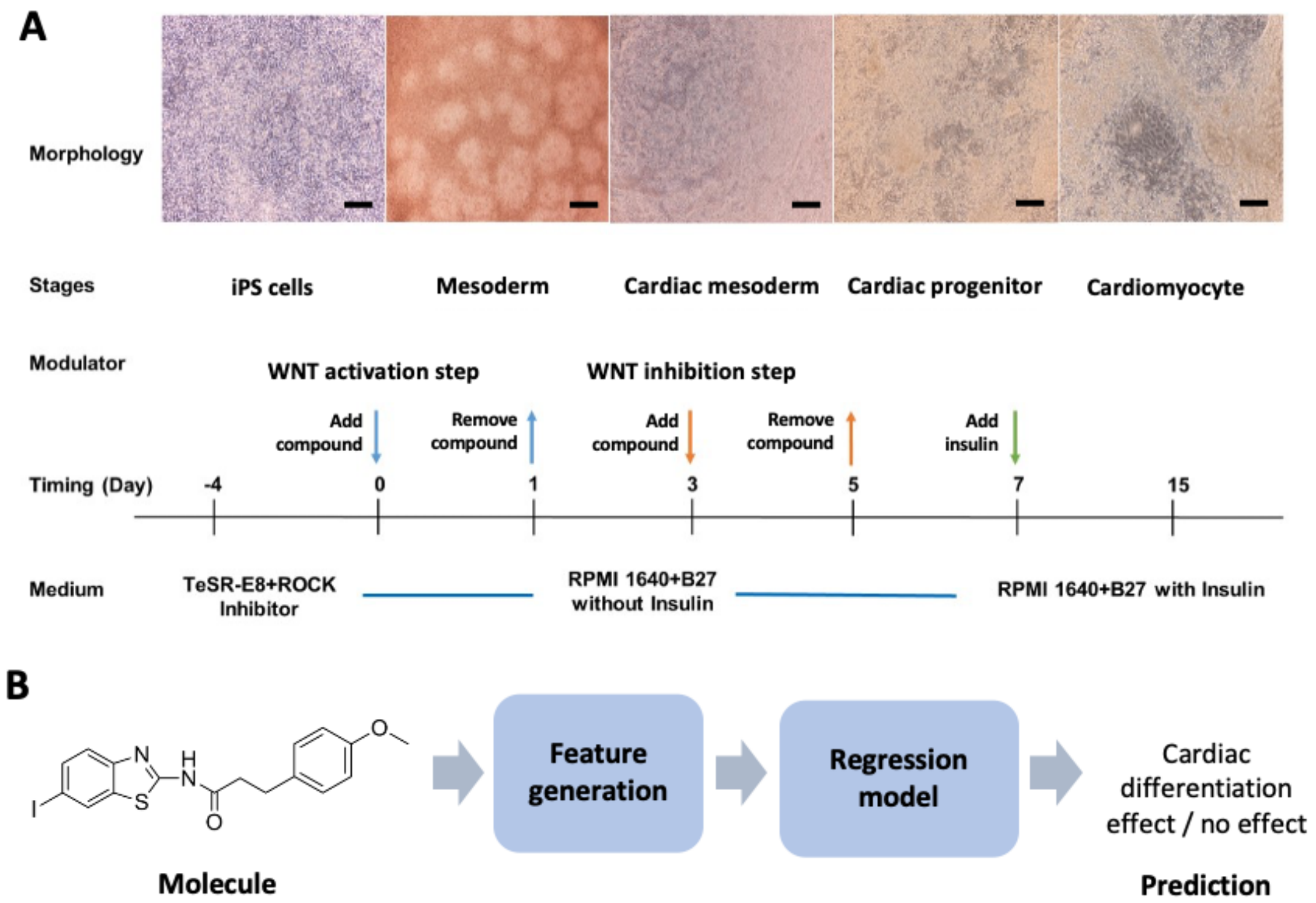

**Figure 1.** (A) Example of a protocol for cardiomyocyte differentiation for iPS stem cells. The row labelled 'morphology' shows images of tissue from an optical microscope. Scale bars = 200 µm. The row labelled 'modulator' shows the chemical compounds added to the iPS culture, and the arrows underneath identify when they are added and removed. The row labeled 'medium' shows the composition of the cell growth medium. (B) Scheme for predicting cardiac differentiation effect of a chemical compound when used during the Wnt inhibition step. The molecule shown is from the data set used in this study.

Data science aims to build regression models that can predict experimental outcomes using an appropriate set of input variables (known as 'features' or 'descriptors'). For these models to be scientifically meaningful, these features should be clearly related to the prediction target. This is where a problem (ii) arises. While the chemistry literature provides numerous choices of features for describing molecules [23 - 25], few of them have an obvious connection with stem cell differentiation. To build a regression model for predicting the occurrence of stem cell differentiation, we should consider features related to the three-dimensional shape of the molecule. This is because the molecule must adopt specific conformations in order to interact with the cellular proteins responsible for the differentiation process [26]. Shape-based molecular descriptors have been widely used in cheminformatics and pharmacology for many years, and mainly define shape in terms of the union of the van der Waals spheres of the atoms in the molecule [27 - 30]. However, for cases in which shape-based descriptors fail, there are few directions for improving them.

In this paper, we demonstrate the design of a new compound for inducing cardiomyocyte differentiation using simple regression models and a small sample of chemical screening data. Our models use *decorated shape descriptors* as inputs, a new molecular feature representation that integrates both shape and hydrophilicity information seemlessly. Two types of regression models are considered: a logistic regression (LR) model and a single-layer feedforward neural network (NN) model. For the case of the LR model, we show that decorated shape descriptors significantly improve predictive performance compared to ordinary shape descriptors, even under the small data conditions considered here. For the case of the NN model, we find that decorated shape descriptors significantly improve robustness to overtraining, which is usually a serious problem for neural networks trained under small data conditions. By use of a conservative molecular design strategy, we identify an entirely new compound that can be used during the Wnt inhibition step for inducing cardiomyocyte differentiation, as predicted by the LR and NN models. This prediction is confirmed by chemical synthesis followed by real-time polymerase chain reaction (RT-PCR) experiments on real human iPS cell lines.

## 2. Method

### *2.1. Training data acquisition*

A sample of 80 compounds were synthesized at Kyoto University (see [6] and references therein). These compounds are derivatives of KY02011, a compound that indirectly modulates Wnt signaling for cardiomyocyte generation [6, 10]. The cardiac differentiation effect of these compounds was tested using the protocol in reference [6]. Depending on the intensity of the fluorescence, the cardiac differentiation effect was classified as 'very weak' (33 compounds), 'weak' (22 compounds), 'medium' (12 compounds), 'strong' (8 compounds) or 'super strong' (5 compounds). For subsequent model fitting, compounds classified as 'medium' or above were labelled as 1, and all others labelled as 0. The final training data set therefore consisted of 25 compounds labelled as 1 and 55 compounds labelled as 0. The full data set is available in Supporting Information 1.

### *2.2. Decorated shape descriptors*

Our decorated shape descriptors were generated according to the scheme shown in Figure 2. For a given molecule this scheme proceeds through six steps. In step (i), a set of conformations for the molecule are generated from the molecule's two-dimensional chemical structure using classical force fields. Concretely, we perform a succession of 10 short molecular dynamics (MD) runs, relaxing the conformation emerging at the end of each run. All MD runs and structural relaxations were performed using the Materials Studio Forcite module [31] and the DRIEDING force field [32] with no solvent. Each MD runs used trajectory lengths of 1.0 ns in total and at 25°C. The cycles were managed using the Pipeline Pilot software [33].

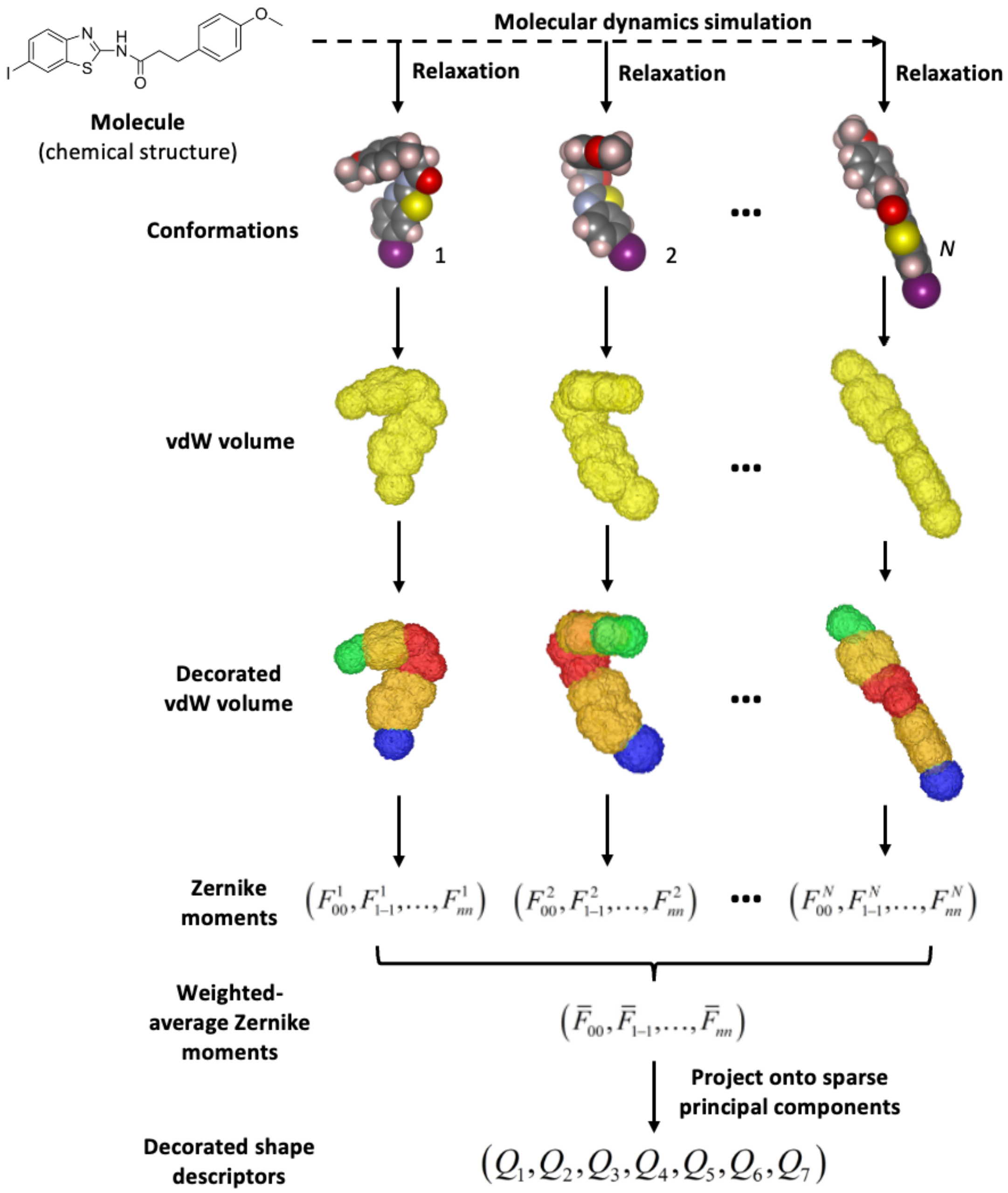

**Figure 2**. Flow diagram for building decorated shape descriptors from a two-dimensional molecule structure. See text for details. In the second row, grey spheres = carbon atoms, white spheres = hydrogen atoms, yellow spheres = sulfur atoms, purple spheres = iodine atoms, red spheres = oxygen atoms, and cyan spheres = nitrogen atoms. Molecule structures drawn with Materials Studio Visualizer [29]. In the fourth row, fragment hydrophilicities are indicated by the blue (most strong), green, orange, and red (most weak) colors.

In step (ii), we compute the van der Waals exclusion volume for each conformation obtained above. The vdW exclusion volume (or simply 'vdW volume') is defined as the unions of the spheres drawn by the vdW radii of each atom in the molecule. Concretely, the vdW volume is

$$\rho(\mathbf{r}) = \begin{cases} 1 & \text{if } |\mathbf{r} - \mathbf{r}_j| < v_j \text{ for any } j \\ 0 & \text{otherwise,} \end{cases} \tag{1}$$

where $j$ refers to atom $j$ in the molecule, $\mathbf{r}_j$ is the position of atom $j$, $|\mathbf{r} - \mathbf{r}_j|$ denotes the distance between $\mathbf{r}$ and $\mathbf{r}_j$, and $v_j$ is the vdW radius of atom $j$. Equation (1) was calculated using a rectangular mesh defined within a unit sphere. The molecular conformation was first rotated and translated so that its center of mass lay at the origin, and the atom coordinates and vdW radii were scaled so that the molecule lay entirely within the unit sphere. Hydrogen atoms were also ignored. The grid mesh was obtained by first creating a 200 x 200 rectangular mesh for the volume $[-1, 1]^3$ and then removing all mesh points lying at distance greater than 1 from the origin.

In step (iii), we 'decorate' the vdW volumes by incorporating hydrophilicity information. Concretely, we compute

$$\tilde{\rho}(\mathbf{r}) = h(\mathbf{r})\,\rho(\mathbf{r}), \tag{2}$$

where $h(\mathbf{r})$ is the hydrophilicity value of the molecular fragment (defined below) closest to point $\mathbf{r}$. The resulting decorated vdW volume resembles a three-dimensional object patterned according to the distribution of hydrophilicity values within the molecule (see the third row of Figure 2).

These molecular fragments were generated prior to feature generation by breaking all ring-connected single bonds in every compound in our training data. The resulting fragments were passivated by adding hydrogen atoms to the undercoordinated atoms. 54 fragments were obtained in total, including various short alkyl chains, alcohols, hydrogen halides, and aromatic rings. Most fragments appeared in several different molecules. For each fragment, a 1 ns-long MD run was performed with an explicit water solvent included. The hydrophilicity value $h$ for the fragment was obtained by calculating the time-averaged interaction energy between the fragment and solvent, normalized by the number of non-hydrogen atoms in the fragment. These MD simulations were performed with the LAMMPS code [34] and the DRIEDING force field [32]. Full details are provided in Supporting Information 2. The full list of molecular fragments and hydrophilicity values are listed in Supporting Information 3. The hydrophilicity values range from -1.8 kcal/mol for strongly hydrophilic fragments (such as HI) through to -0.05 kcal/mol for weakly hydrophilic ones (such as $CH_3SO_3NH_2$).

In step (iv), we expand the decorated vdW volumes into a basis of 3D Zernike polynomials:

$$\tilde{\rho}(\mathbf{r}) = \sum_{n=0}^{\infty} \sum_{l=0}^{n} \sum_{m=-l}^{l} c_{nlm} Z_{nlm}(\mathbf{r}), \tag{3}$$

where $c_{nlm}$ is a complex coefficient and $Z_{nlm}(\mathbf{r})$ represents a 3D Zernike polynomial. The 3D Zernike polynomial is defined as $Z_{nlm}(\mathbf{r}) = R_{nl}(r)Y_{lm}(\theta, \phi)$, where $R_{nl}(r)$ is a radial factor and $Y_{lm}(\theta, \phi)$ a spherical harmonic (see [28] for details). Note that equation (3) only applies for functions defined within the unit sphere. From equation (3), a set of so-called *Zernike moments*, defined as

$$F_{nl} = \left|c_{n,l,-l}\right|^2 + \left|c_{n,l,-l+1}\right|^2 + \cdots + \left|c_{n,l,l}\right|^2, \tag{4}$$

are computed, where $|c|^2$ denotes the square modulus of $c$. Zernike moments are invariant under rotations, which circumvents ambiguities arising from molecular orientation within the unit sphere. In this work, we computed the Zernike moments for $n = 0$ up to $n = 25$, yielding 351 Zernike moments in total. This work used a custom C++ code to compute the Zernike moments, along with the BOOST library to compute the spherical harmonics [35]. The coefficients were computed as

$$c_{nlm} = \int \tilde{\rho}(\mathbf{r})\, Z^*_{nlm}(\mathbf{r})\, d\mathbf{r}, \tag{5}$$

using the rectangular mesh described above, where the asterisk denotes the complex conjugate.

In step (v), we thermodynamically average the Zernike moments across the 10 conformations. This yields

$$\bar{F}_{nl} = \sum_k p_k F^k_{nl}, \tag{6}$$

where $F_{nl}^k$ indicates the Zernike moment obtained from the $k$th molecular conformation, and

$$p_k = \frac{e^{-\epsilon_k/k_B T}}{\sum_j e^{-\epsilon_j/k_B T}}, \tag{7}$$

denotes the thermodynamic probability of conformation $k$. In equation (7), $\varepsilon_k$ denotes the energy of configuration $k$ (as obtained in step (i)), $k_B$ denotes the Boltzmann constant, and $T = 25$ °C denotes temperature.

Having computed the conformation-averaged Zernike moments, we obtain an 80 x 351 data matrix, where each row represents one compound from the training data and each column represents an averaged Zernike moment. In step (v), sparse principal component analysis is applied to this matrix to obtain our final set of features. In this work, we retain 7 sparse principal components (PCs) to use as features for machine learning. These 7 sparse PCs account for 90 % of the variation of the original data matrix. The "sparsepca" package for R was used to perform the sparse principal component analysis [36, 37]. The final set of features (decorated shape descriptors) for a particular compound are denoted as $\mathbf{Q} = (Q_1, Q_2, \ldots, Q_7)$.

For the purpose of comparison, we also consider the case of ordinary shape descriptors. These can be generated according to the above scheme by simply bypassing step (iii).

*2.3.Regression model training and validation*

In this work, two types of regression models were developed for predicting compound labels: a logistic regression (LR) model and a neural network (NN) model. These models represent two extremes. On the one hand, the simple functional form of the LR model will limit its accuracy, but also makes it robust to overtraining. On the other hand, the highly flexible functional form of the NN model will result in a higher accuracy in principle, but also makes it more susceptible to overtraining.

The LR model takes the form

$$\frac{p\left(\mathbf{Q}\right)}{1-p\left(\mathbf{Q}\right)} = \exp\left(\beta_0 + \boldsymbol{\beta}\cdot\mathbf{Q}\right), \tag{8}$$

where $\mathbf{Q}$ denotes the feature vector for a compound, $(\beta_0, \boldsymbol{\beta})$ is a vector of parameters, and $p(\mathbf{Q})$ is the probability of the compound being labeled 1 (where label 1 indicates that the compound is an effective during the Wnt inhibition step for cardiomyocyte differentiation). In this work, the logistic model was fit using the 'glm' function in the R 'stats' package [36]. A threshold value of $p(\mathbf{Q}) = 0.5$ was used to assign labels to predictions. The LR model was built using the entire training data. Wald and chi-square tests were performed on the final model to determine the statistical significance of the fitted parameters. The robustness of this model architecture to overtraining was tested using five-fold cross validation.

The NN model consisted of a single hidden layer with 5 nodes and logistic activation functions. The neural network was fit using the R 'neuralnet' package [38], using the resilient backpropagation algorithm with learning rates between 0.5 and 1.2. A threshold of 0.01 for the gradient of the error function was used to specify convergence. The sum-square of errors was used as the error function. The NN model was built using the entire training data, and six-fold cross validation was used to test the robustness of the model architecture to overtraining.

*2.4. Random sensitivity analysis*

The Random Sensitivity Analysis (RSA) provides a further diagnostic for model overtraining. Our approach follows the methods presented in references [39] and [40], with some modification.

Consider a molecule composed of $n$ fragments, as described section 2.2. Let $\mathbf{v} = (v_1, v_2, \ldots, v_n)$, where $v_i$ is the hydrophilicity value for the $i^{\text{th}}$ fragment. We define the random variable $\mathbf{U} \sim N(\mathbf{v}, \sigma^2\mathbf{I})$, where $\mathbf{I}$ is an $n \times n$ identity matrix. $\mathbf{U}$ is a vector of perturbed hydrophilicity values. To perform RSA on the molecule, we first generate a sample of independent realizations of $\mathbf{U}$. For each of these realizations, we then compute the feature vectors $\mathbf{Q}$, and then predict the molecule's label using the trained regression model. The sample of realizations is then partitioned into two groups according to the predicted labels. Finally, we compute the so-called $F$-value, which is defined as the fraction of realizations which result in the prediction of the label 1. If the model prediction with unperturbed features is 1 (resp. 0), and the $F$ value is also close to 1 (resp. 0), then the model's predictions are said to be robust with respect to hydrophilicity perturbations. Conversely, if the $F$ value is close to 0 (resp. 1), then the model's predictions are said to be sensitive to hydrophilicity perturbations. The presence of multiple molecules with such $F$ values is a symptom of model overtraining.

In this work, the random sample was generated using Latin hypercube sampling, as implemented in the R package "lhs" [41]. The sample size was set to 25 and the standard deviation $\sigma$ set to 0.25 kcal/mol. RSA was performed on 10 molecules. Computational demands make it difficult to consider larger numbers of molecules. We only consider molecules with training labels of 1, as are primarily interested in model sensitivity when predicting these cases.

### 2.5. *Molecular design scheme*

New molecule structures were designed by an original scheme which combines fragments from the fragment library (section 2.2). By designing new molecules on the basis of fragments from the training data, we can ensure that they will not deviate far from the molecules present in the training domain of the regression models. This conservative strategy therefore minimizes the risk of designing a molecule with a false-positive predictions. A low-risk strategy is appropriate given the high costs of synthesizing and testing new molecules in a real stem cell differentiation protocol

The scheme is illustrated in Supporting Information 4. It starts by noting that all molecules in the training data can be represented by the pattern "X-benzothiozole-linker-aryl-YZ". In words, a modifier group X is connected to a benziothiozole, which is connected via an alkyl linker to an aryl ring, which is connected to two more modifiers Y and Z. All fragments in the database are first classified as X, benzothiozole, linker, benzene, Y, or Z, according to the pattern above. The molecule design begins with the benzothiozole and aryl (benzene) fragments. A linker fragment is then selected and used to connect the benzothiozole and aryl fragments. Finally, a functional group fragments X is selected and connected to the benzothiozole, and two more functional group fragments Y and Z are then selected and added to the aryl fragment. The fragments are added by removing the passivating hydrogen atoms (added during the fragmentization process in section 2.2) and drawing a bond between those positions. Predictions are made on the designed molecule by computing their feature vectors (section 2.2) and then using the trained models.

### 2.6. *Experimental verification*

*Synthesis.* The new molecule (ID 5 in Table 3) was synthesized using a two-step reaction from commercial starting materials. The integrity of the final product was confirmed by $^{1}$H-NMR spectroscopy. See Supporting Information 5 for further details.

*Human iPS stem cell maintenance and culture.* The human iPS cells (253G1 cell line) were supplied by RIKEN Bio Resource Center and cultured as described previously [42]. Briefly, feeder-free human iPS cells were cultured with TeSR-E8 culture medium in a humidified atmosphere of 95% air and 5% $CO_2$ at 37 °C. The cells were then passaged upon reaching 70% confluency and the medium was changed daily. A NucleoCounter NC-200 automated cell counter was used to analyze the cells *via* a Via1-Cassette (ChemoMetec, Copenhagen, Denmark) to determine cellular proliferation and viability. Furthermore, cellular adhesion and density on the culture plate were visually inspected by an IX71 inverted microscope (Olympus, USA) equipped with a DP71 camera (Olympus).

*Cardiomyocyte differentiation.* Differentiation of human iPS cells to cardiomyocytes was accomplished using the protocol described in Figure 1 [5]. In brief, the iPS cells were induced to the mesoderm step by adding a medium consisting of B-27 supplemented-RPMI (containing B-27 minus insulin supplemented-RPMI, Thermo Fisher Scientific, NY, USA) and 12 μM CHIR99021 (Wako, Japan) upon reaching 90–100% confluence. After one day after the CHIR99021 treatment, the CHIR99021 was removed. On day 3 of differentiation, cardiac mesoderm induction was started by adding Wnt inhibitor (either the reference case of 2 μM XAV939 (Wako) or the molecule predicted above) for 48 hours. On day 7 of differentiation, insulin (1μg) was added and the medium was changed every other day. Beating was expected to be observed during days 7-14 of differentiation. Video of the beating was captured with an IX71 inverted microscope (Olympus) equipped with a DP71 camera (Olympus).

*Gene expression analysis.* Gene expression levels were analyzed using quantitative polymerase chain reaction (RT-qPCR) performed with the Applied Biosystems 7900HT Fast Real-Time PCR System. RNA extraction and cDNA synthesis were performed according to the manufacturer's instructions, RNeasy Mini Kit (QIAGEN, 74104) and cDNA synthesis kit (QIAGEN, 205311). Consequently, RT-qPCR was accomplished with Applied Biosystems™ Power SYBR™ Green PCR Master Mix (UK). Relative gene expression profiles were analyzed using the REST2009 software (Qiagen Inc.) by considering three replicates of samples in each group. The primer information was listed in Supporting Information 6.

*Immunofluorescent staining of the cell culture.* For protein expression characterization, the samples were fixed with 4% (w/v) paraformaldehyde (PFA, Sigma, USA) in PBS for 30 minutes, followed by permeabilization with 0.1% (v/v) Triton X-100 (USB, USA) in PBS for 10 minutes. Subsequently, we performed blocking with 1% (v/v) bovine serum albumin (BSA, Sigma-Aldrich) in PBS for 30 minutes. To evaluate the pluripotency of our iPS cells, the samples were incubated with mouse monoclonal anti-Oct-3/4 antibody (clone C-10, Santa Cruz) and rabbit polyclonal anti-NANOG antibodies (H-155, Santa Cruz) and incubated with Goat anti-mouse and anti-rabbit secondary antibodies described in Supporting Information 6. To investigate cardiac differentiation, the substrates were incubated with mouse monoclonal cTnT primary antibody (MAB 1874, R&D Systems) and rabbit monoclonal MYH7 primary antibody (MAB 1874, R&D Systems) and incubated with Goat anti-mouse and anti-rabbit secondary antibodies, under the conditions described in Supporting Information 6. The nuclei were stained by DAPI (ProLong Gold Antifade Mountant, Invitrogen, Thermo Fisher Scientific). The stained samples were analyzed with inverted fluorescence microscopy (IX71, Olympus) and confocal microscopy (FV10, Olympus). The primary and secondary antibody information and incubation conditions are described in Supporting Information 6.

## 3. Results

### *3.1.Dataset visualization*

Our data set is plotted in Figure 3A using the tSNE method [43, 44]. Each point corresponds to one molecule, and the points are embedded according to their decorated shape descriptors. The molecules are concentrated into six clusters, labelled *a* - *f*. Importantly, the positive-hit compounds (red points) tend to appear in the same clusters. This is most obvious for the clusters *a* and *b*, which both contain 10 compounds in total and 7 and 5 positive hits, respectively. Moreover, the positive-hit compounds are clearly localized in the same areas of these clusters. Similar remarks hold for cluster *d*, although it is much smaller (3 compounds in total, 2 positive hits). Cluster *c* contains 21 compounds in total and 9 positive hits. This cluster is more diffuse than clusters *a* and *b*, which reflects the wider range of structures contained in it. Consequently, the positive-hit compounds in cluster *c* are also less localized, although they still tend to neighbors. Cluster *f* is the most diffuse cluster, containing 30 compounds in total and only 3 positive hits. This is the only cluster in which the positive-hit compounds are not clearly located together. Cluster *e* is relatively compact, indicating a high degree of structural uniformity, but contains no positive-hit compounds. The clear assignment of positive-hit compounds to specific clusters, as well as the tendency for these compounds to locate together within the same clusters, suggests that decorated shape features are effective at separating compounds on the basis of their cardiac differentiation effect. In Supporting Information 7, we select a few compounds from each cluster and show their typical three-dimensional conformations. The various clusters appear to be associated with compounds with specific combinations of structural features. This suggests that decorated shape descriptors are effective at separating compounds on the basis of their structural properties.

### *3.2.Regression model performance*

In Figure 3B, we plot the same data again, this time coloring the points according to the predictions of the LR model. The LR model achieves a high overall prediction accuracy of 84 % compared to the experimental dataset. The confusion matrix shows a true-positive rate (TPR) of 64 % (Figure 3B insert). The LR model performs well for clusters *a*, *b*, and *d*, where the positive hit compounds tend to be located together. The LR model has a higher fail rate in clusters *c* and *f*, which contains a larger diversity of structures. The failure of the model in cluster *f* is not surprising considering the small fraction of positive hit compounds contained in it. The LR model performs very well when predicting true negatives, achieving a true negative rate (TNR) of 93 % and reproducing well the clusters dominated by negative-hit compounds.

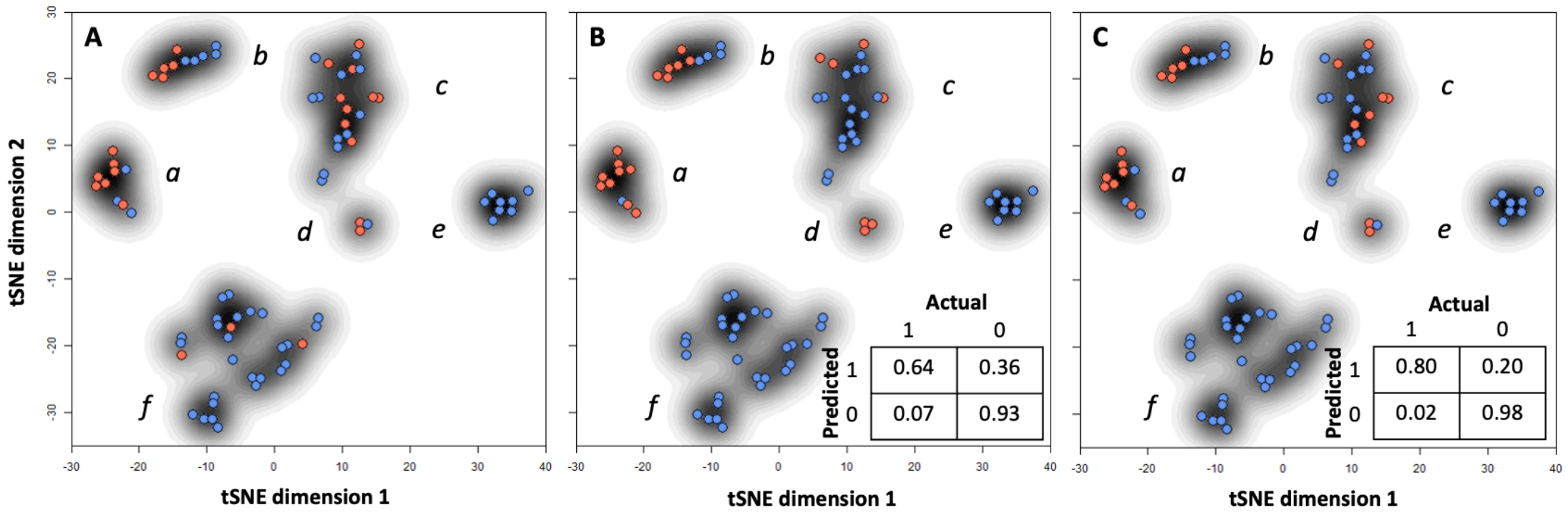

**Figure 3.** (A) Plot of the data using the t-distributed stochastic neighbor embedding (tSNE) method. Blue (red) points correspond to compounds with poor (strong) cardiac differentiation ability when used during the Wnt inhibition step in the protocol shown in Figure 1. The back background highlights the spatial distribution of the data as computed using kernel density estimation. Clusters are identified by the lower-case letters $a - f$. (B) As for A, but points colored according to the predictions of the logistic regression model. (C) As for A, but points colored according to the predictions of the neural network model. Model confusion matrices are shown in the inserts.

Hypothesis testing confirms that the fitted model parameters ($\beta_0$ and $\boldsymbol{\beta}$ in equation (8)) are statistically meaningful. Using the Wald test, the intercept parameter $\beta_0$ and two of the parameters in $\boldsymbol{\beta}$ are found to be highly significant ($p$ = 0.005, 0.01, and 0.006, respectively; see Supporting Information 8). One other parameter in $\boldsymbol{\beta}$ is found to have borderline significance ($p = 0.06$). These findings are confirmed by a chi-square test, which again identifies two highly significant parameters in $\boldsymbol{\beta}$ ($p$ = 0.008 and 0.001, respectively) and two with borderline significance ($p$ = 0.08 and 0.09, respectively). Overall, these results confirm that the LR model harnesses shape and hydrophilicity information in a non-trivial way to make predictions.

In Figure 3C, the points are colored according to the predictions of the NN model. The NN model apparently outperforms the LR model, achieving an overall prediction accuracy of 93 %, a TPR of 80 %, and a TNR of 98 %. Like the LR model, it performs well for clusters *a*, *b*, and *d*. It also performs better in cluster *c*, correctly predicting some positive hits in the bottom part of the cluster. Like the LR model, it fails to predict the positive-hit compounds in the diffuse cluster *f*.

In order to diagnose overtraining in the LR model and NN models, we performed 5-fold cross-validation. Cross-validation was performed by ensuring that each fold contained the same proportion of positive- and negative-hit compounds as the original dataset. Figure 4A plots receiver operator characteristic (ROC) curves computed *via* cross-validation for the LR model. The light grey curves correspond to one realization of the ROC curve for a specific test data fold, and the dark black curves correspond to the average over the test folds. The fold-averaged area-under-the-curve (AOC) value of 0.76 (standard error 0.05). The outward bulge of the ROC curves suggests that, on average, the LR model architecture performs better than a random classifier for most threshold choices. The relatively small spread of the test-fold ROC curves (grey lines) from the fold-averaged ROC curve (black line) suggests robustness of the model architecture to overtraining. For the threshold value of 0.5, the LR model achieves a fold-averaged TPR

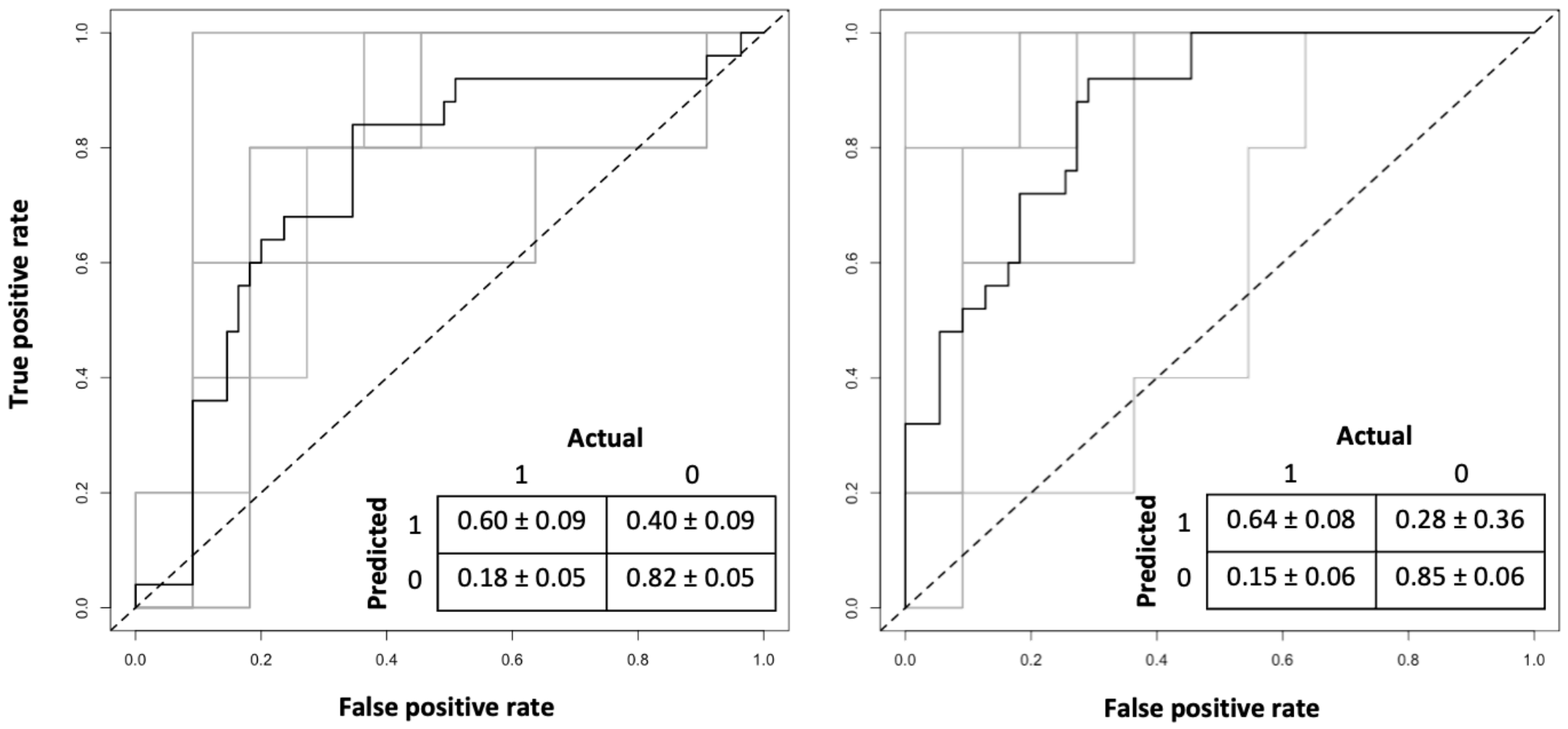

**Figure 4.** Receiver operator characteristic (ROC) curves under five-fold cross validation. (A) Logistic regression model. (B) Neural network model. Thin grey lines are for the folds. Thick black line is the fold average. Inserts show the confusion matrices (values reported as fold-averages ± fold standard errors)

of 60 % (standard error 9 %) and a fold-averaged TNR of 81 % (standard error 5 %). The TPR and TNR values are decreased by roughly 4 % and 12 %, respectively, compared to the LR model trained with all data, which is partly expected given the smaller amount of training data available during cross-validation. Overall, the performance decreases and standard errors are not large, suggesting that the LR model is not seriously overtrained.

Cross-validation for the case of the NN model suggests a slightly higher degree of overtraining compared to the LR model (Figure 4C). While the outward bulge of the fold-averaged ROC curve is more prominent for this case, achieving a larger AOC value of 0.80 (standard error 0.07), the test-fold ROC curves are spread more somewhat widely about the fold average. At the threshold value of 0.5, the NN model architecture achieves a fold-averaged TPR of 64 % (standard error 9 %) and a fold-averaged TNR of 85 % (standard error 6 %). Again, the performance decrease compared to the NN model trained with all data is partly expected due to the smaller training data size used for cross-validation. The decrease in both the fold-averaged TPR and TNR in the order of 15 %. The larger decrease in the TPR for the NN model compared to the LR model suggests that the former architecture is less robust to changes in the training data.

We now compare the above results to the case where the models are trained using ordinary (non-decorated) shape descriptors. For the case of an LR model trained using the entire data set, we obtain a lower overall accuracy of 69 % and a decidedly lower TPR of only 16 %. Interestingly, this model achieves a high TNR of 93 %, suggesting that hydrophilicity information is not useful for identifying negative-hit cases. However, hypothesis testing raises doubts about whether shape information is truly harnessed by the model when forming its predictions: both the Wald test and chi-square test identify only two parameters in $\boldsymbol{\beta}$ which are statistically significant, but with somewhat borderline *p* values ($p = 0.04$ and 0.03, respectively, for the Wald test; $p = 0.06$ and 0.02, respectively, for the chi-squared; see Supporting Information 8). These results confirm that decorated shape descriptors lead to LR models with higher accuracies than do ordinary shape descriptors. Five-fold cross-validation results in a fold-averaged TPR of 24 % (standard error 7.4 %) and a fold-averaged TFR of 85 % (standard error 6.1 %). The relatively small change in TPR and TNR values under cross-validation compared to the model trained with all data suggests that this LR model is also robust to overtraining, despite its poor accuracy overall.

A different story emerges for the case of an NN model trained using ordinary shape descriptors. Using the entire data set, we obtain an overall accuracy of 99 %, a TPR of 100 %, and a FPR of 98 %. Such large values are almost certainly the result of overtraining, especially given the small training data sizes considered here. This point is confirmed by five-fold cross-validation, which results in a fold-averaged TPR of 52 % (standard error 4.9 %) and a TFR of 80 % (standard error of 4.4 %). The massive drop in the TPR under cross-validation demonstrates a high sensitivity of model performance to the training data, and, hence, severe model overtraining.

### *3.3.Random sensitivity analysis*

Having established the superiority of the LR model built using decorated shape descriptors, we now examine its sensitivity under hydrophilicity perturbations using Random Sensitivity Analysis (RSA). The results of RSA confirm that the LR model is not seriously overtrained (Table 1). For 7 of the 10 molecules tested, model predictions were insensitive to fragment hydrophilicity perturbations. Four of these molecules (IDs 1, 4, 37, and 65) have unperturbed model predictions of 1, and $F$ values between 0.88 and 0.96, indicating that only 4 – 12 % of the applied perturbations resulted in a change in model prediction for these cases. The remaining three molecules (IDs 17, 73, and 75) have unperturbed model predictions of 0 and $F$ values between 0.00 and 0.04, indicating that only 0 – 4 % of the applied perturbations resulted in a change in model predictions for these cases. Among the remaining molecules tested, only one of them shows an $F$ value equal to exactly 0 despite an unperturbed model prediction of 1 (ID 6, identified by the bold red text). This means that all perturbations applied to this molecule resulted in a change in model prediction, implying a severe sensitivity of the model to this case. The remaining two cases resulted in $F$ values close to 0.5, indicating that these are moderately sensitive cases where around half of the applied predictions resulted in a change in model prediction (IDs 24 and 39, highlighted in bold blue text). The observation of only one strongly sensitive case and two moderately sensitive cases among the 10 examined suggests that the LR model is not seriously overtrained overall.

For cases where $F$ is between 0.2 and 0.8, we perform a series of hypothesis tests to identify the fragments associated with significant changes in model predictions. Specifically, for a given molecule, a $t$-test was performed for each fragment. Each $t$-test compares the means of two samples: the sample of hydrophilicity perturbations which resulted in a model prediction of 0, and the sample of perturbations which resulted in a model prediction of 1. We restrict this test to cases where $0.2 < F < 0.8$, to ensure that the two samples have sufficient sizes. All $t$-tests were performed using the 't.test' function in the R base package [36].

| ID | Label | Prediction | F | Structure |
|---|---|---|---|---|
| 1 | 1 | 1 | 0.88 | 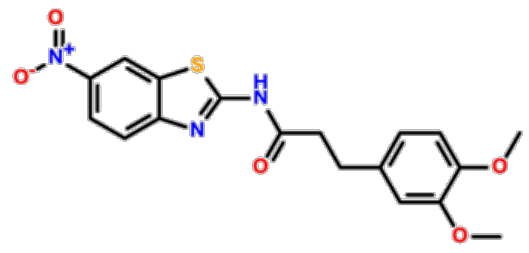 |
| 4 | 1 | 1 | 0.96 | 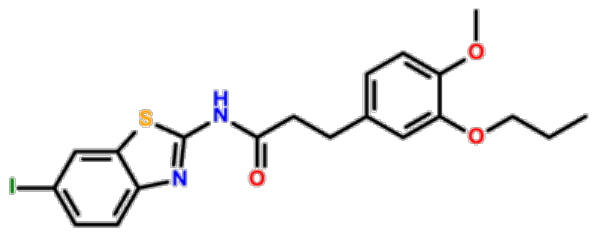 |
| 6 | 1 | 1 | 0.00 | 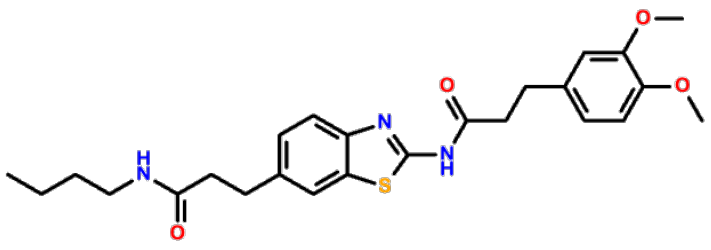 |
| 17 | 1 | 0 | 0.00 | 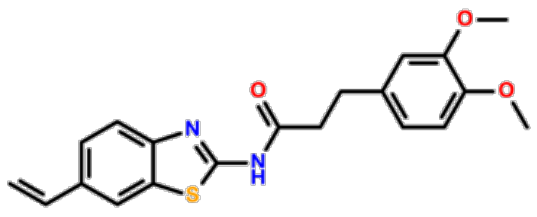 |
| 24 | 1 | 1 | 0.52 | 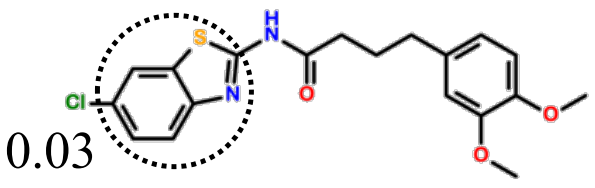  |
| 37 | 1 | 1 | 0.88 | 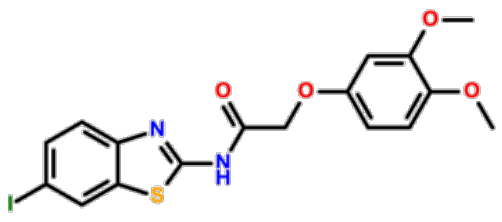 |
| 39 | 1 | 1 | 0.56 | 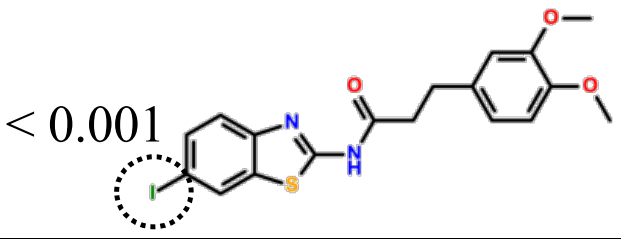  |
| 65 | 1 | 1 | 0.96 | 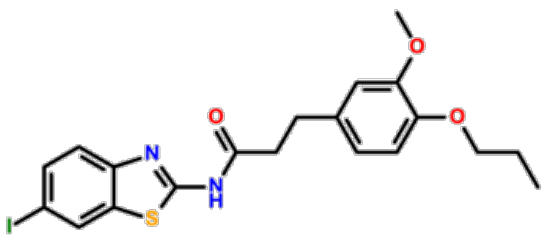 |
| 73 | 1 | 0 | 0.00 | 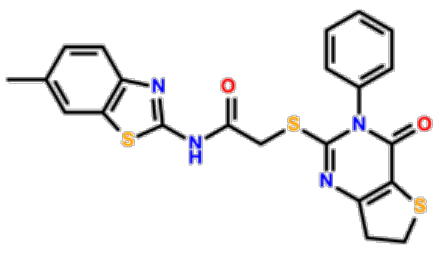 |
| 75 | 1 | 0 | 0.04 | 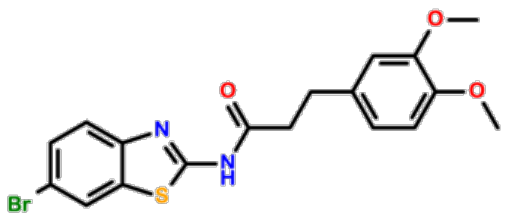 |

**Table 1.** Random sensitivity analysis applied to ten molecules from the data set. 'Prediction' refers to the prediction by the logistic regression model. $F$ is the fraction of perturbations that resulted in a positive label. Cases where $F$ differs significantly from the prediction are highlighted in red. The dotted circles identify molecular fragments associated with statistically significant changes in model predictions under random hydrophilicity perturbations (only shown for cases where $0.2 < F < 0.8$, highlighted in blue). The numbers above correspond to $p$-values from a $t$-test. See Section 3.3 for details.

Only two cases meet the criterion $0.2 < F < 0.8$ (ID 24 and 39, highlighted in blue). Fragments returning test statistics with $p$-values less than 0.05 are indicated by the dotted circles. Only two fragments are identified from this analysis: the benzothiazole fragment (for the case of ID 24) and the iodine atom fragment attached to the benzothiazole (ID 39). The direct implication of this result is that the training data lacks sufficient structural variation around the benzothiazole region are to achieve a model with absolute robustness. An indirect implication is that the benzothiazole regions may play a special biochemical role during cardiomyocyte differentiation.

| ID | Label | Prediction | F | Structure |
|---|---|---|---|---|
| 1 | 1 | 1 | 0.48 | 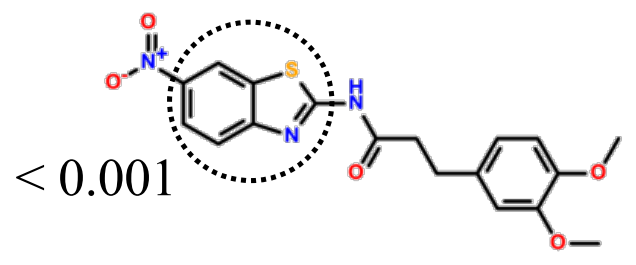  |
| 4 | 1 | 1 | 0.56 | 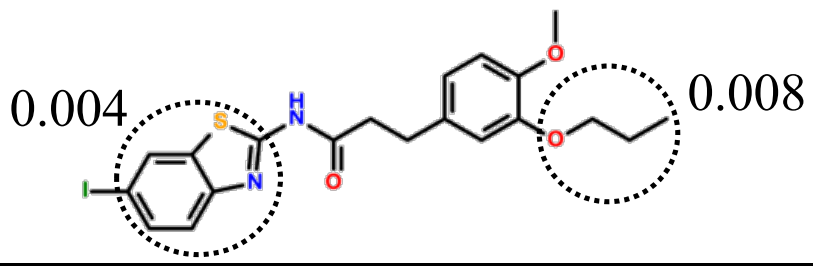  |
| 6 | 1 | 1 | 0.00 | 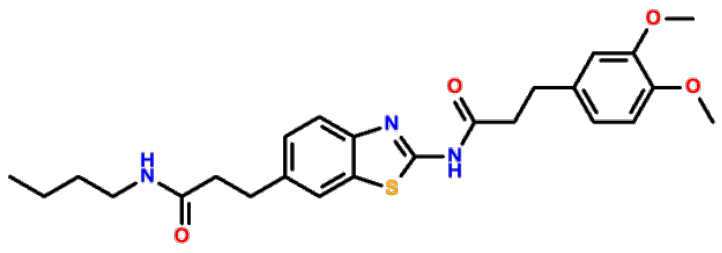 |
| 17 | 1 | 0 | 0.00 | 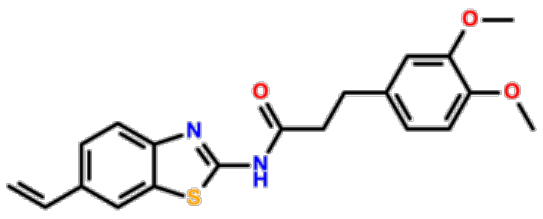 |
| 24 | 1 | 1 | 0.84 | 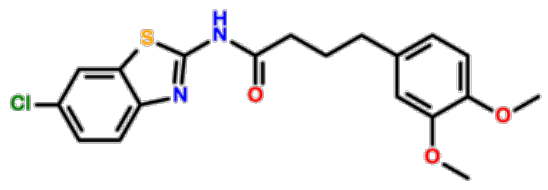 |
| 37 | 1 | 1 | 0.60 | 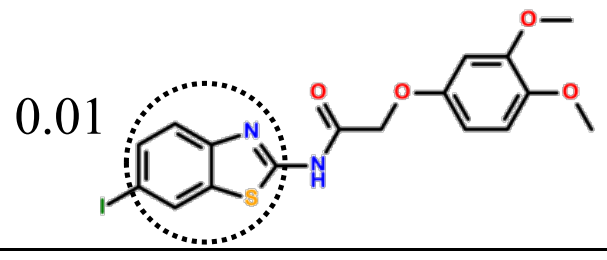  |
| 39 | 1 | 1 | 0.08 | 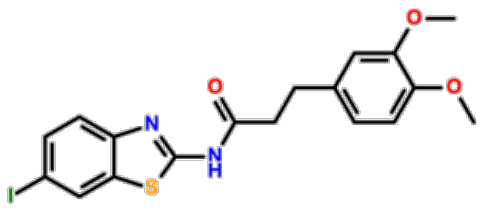 |
| 65 | 1 | 1 | 0.68 | 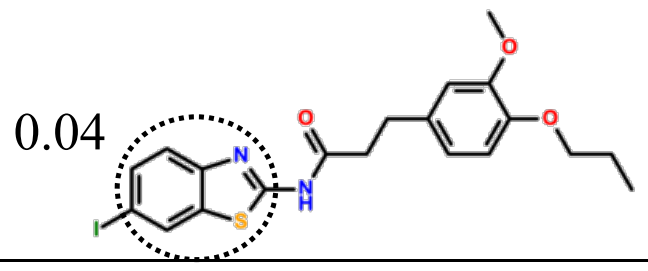  |
| 73 | 1 | 0 | 0.00 | 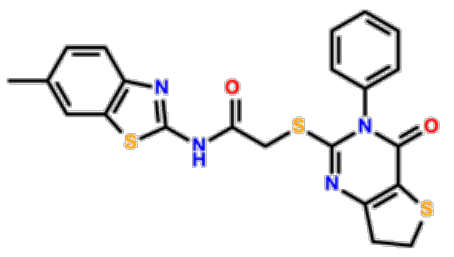 |
| 75 | 1 | 1 | 0.00 | 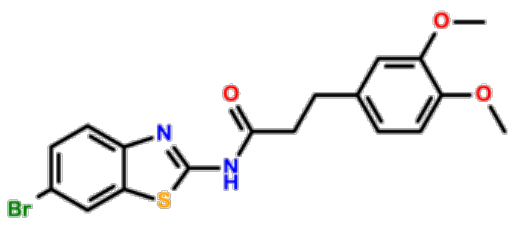 |

**Table 2.** As for Table 1, but with the neural network model used for prediction.

The results of the RSA applied to the case of the NN model trained using decorated shape descriptors are shown in Table 2. The results are quite different: model predictions were insensitive for only three of the molecules tested (IDs 17, 24, and 73). An *F* value of 0.00 was observed for molecules 17 and 73, where the unperturbed model prediction is also 0. Thus, none of the applied perturbations resulted in a change in model predictions for these cases. An *F* value of 0.84 was observed for molecule 24, where the model prediction is 1. This means that only 16 % of the perturbations applied to this case resulted in a change in model prediction. Of the remaining seven molecules tested, model predictions were severely sensitive for three cases (IDs 6, 39, and 75, bold red text). Unperturbed model predictions for these cases were all 1, however *F* values between 0.00 and 0.08 were observed, indicating that 92 – 100 % of the applied perturbations resulted in a change model prediction for these cases. The remaining four cases (IDs 1, 4, 37, and 65, bold blue text) all had unperturbed model predictions of 1 and *F* values between 0.48 – 0.68, indicating that around 30 – 50 % of the applied perturbations resulted in a change in model predictions for these cases. Thus, of the 10 cases examined, we observe three strongly sensitive cases and four moderately sensitive cases. While clearly more reliable than NN model predictions obtained using ordinary shape descriptors, this NN model is clearly more overtrained than the LR model above.

The four moderately sensitive cases described above all meet the criterion $0.2 < F < 0.8$. For these molecules, the fragments associated with statistically significant changes in model predictions are again identified by the dotted circles. Again, the benzothiazole fragment features prominently, reinforcing the conclusion that the training data does not contain sufficient variation at this important region of the model. For the case of molecule 4, a propanol fragment at the opposite end of the molecule is also identified, suggesting that more structural variation at the aryl-region of the molecules would also be helpful in order to train a neural network model with higher robustness.

### *3.4.Molecular design and experimental verification*

In order to design a molecule that is straightforward to synthesize, we limited our molecular design algorithm to the case where X is a halide atom, Y and Z are short alcohol fragments, and the linker fragment is one of several simple types (see Supporting Information 4). Permuting and combining all such fragments in our dataset, we generated 9 new compounds outside of our original training set. Furthermore, we confirmed that these molecules have never been published or patented before using the structure search feature in ChemSpider [45]. The compounds are listed in Table 3.

Due to resource constraints, we selected only one compound for synthesis and testing. This selection was made using two criteria: (i) whether the compound was predicted to be a positive hit on the basis of both regressions, using a strict decision boundary of 0.7 for both the LR and NN models obtained above, and (ii) whether the compound locates close to clusters *a* or *b* in the tSNE plot of the training data. These criteria serve as precautions against false-positive predictions due to overtraining. In particular, step (ii) ensures that the new molecule lies near the training domain of the models, especially the part where the models perform well. The coordinates of the new molecule in the tSNE plot were predicted using linear regression models trained on the molecule feature vectors (see Supporting Information 9). The model predictions and tSNE coordinates of these new molecules are shown in Figure 5. Of the 9 molecules, only 5 of them pass criterion (i). Furthermore, only one of them (Molecule 5, M5) clearly satisfies criterion (ii). M5 was therefore selected for synthesis and experimental testing.

| ID | Structure | LR prediction | NN prediction |
|---|---|---|---|
| 1 | 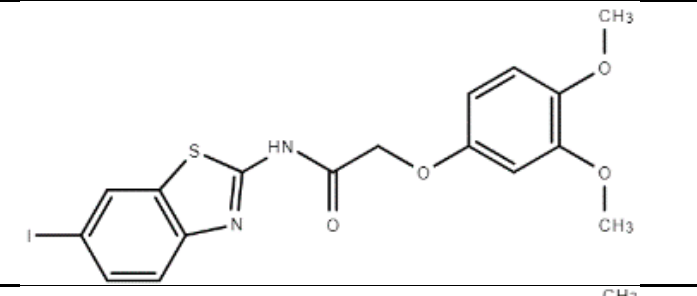 | 0.57 | 0.16 |
| 2 | 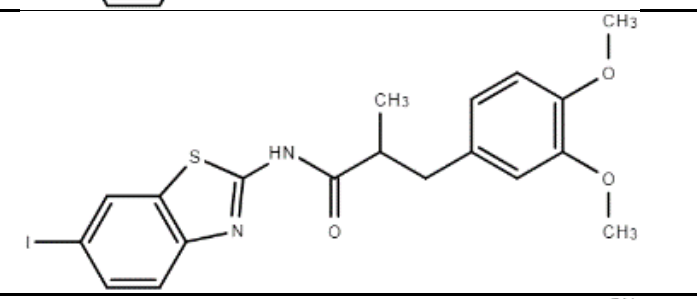 | 0.67 | 0.74 |
| 4 | 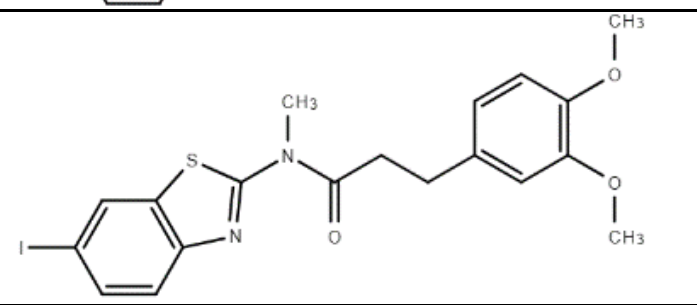 | 0.63 | 0.21 |
| 5 | 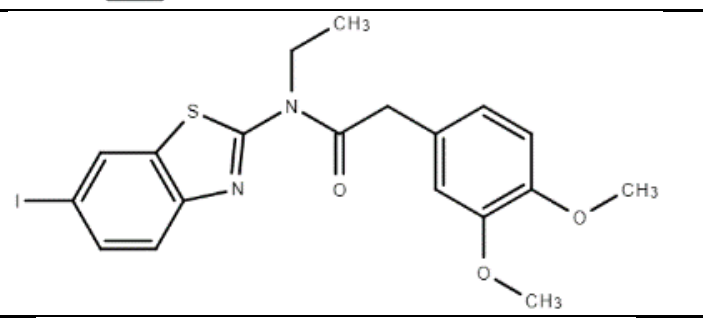 | 0.75 | 0.92 |
| 6 | 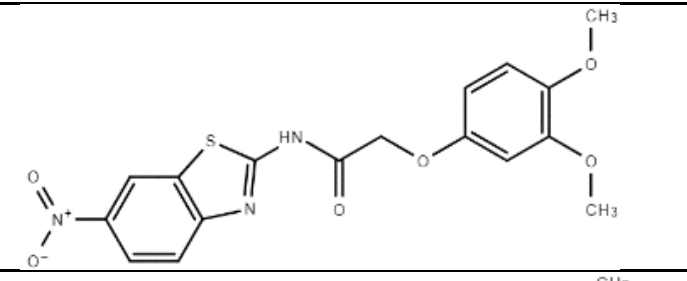 | 0.74 | 0.99 |
| 7 | 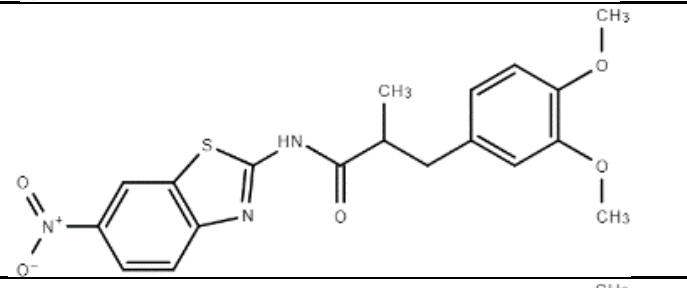 | 0.70 | 0.94 |
| 8 | 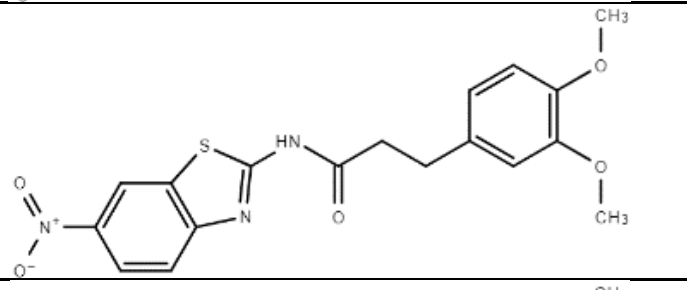 | 0.59 | 0.37 |
| 9 | 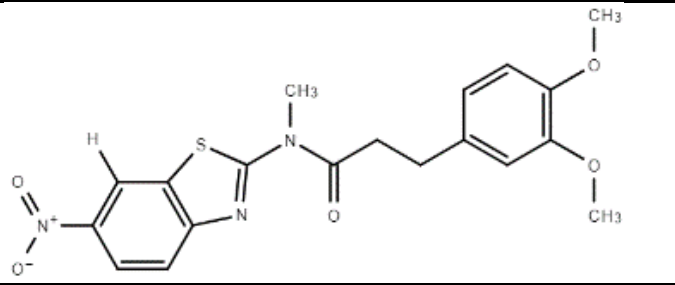 | 0.81 | > 0.99 |
| 10 | 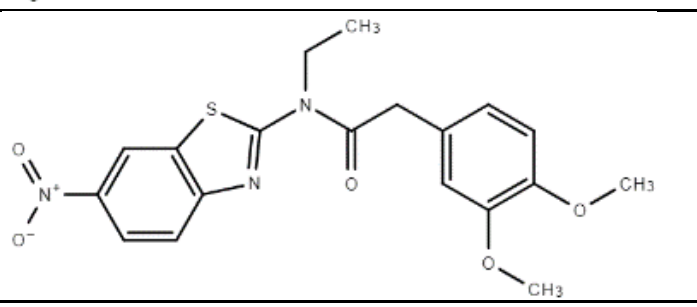 | 0.89 | > 0.99 |

**Table 3.** Compounds obtained with our molecular design protocol. "LR prediction" refers to the prediction of the logistic regression (LR) model. "NN prediction" refers to the prediction of the neural network model.

M5 was synthesized as described in section 2.6. verified by nuclear magnetic resonance (NMR) spectroscopy. To test whether M5 can be used as during the Wnt inhibition step for cardiac cell differentiation, we used the iPS cell differentiation protocol described in Section 2.6. The protocol was optimized for our experimental condition and the human iPS cell line XXX253G1 by using the compounds XAV939, IWP2, and KY02111, which are known effective Wnt inhibitors of cardiomyocyte differentiation [5-7] (Supporting information 10). We then performed a real-time quantitative polymerase chain reaction (RT-qPCR) assay on the cells treated with three different concentrations of M5 to assess the effect on the endogenous expression level of well-established cardiac marker genes NKX2-5, TNNT2, and MYH7 [8]. The expression profile of these genes was compared to two controls: a negative control using the solvent dimethyl sulfoxide (DMSO), and a positive control using XAV939, an established Wnt inhibitor for cardiac differentiation. At a low concentration (2 µM), the compound M5 achieved similar expression levels of the marker genes NKX2-5, TTNT2, and MYH7 as the positive control (Figure 6). The effect was less significant at higher concentrations, with 10 µM M5 being comparable to the negative control. Immunostaining of the iPS cells at day 8 of cardiac differentiation shows TNNT2 signals detected in the M5-treated cells (Figure 7). These results suggest that M5 could be used during the Wnt inhibition step during cardiac differentiation, consistent with the predictions of the LR model.

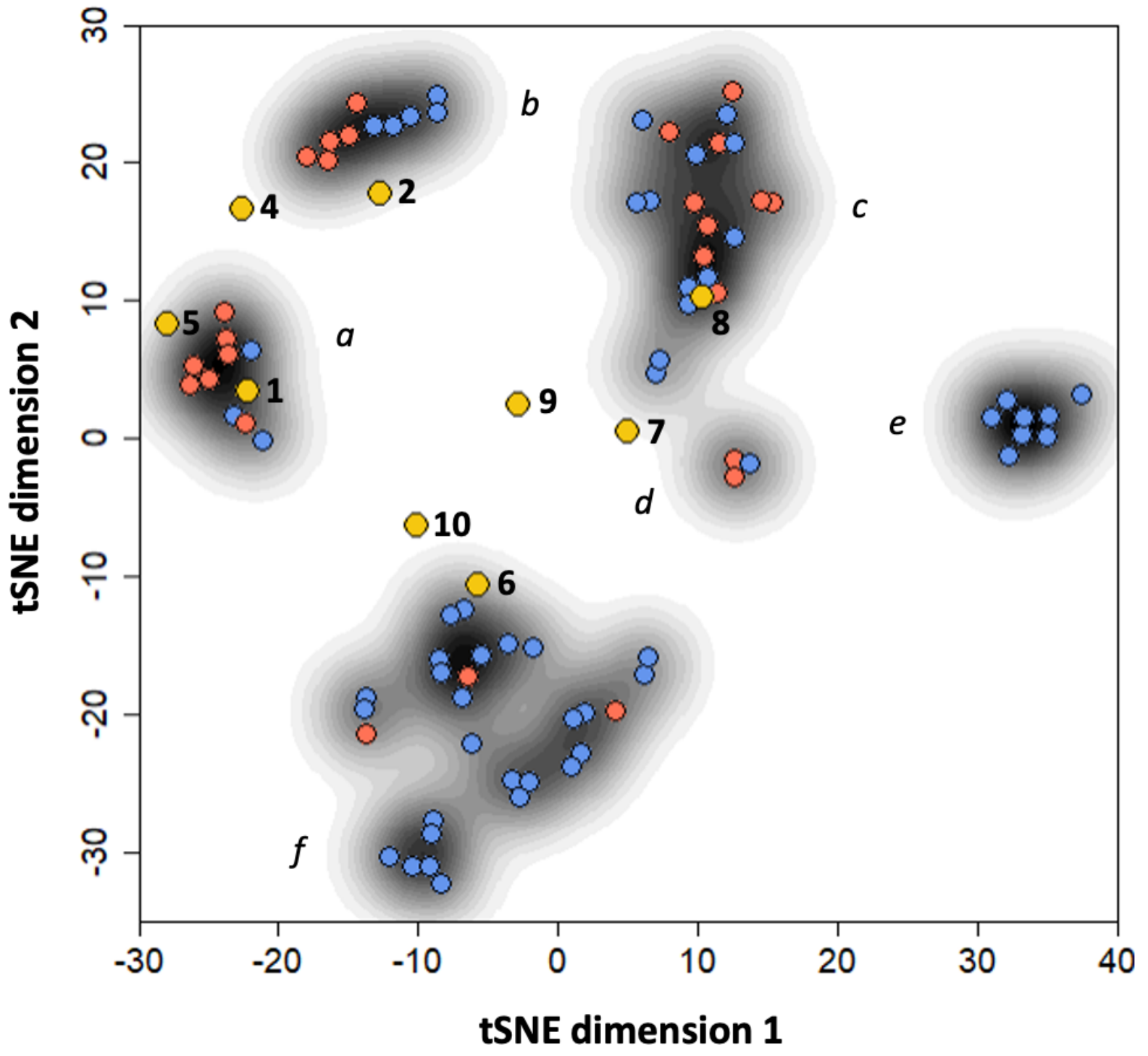

**Figure 5.** As for Figure 3, but with the locations of the predicted molecules included (yellow points). See Section 3.4 for details.

## 4. Discussion

Several directions for future work are suggested. Like ordinary shape descriptors, decorated shape descriptors can require considerable computational time. This is because the integrals in equation (5) need to be computed numerically over a fine, three-dimensional grid. In this work, these integrals were computed directly, and moreover they were computed for every coefficient in the expansion in equation up to $n$ = 25, which required a few hours in total on our hardware. While undesirable, this computational burden was tolerable for our work, which only required computing Zernike moments for a small number of molecules. It is well-known that Zernike moments can be time-consuming to calculate by direct integration, however several authors have reported accelerations using special algorithms (albeit for the cases of 2D Zernike polynomials; see [46 - 48]). The further development of such algorithms, as well as modern hardware such as graphics processing units, is essential to ensure the widespread adoption of these methods. The need to compute hydrophilicities from molecular dynamics simulation is another time-consuming point, requiring nanosecond-long molecular dynamics trajectories with explicit inclusion of solvent molecules. However, these calculations can be performed prior to the actual computation of a decorated shape descriptor, and the results are stored in a look-up table. Thus, they do not directly contribute to the calculation time of a decorated shape descriptor. Molecular dynamics simulations can be performed quite quickly using modern hardware and codes and could be more widely explored for building descriptors in future research in the chemical biology field.

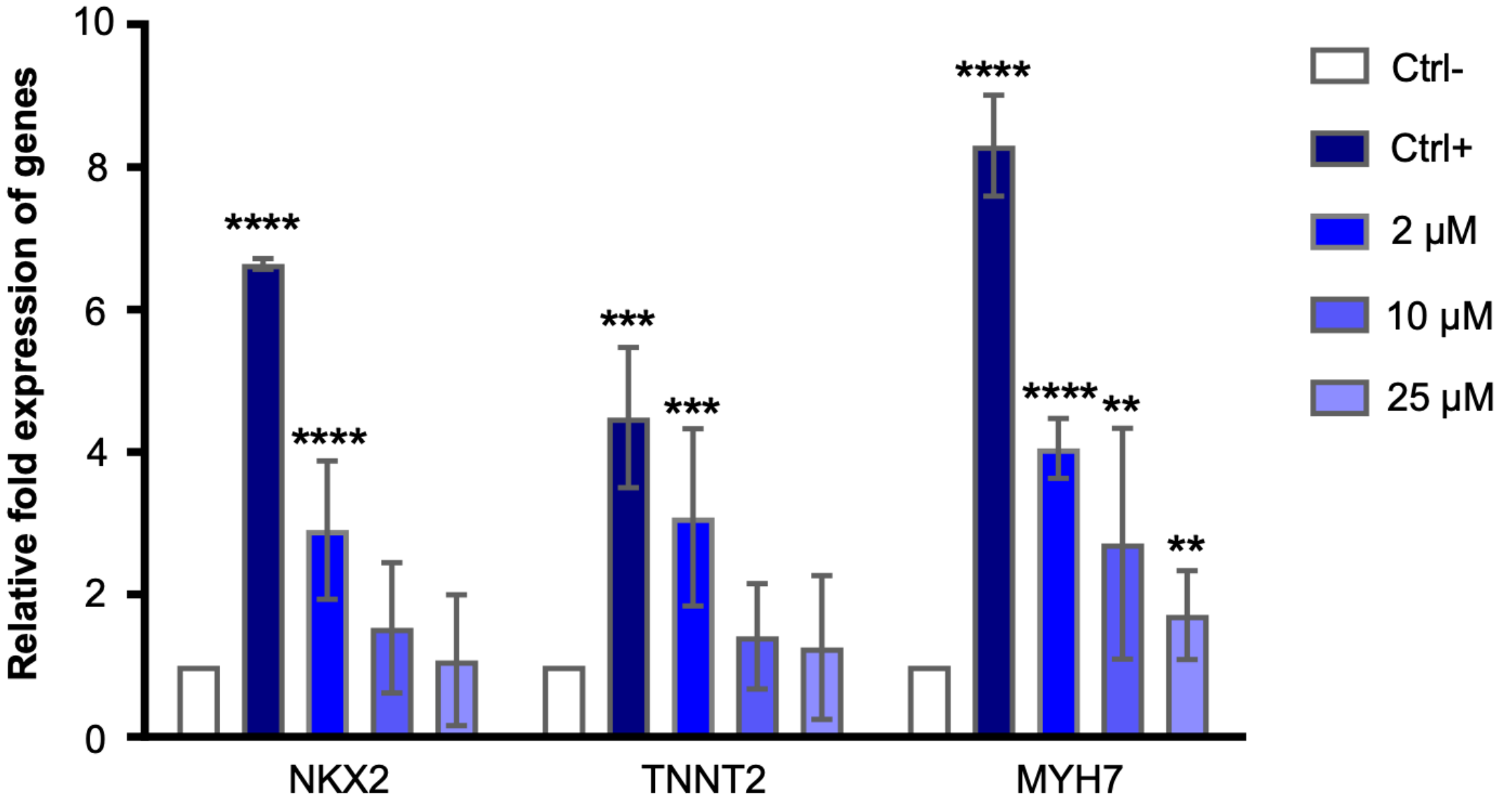

**Figure 6.** Comparison of gene expression profiles associated with cardiomyocyte differentiation on day 8 of differentiation between a positive control, negative control, and M5 when used during the Wnt inhibition step. NKX2, TNNT2, and MYH7 are cardiomyocyte marker genes. The vertical axis measures gene expression relative to the negative control. The error bars correspond to the means ± standard deviation (sample size = 4). Asterisks indicate significance as determined from one-way ANOVA test comparing to the negative control (**** $p < 0.001$, *** $p < 0.01$, ** $p < 0.1$)

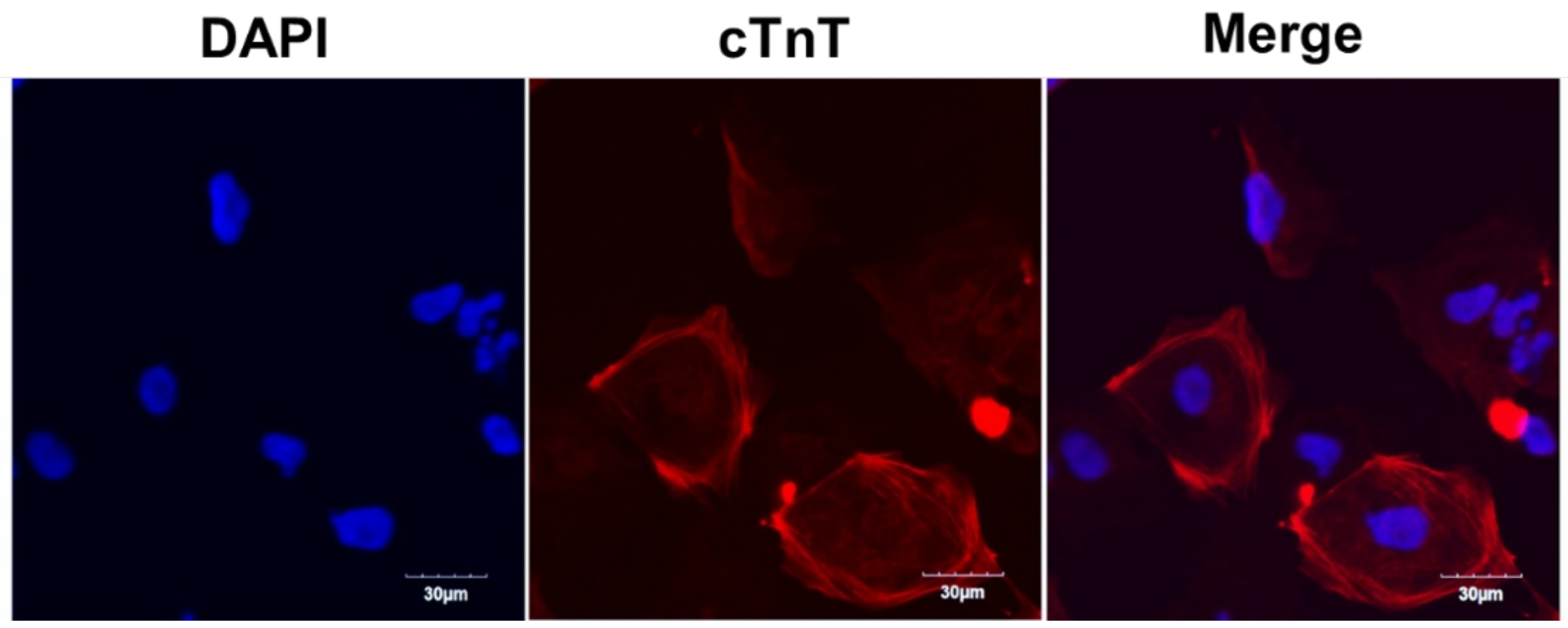

**Figure 7.** Immunostaining analysis of iPS cells for cTnT expression on day 8 of differentiation to cardiomyocytes using the M5 molecule for the Wnt inhibition step. Images were obtained by optical microscopy. Bars correspond to 30 μm.

Another point to consider in future research is why decorated shape descriptors are more effective than non-decorated ones. It is obvious that decorated shape descriptors carry more chemical information into the regression model than do ordinary ones, and moreover a molecule's hydrophilicity distribution is known to be decisive when it interacts with cellular proteins. However, these facts alone provide little guidance for carrying the strategy further. A mathematical theorem that explains how such decorations (hydrophilicities or otherwise) enrich a feature representation would be desirable, perhaps in the form of a statement about an effective reduction in training data requirements. The possibility of using other types of decorations could be considered as well. For example, the conformational flexibility of a molecule is also known to be important when interacting with cellular proteins. Could such kinetic (or entropic) information be used as a basis for future shape decorations as well? This would be an interesting topic for future research.

'Model overtraining' is a widely used and somewhat vague term. It is generally understood as a model being excessively sensitive to small perturbations in the input. From this point of view, decorated shape descriptors are useful for diagnosing overtraining, as they can be perturbed in a straightforward manner. The same cannot be said for ordinary shape descriptors and many other types of molecular feature representations. However, our random sensitivity analysis would benefit from further mathematical elaboration. How many molecules should be tested before one can confidently claim that the model is not overtrained? Moreover, is Latin hypercube sampling an efficient way to generate a sample of perturbed molecular descriptors, or is another method more appropriate? Future work could clarify these points, and perhaps develop a formal hypothesis test for diagnosing model over training with this method.

Our molecular design method used a conservative strategy in which candidate molecules were created by combining molecular fragments from the training data. This is sensible when designing molecules on the basis of regression model predictions, as it prevents us from designing molecules that differ wildly from those in the model's training domain, where the risk of false-positive predictions becomes large. We emphasize that this strategy is simply designed to minimize risk. It is in not necessitated by the small size of our data set. All regression models will fail to make reliable predictions when employed far from their training domains, regardless of the amount of data they were trained with. In future work, our design strategy could be broadened to include a greater element of risk, perhaps by using Bayesian search techniques coupled with evolutionary algorithms to systematically increase the range of available molecular fragments.

## 5. Conclusions

To compensate for the lack of predictive power arising from simple regression models,

the inputs into the model must be expressed using an informative set of features. This work successfully demonstrated the application of decorated shape descriptors to build regression models to predict the ability of small chemical compounds to promote cardiac differentiation of iPS cells. Our simple logistic regression model was shown to achieve superior performance compared to one developed using ordinary shape descriptors. Decorated shape descriptors also decreased the degree of overtraining in neural network models compared to ones using ordinary shape descriptors. Our models were used to design an entirely new compound capable of inducing cardiomyogenesis, which showed notable expressions of marker genes associated with cardiomyocytes when applied in an actual stem cell differentiation experiment. Overall, our work demonstrates a viable strategy for designing new compounds for stem cell differentiation protocol, and will be useful in situations where training data is limited due to the resource constraints of chemical screening.

**Acknowledgements**

This work was supported by JSPS (20KK0160 to D.P. and 22H00350 to M.U.)

**Supporting Information 1**

List of training data. 'Strength' refers to cardiac differentiation effect when used as Wnt inhibitor. Strength was measured according to fluorescence intensity (see reference [6] of the main paper). 'Label' refers to the label assigned for model training (1 for Medium Strong, and Super strong; 0 for Very weak or weak).

| **Index** | **Chemical structure** | **Strength** | **Label** |
|---|---|---|---|
| 1 | 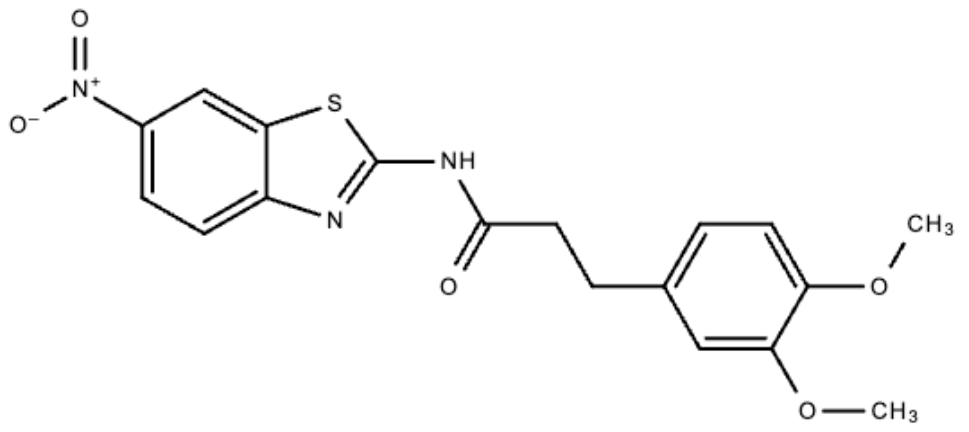 | Strong | 1 |
| 2 | 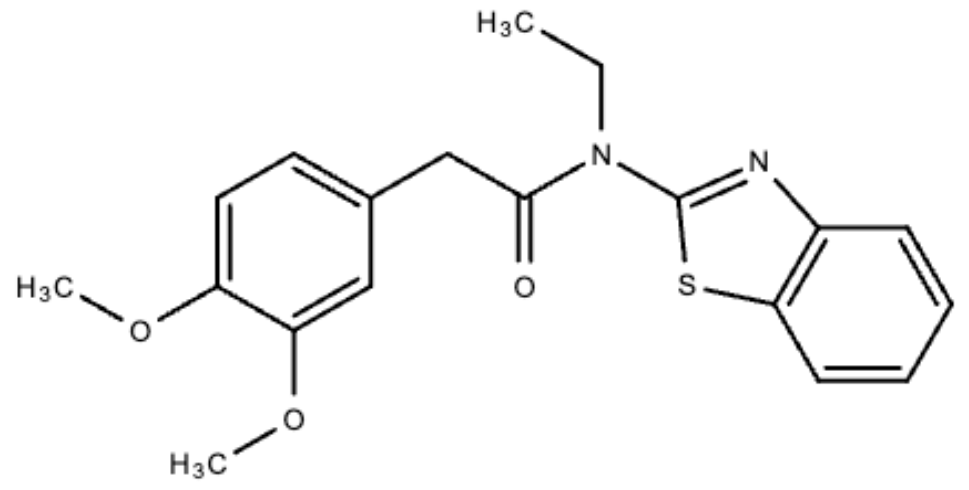 | Very weak | 0 |
| 3 | 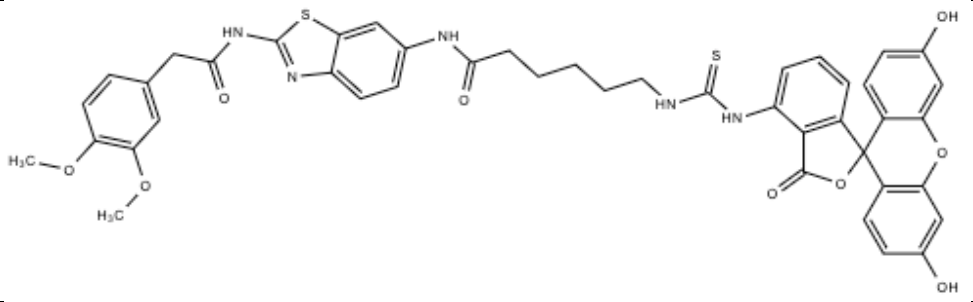 | Weak | 0 |
| 4 | 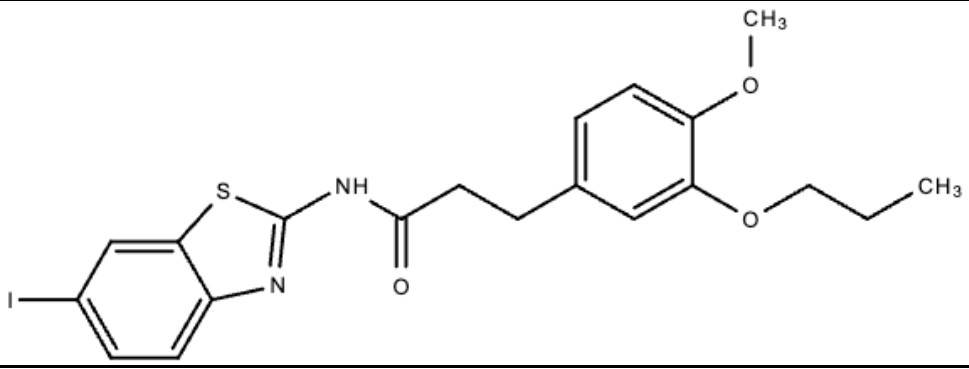 | Strong | 1 |
| 5 | 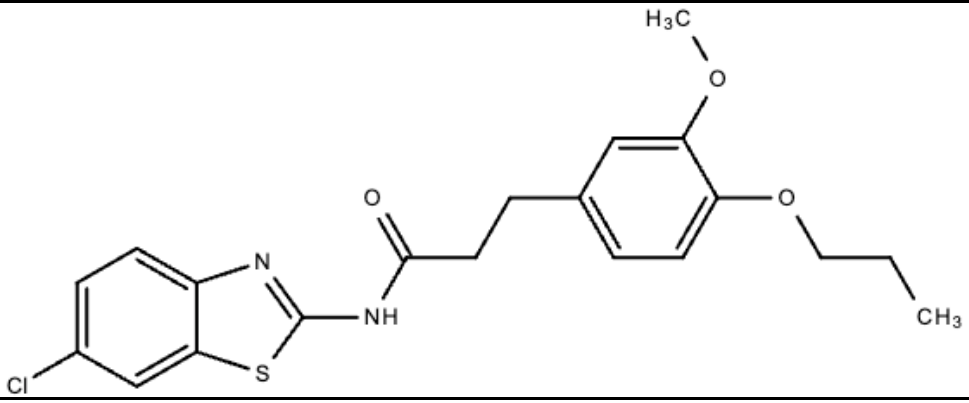 | Weak | 0 |
| 6 | 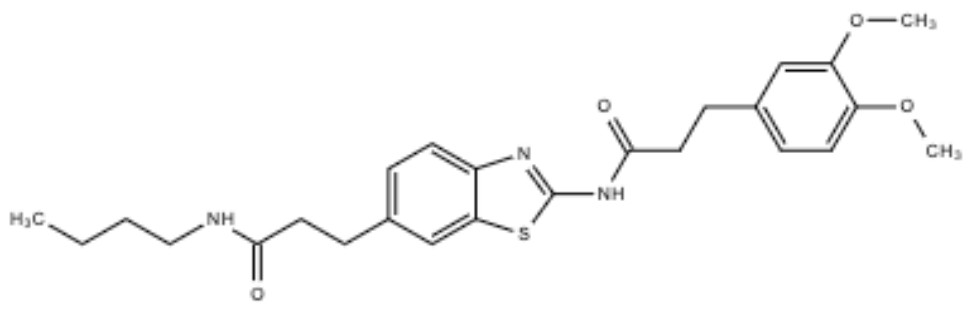 | Medium | 1 |
| 7 | 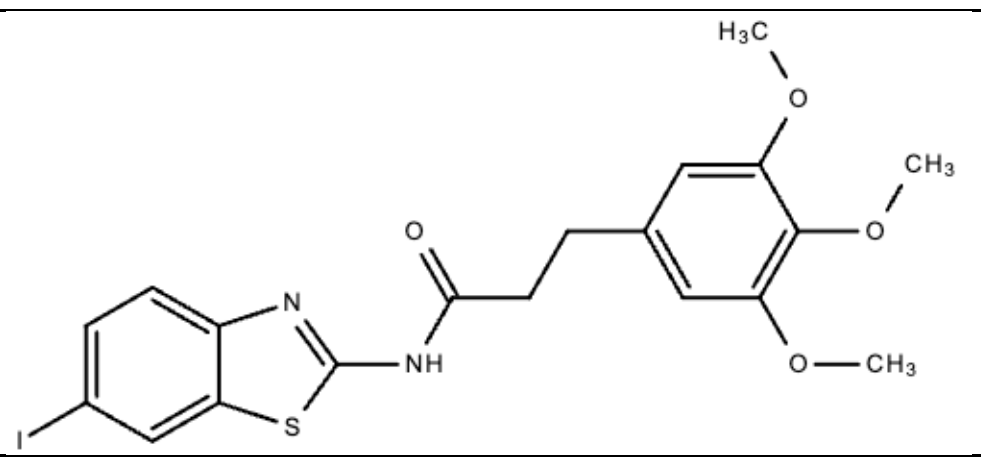 | Medium | 1 |
| 9 | 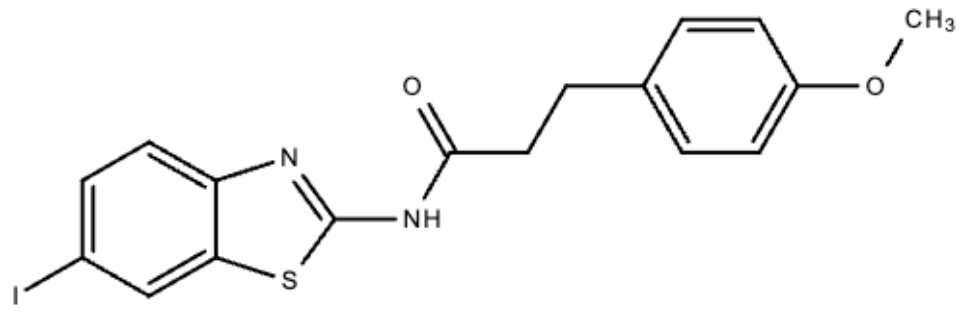 | Weak | 0 |
| 10 | 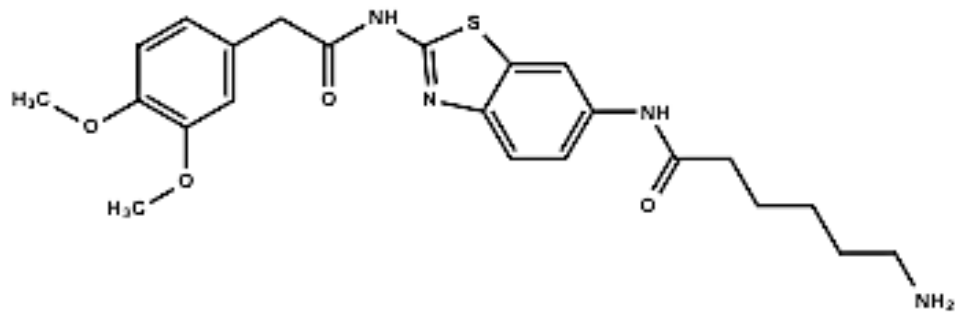 | Very weak | 0 |
| 11 | 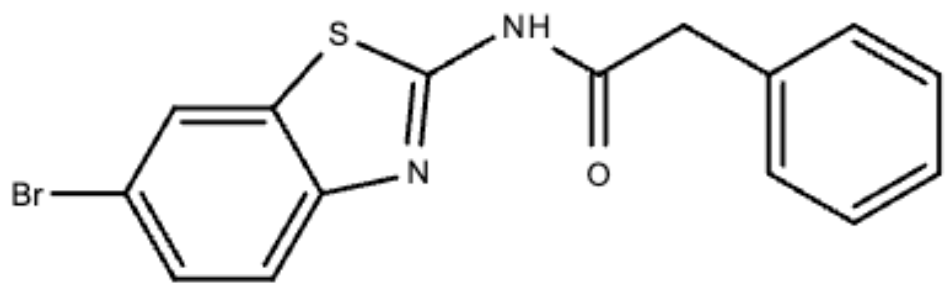 | Weak | 0 |
| 12 | 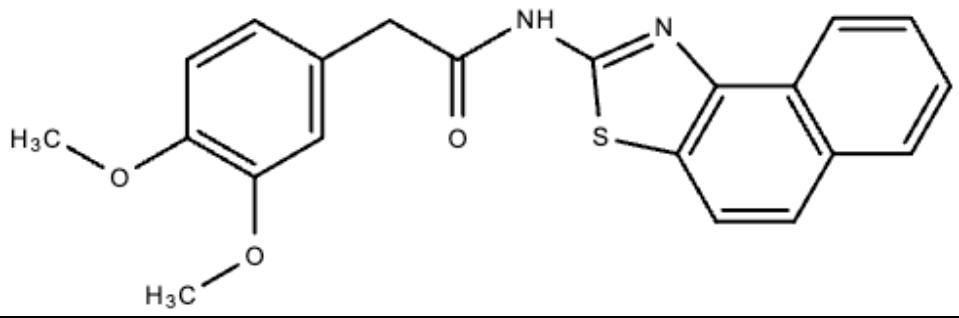 | Very weak | 0 |
| 13 | 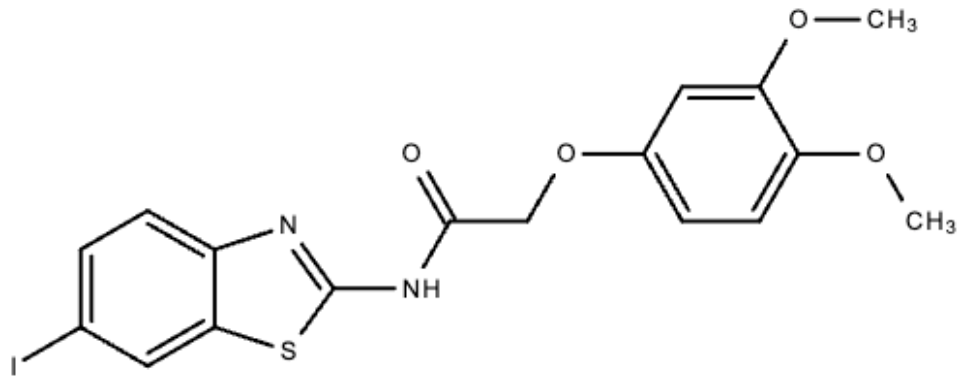 | Super strong | 1 |
| 14 | 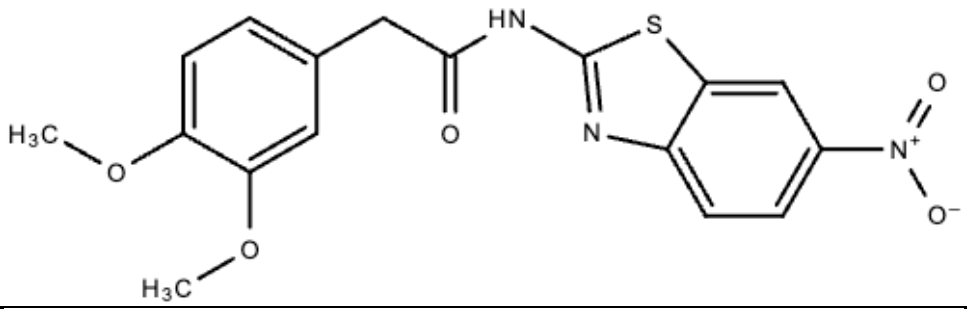 | Strong | 1 |
| 15 | 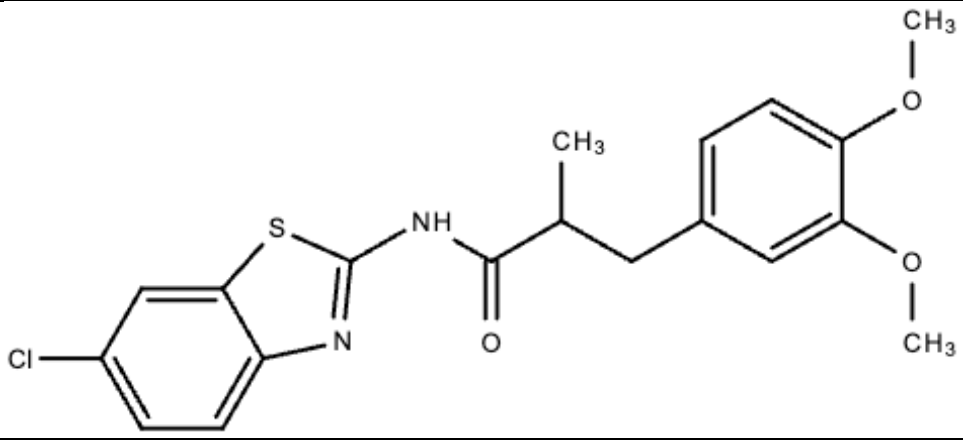 | Weak | 0 |
| 16 | 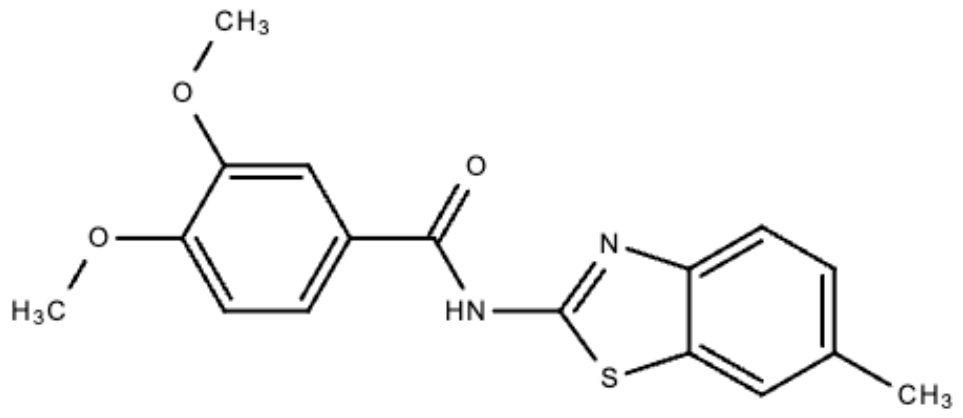 | Very weak | 0 |
| 17 | 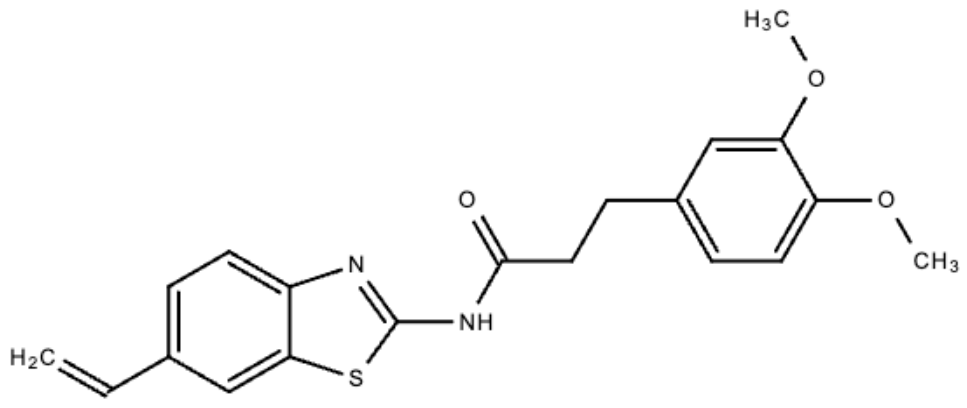 | Strong | 1 |
| 18 | 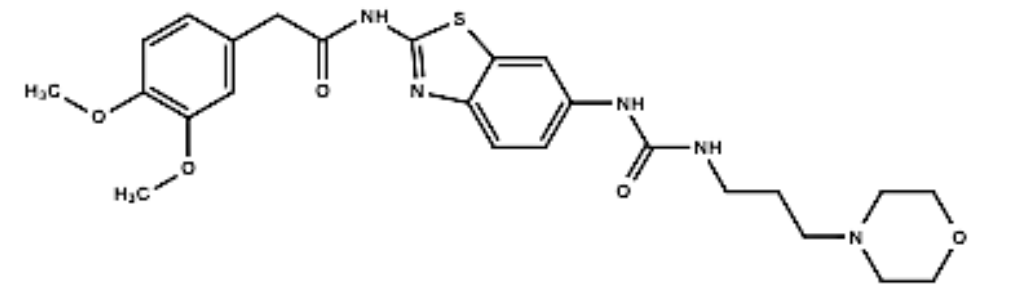 | Very weak | 0 |
| 19 | 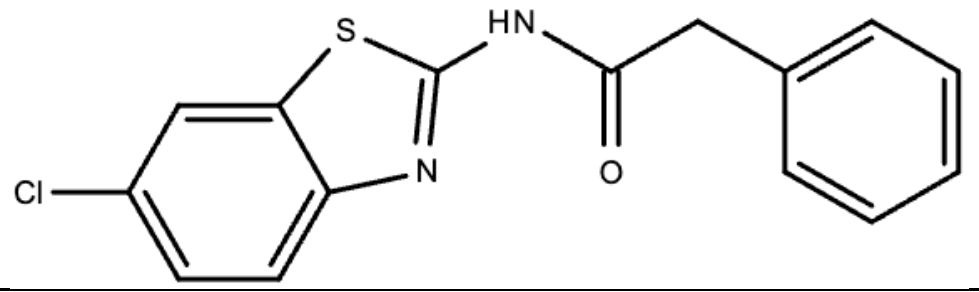 | Very weak | 0 |
| 20 | 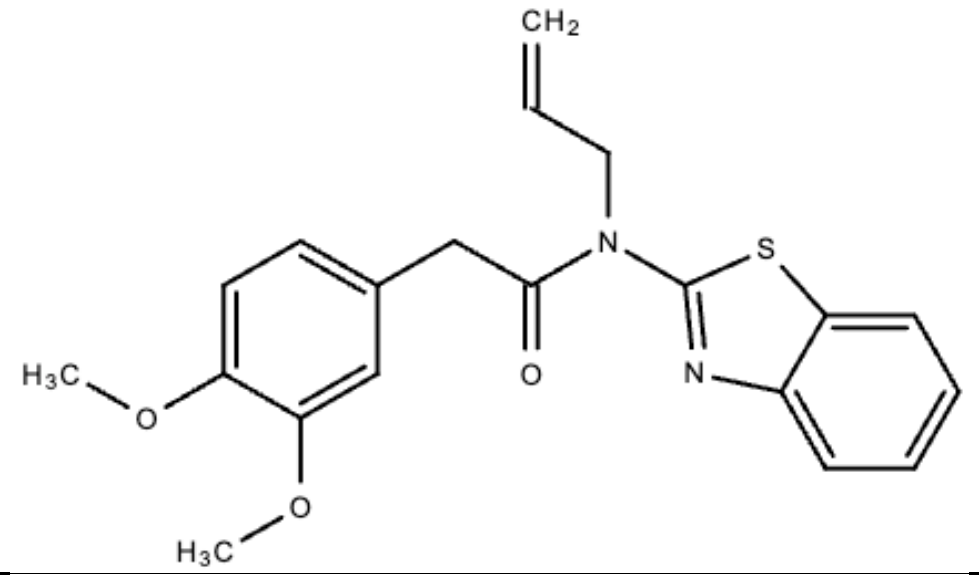 | Very weak | 0 |
| 21 | 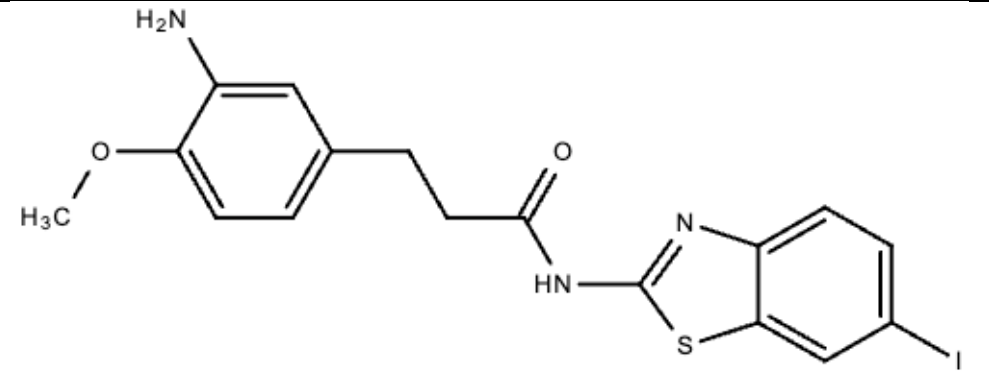 | Weak | 0 |
| 22 | 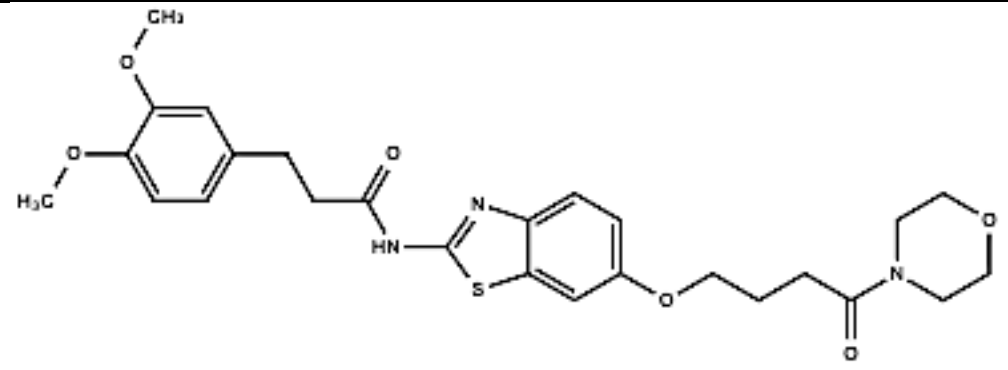 | Weak | 0 |
| 23 | 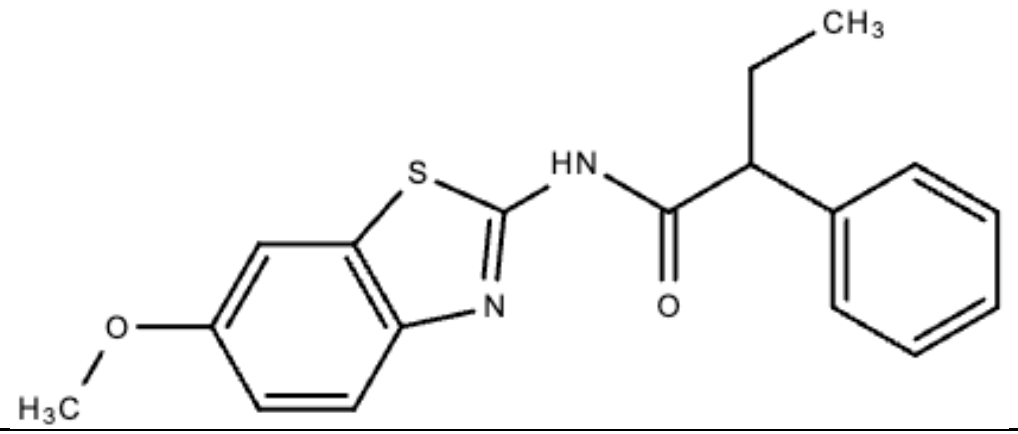 | Very weak | 0 |
| 24 | 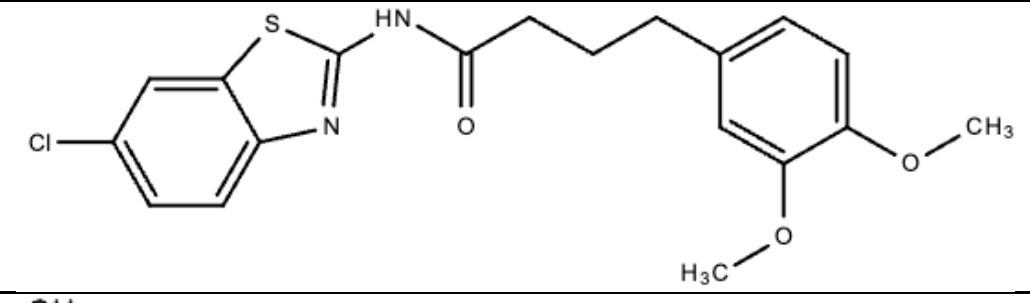 | Strong | 1 |
| 25 | 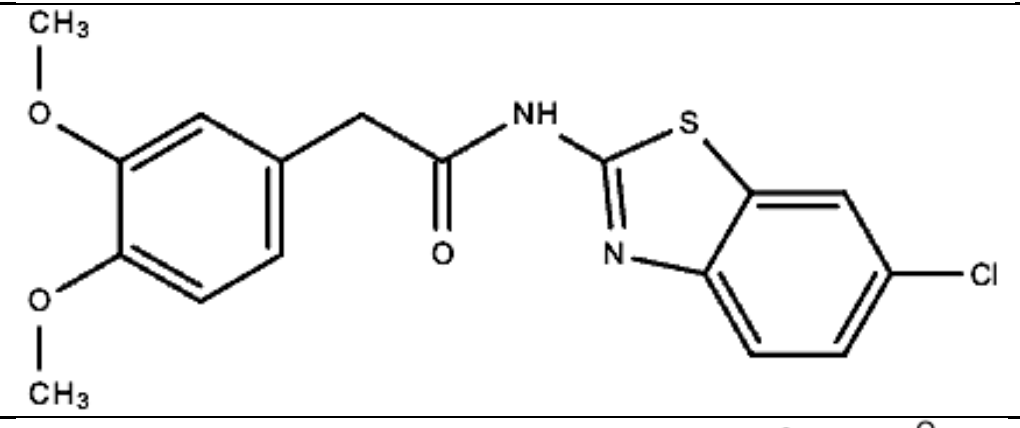 | Medium | 1 |
| 26 | 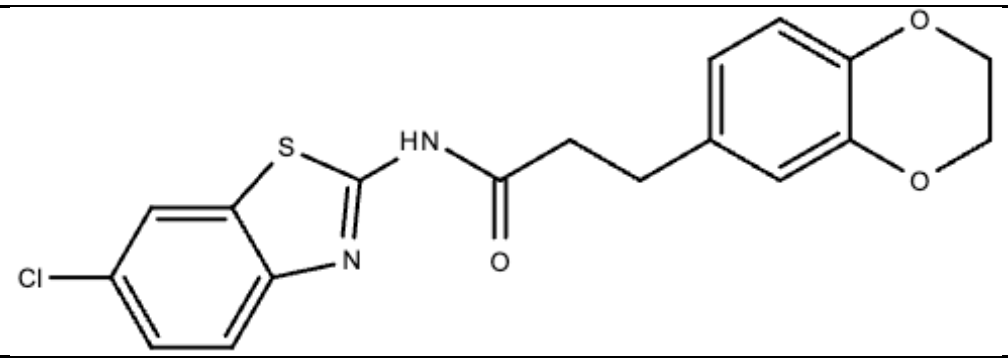 | Very weak | 0 |
| 27 | 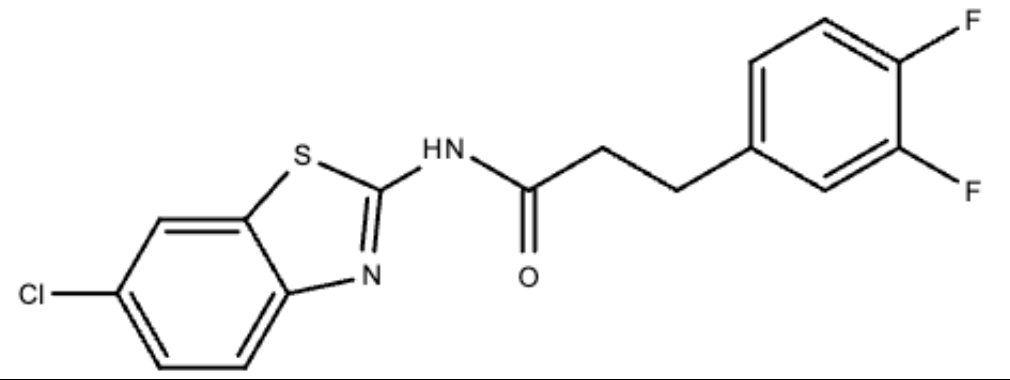 | Very weak | 0 |
| 28 | 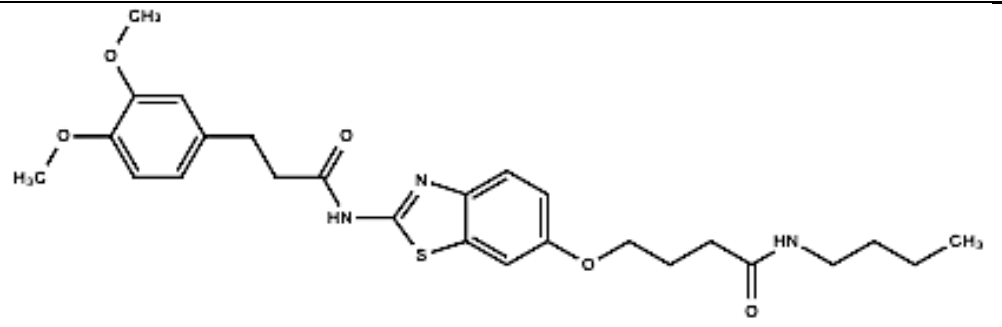 | Very weak | 0 |
| 29 | 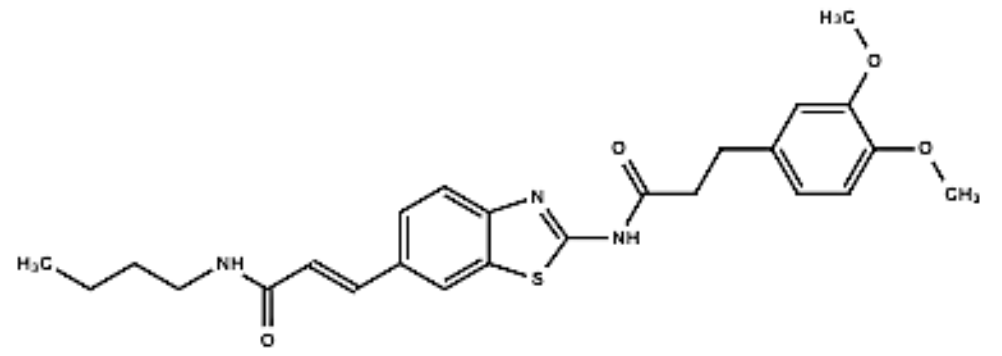 | Medium | 1 |
| 30 | 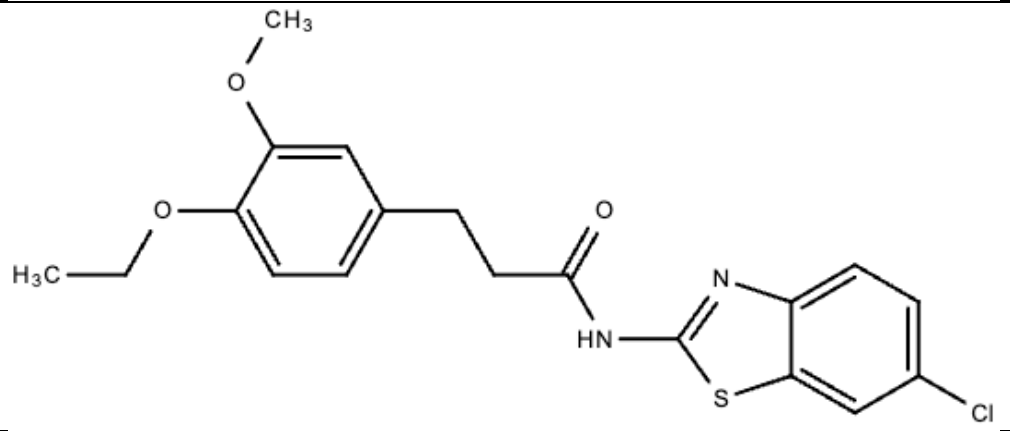 | Weak | 0 |
| 31 | 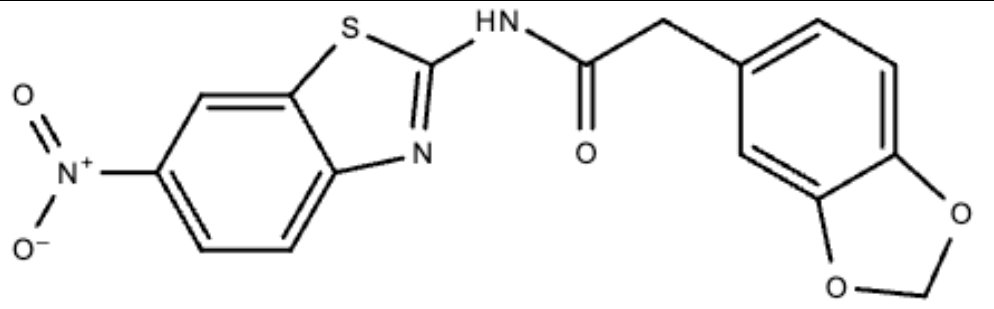 | Very weak | 0 |
| 32 | 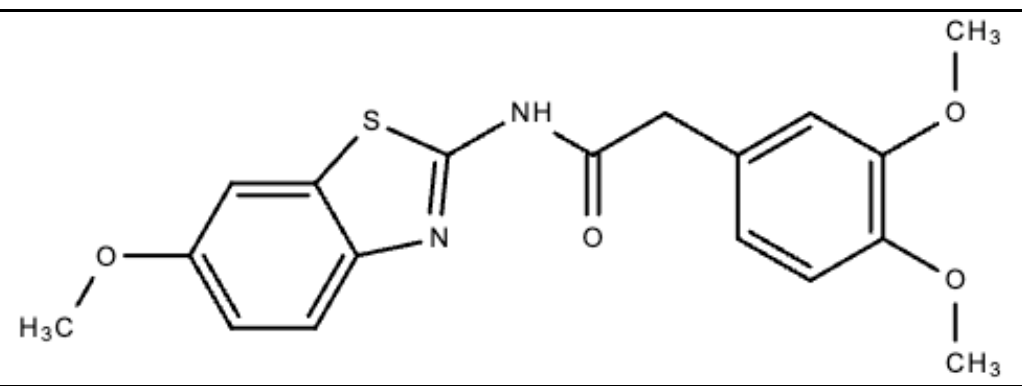 | Weak | 0 |
| 33 | 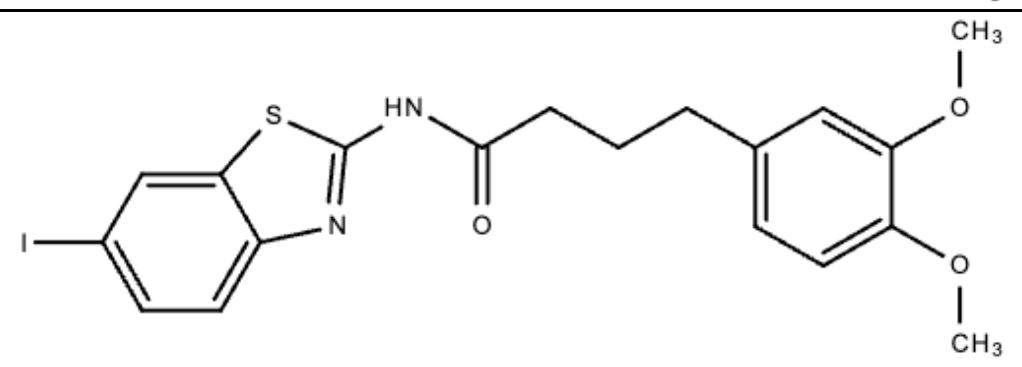 | Super strong | 1 |
| 34 | 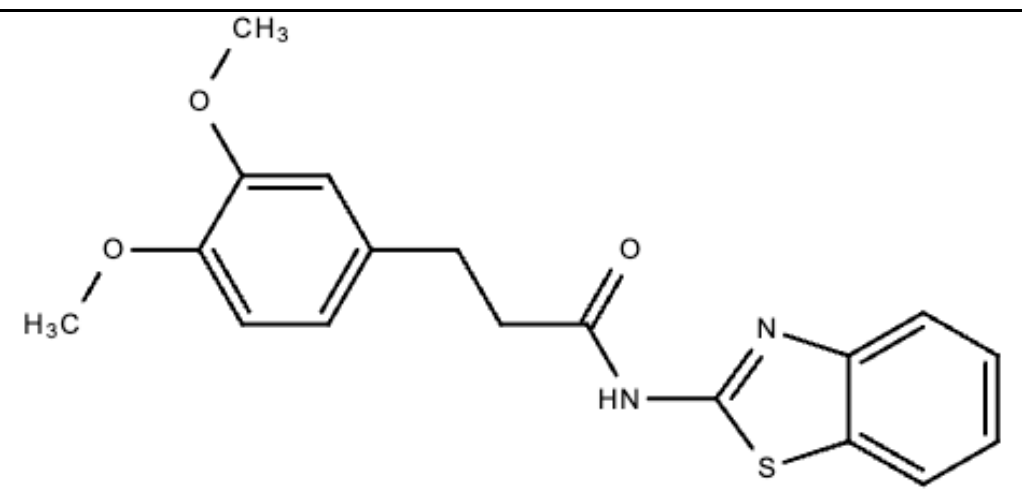 | Weak | 0 |
| 35 | 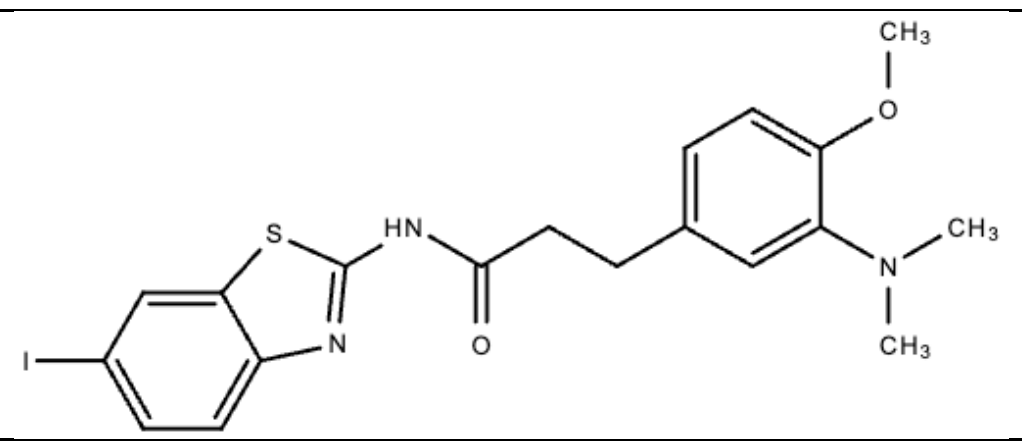 | Medium | 1 |
| 36 | 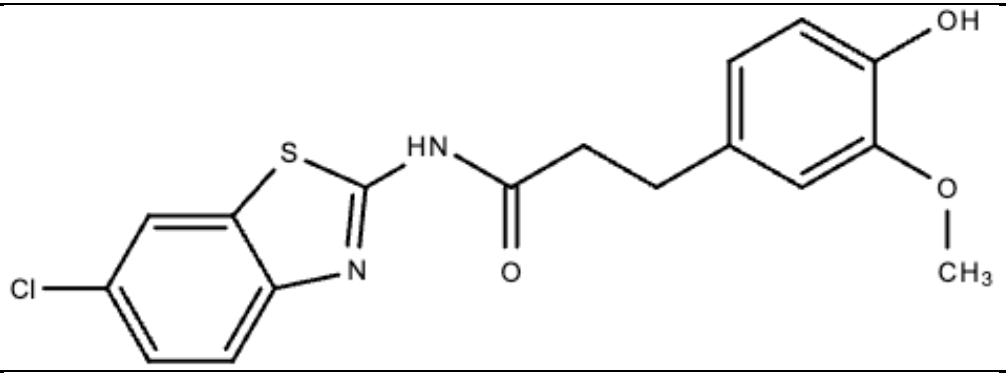 | Very weak | 0 |
| 37* | 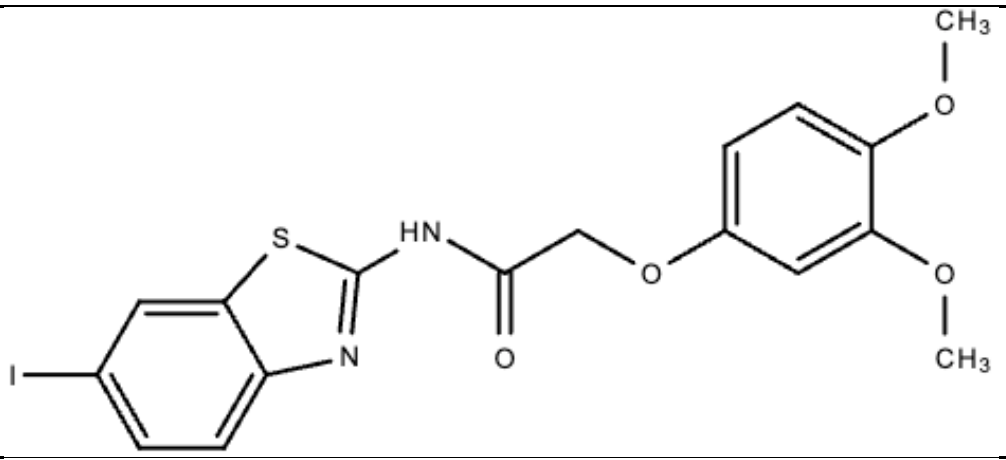 | Medium | 1 |
| 38 | 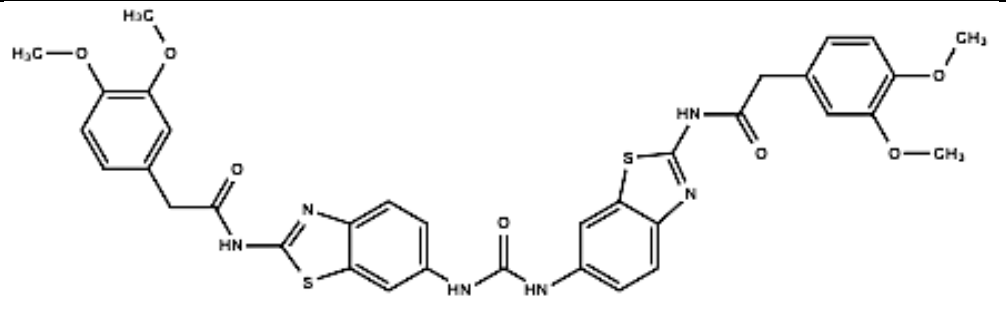 | Very weak | 0 |
| 39 | 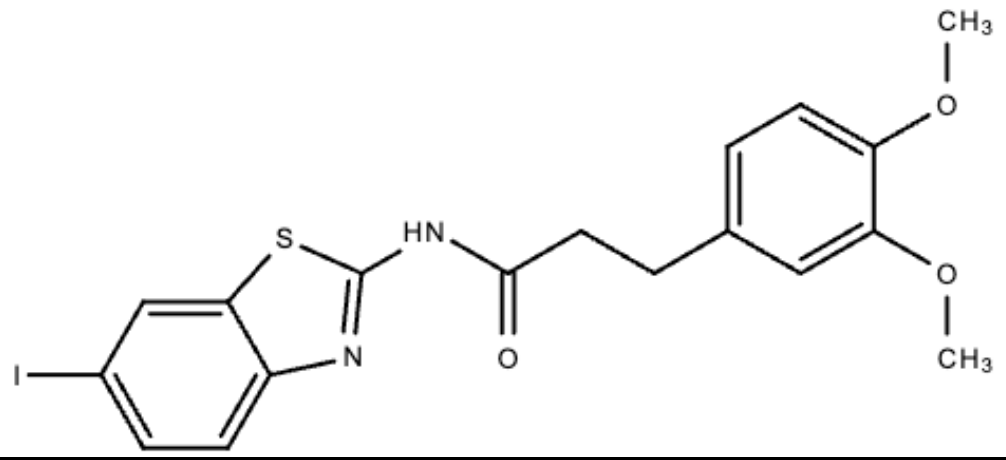 | Super strong | 1 |
| 40 | 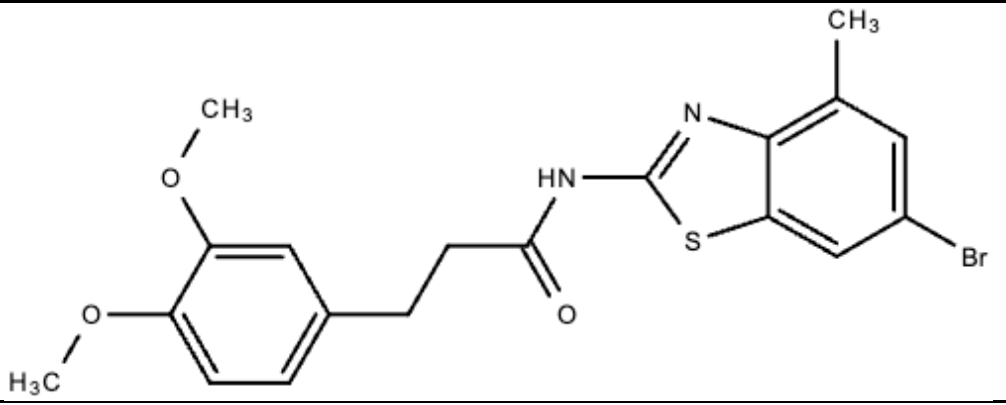 | Medium | 1 |
| 41 | 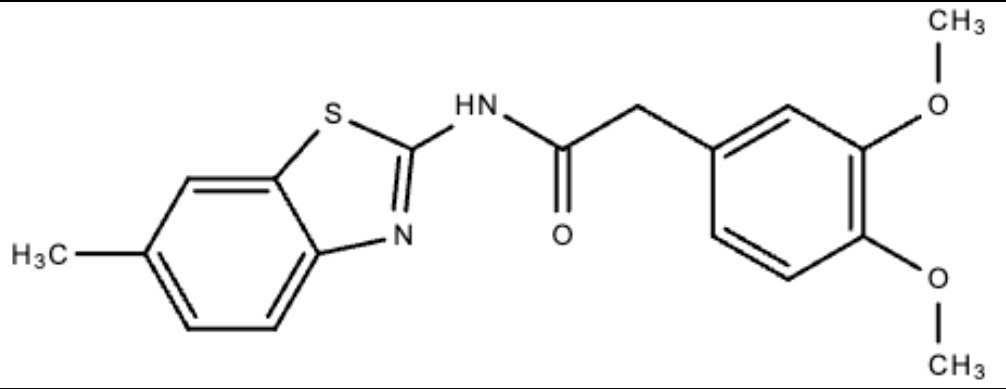 | Weak | 0 |
| 42 | 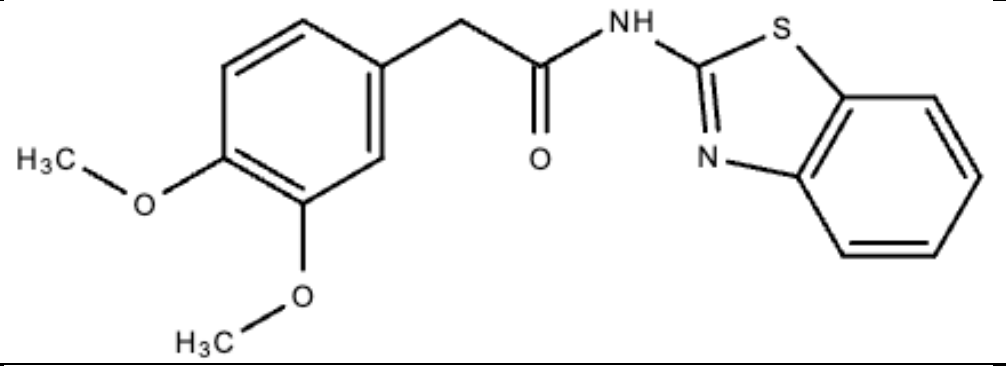 | Very weak | 0 |
| 43 | 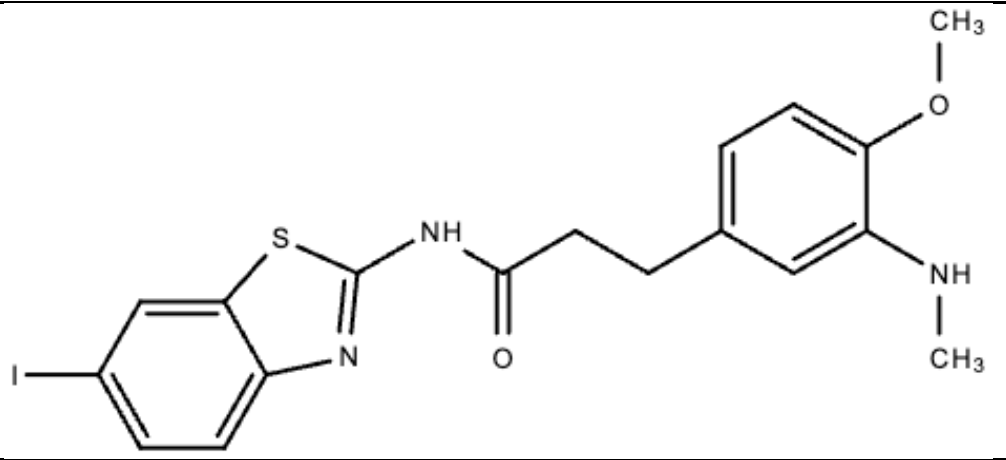 | Weak | 0 |
| 44 | 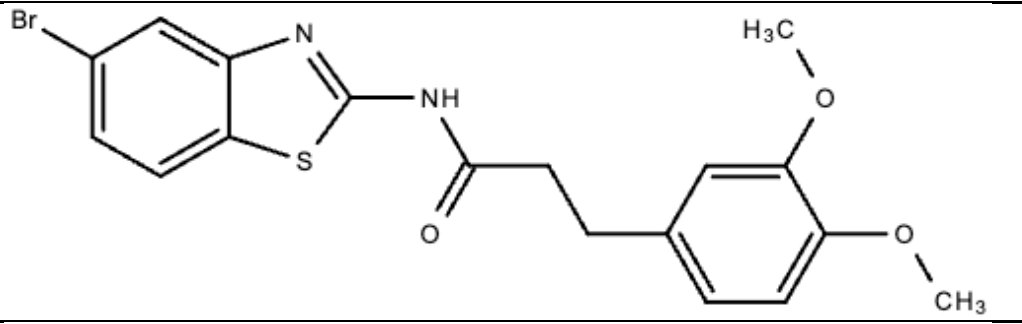 | Weak | 0 |
| 45 | 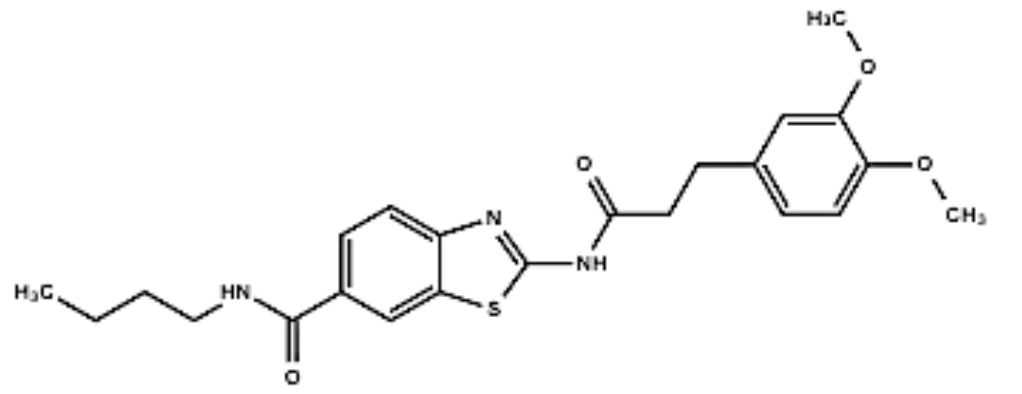 | Very weak | 0 |
| 46 | 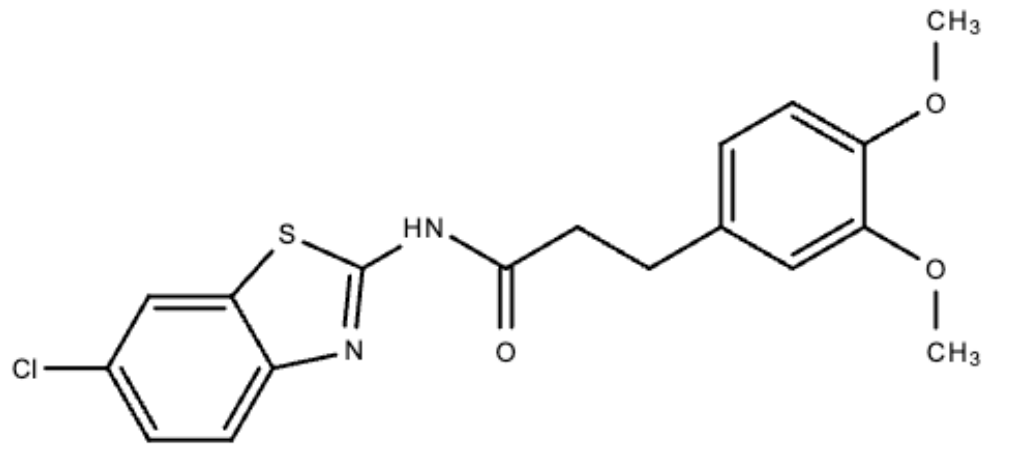 | Strong | 1 |
| 47 | 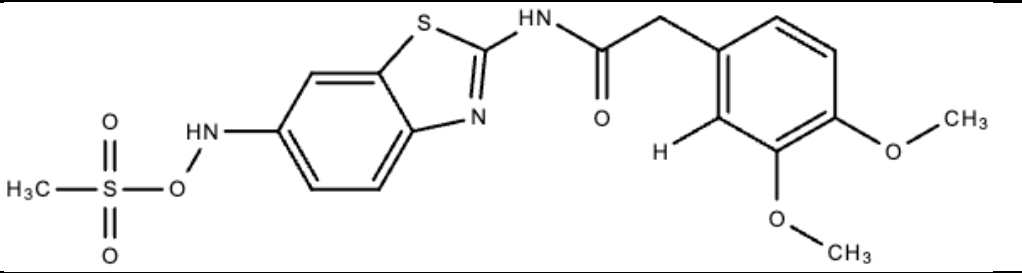 | Very weak | 0 |
| 48 | 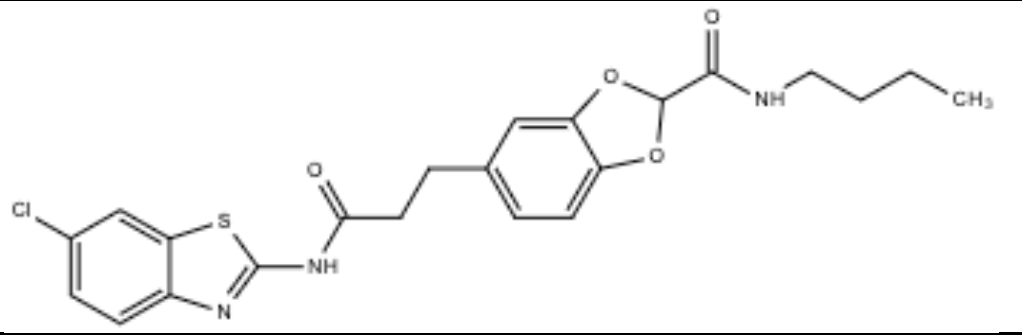 | Very weak | 0 |
| 49 | 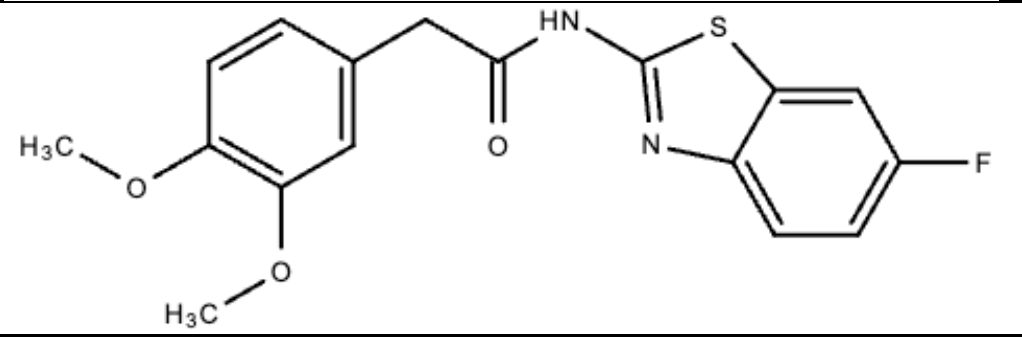 | Weak | 0 |
| 50 | 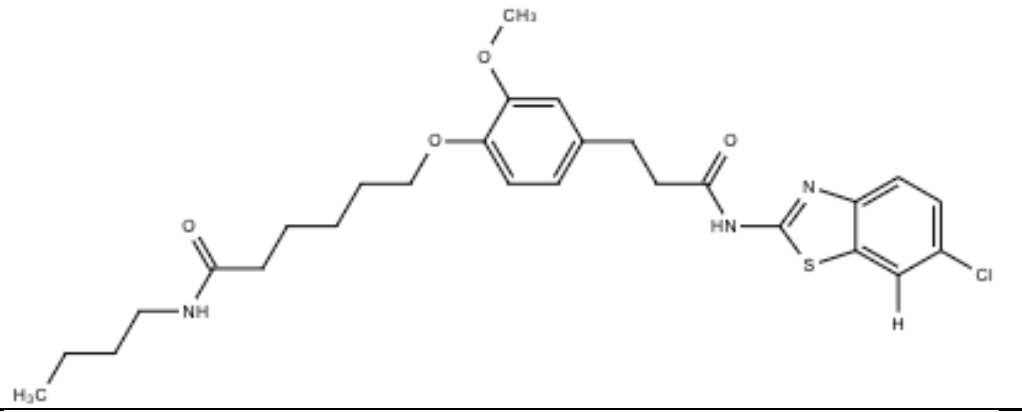 | Very weak | 0 |
| 51 | 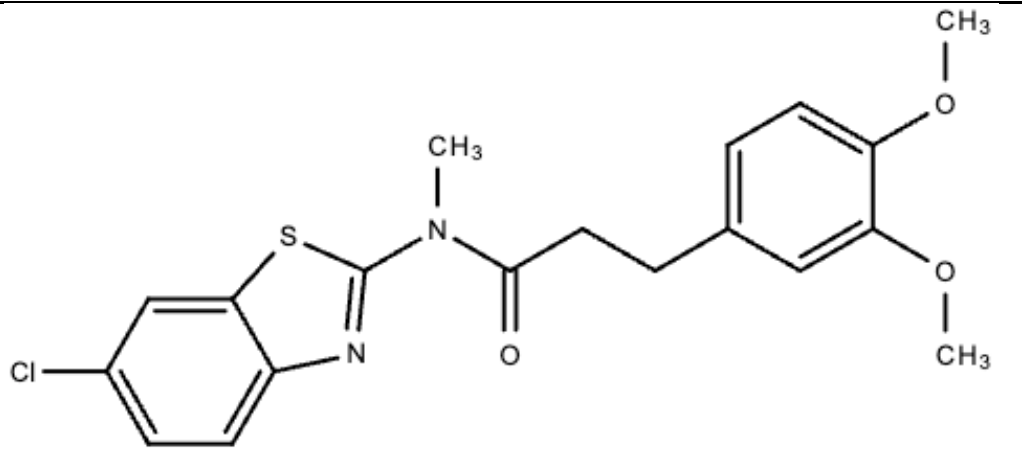 | Very weak | 0 |
| 52 | 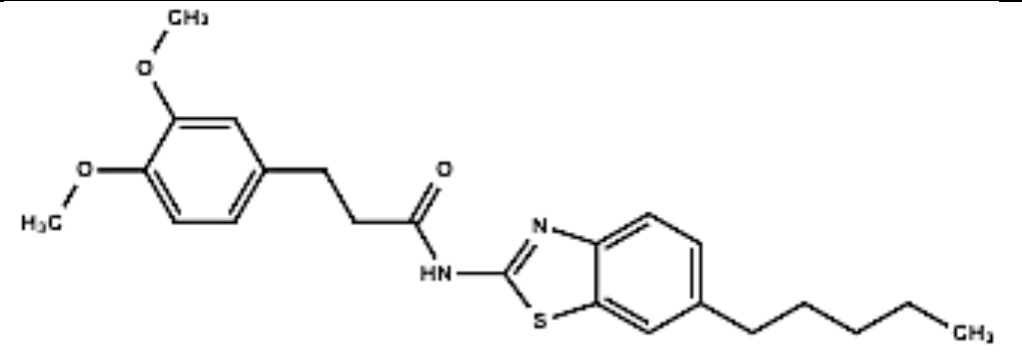 | Weak | 0 |
| 53 | 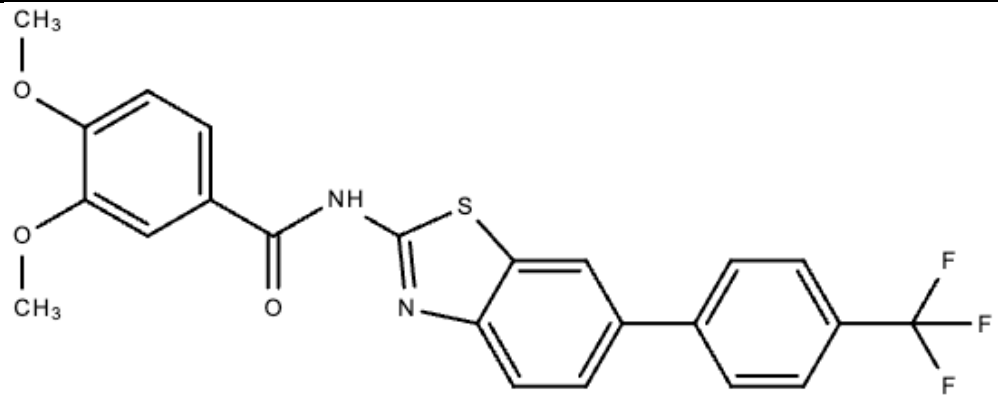 | Very weak | 0 |
| 54 | 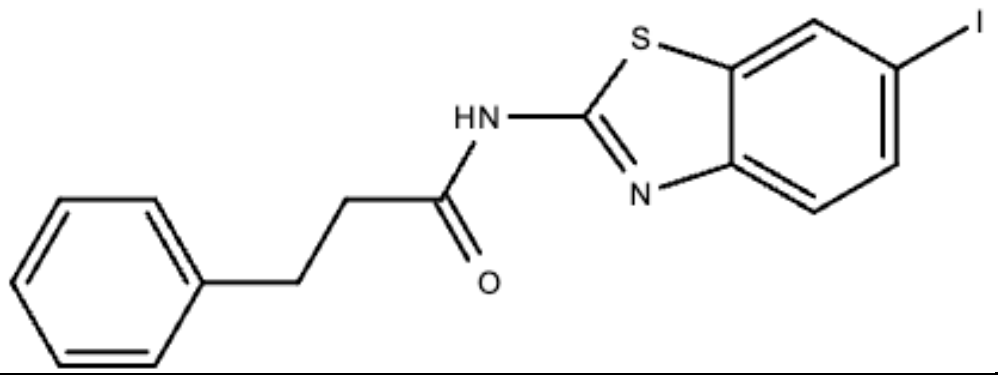 | Weak | 0 |
| 55 | 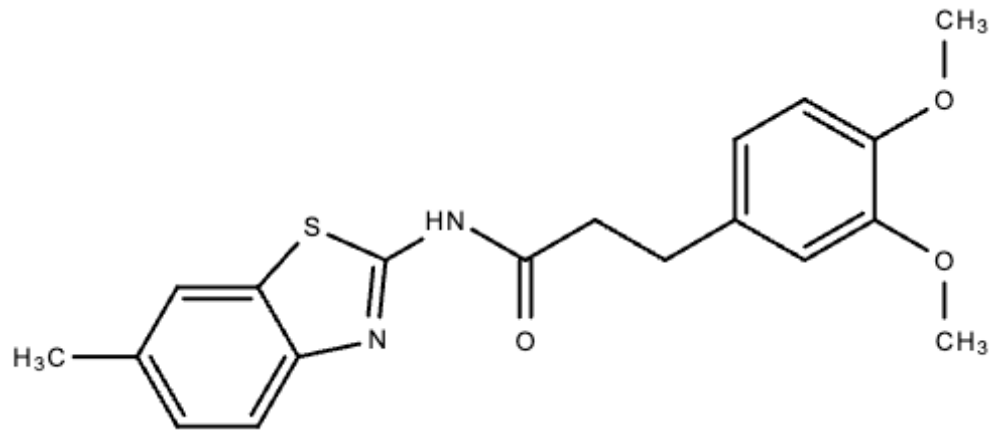 | Medium | 1 |
| 56 | 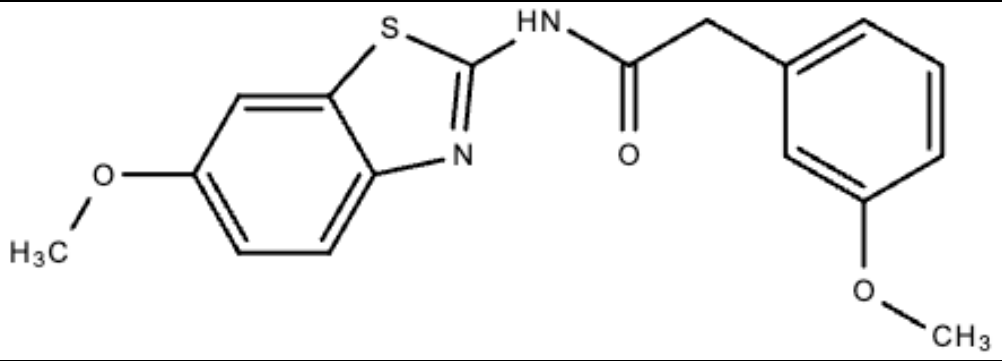 | Very weak | 0 |
| 57 | 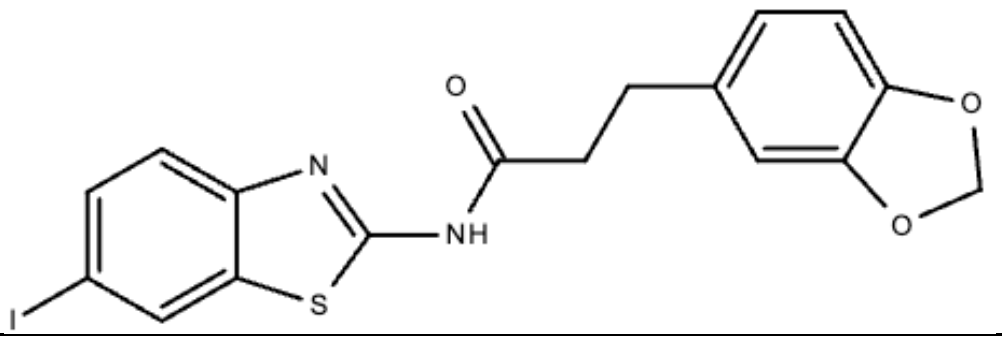 | Weak | 0 |
| 58 | 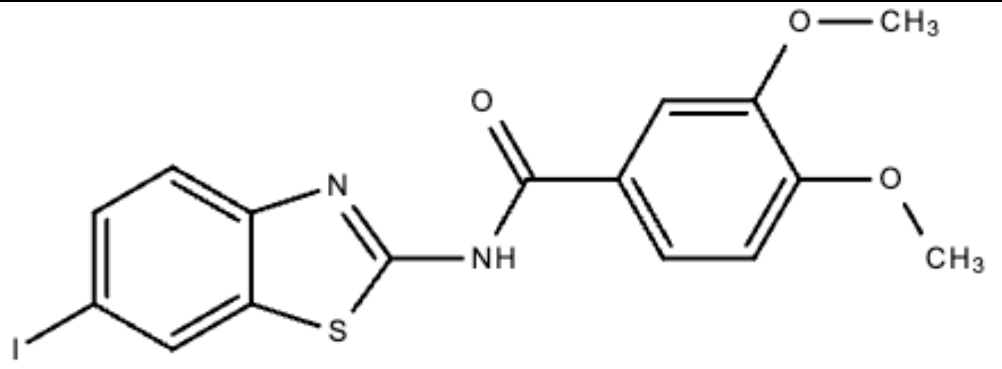 | Very weak | 0 |
| 59 | 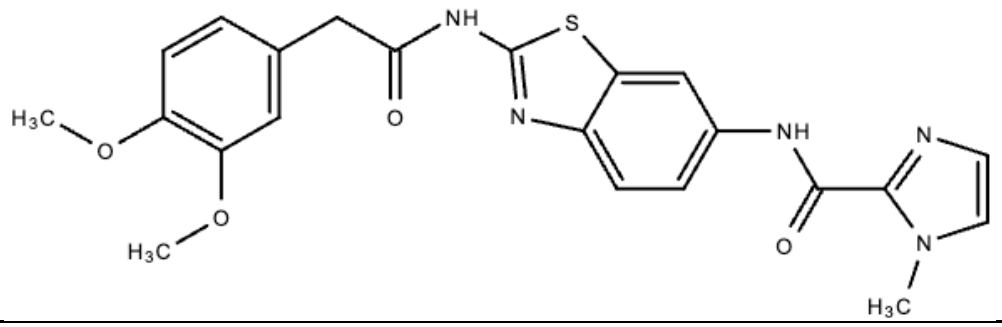 | Very weak | 0 |
| 60 | 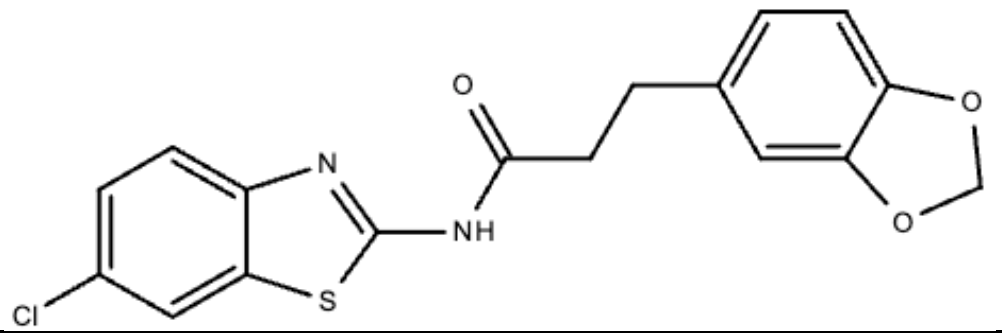 | Very weak | 0 |
| 61 | 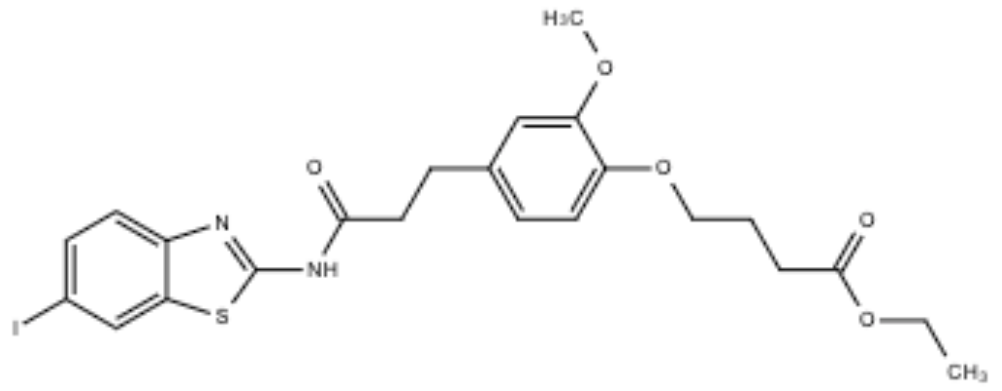 | Medium | 1 |
| 62 | 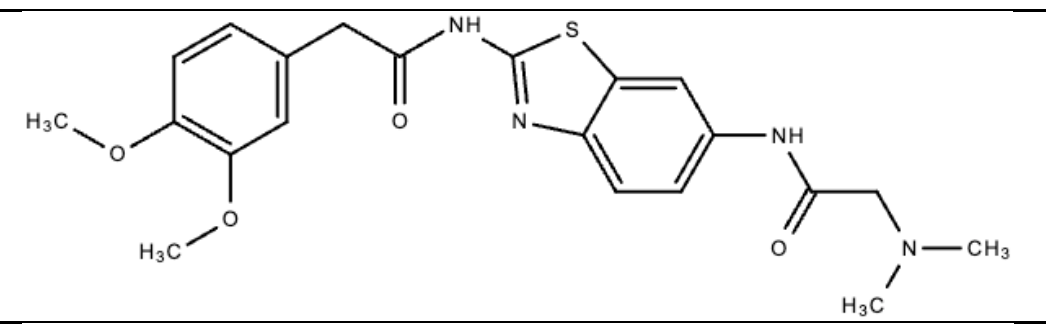 | Very weak | 0 |
| 63 | 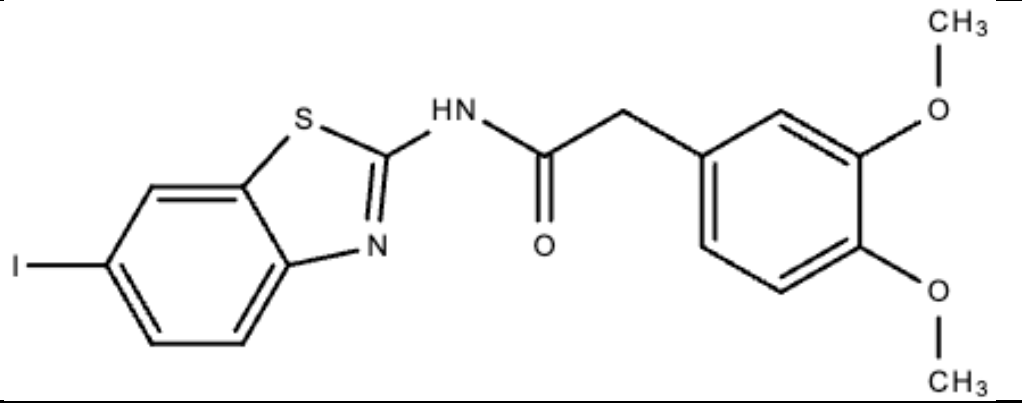 | Medium | 1 |
| 64 | 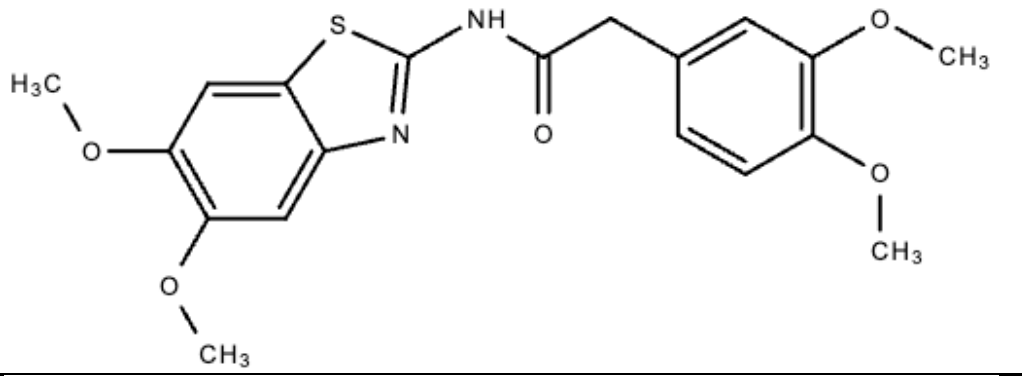 | Weak | 0 |
| 65 | 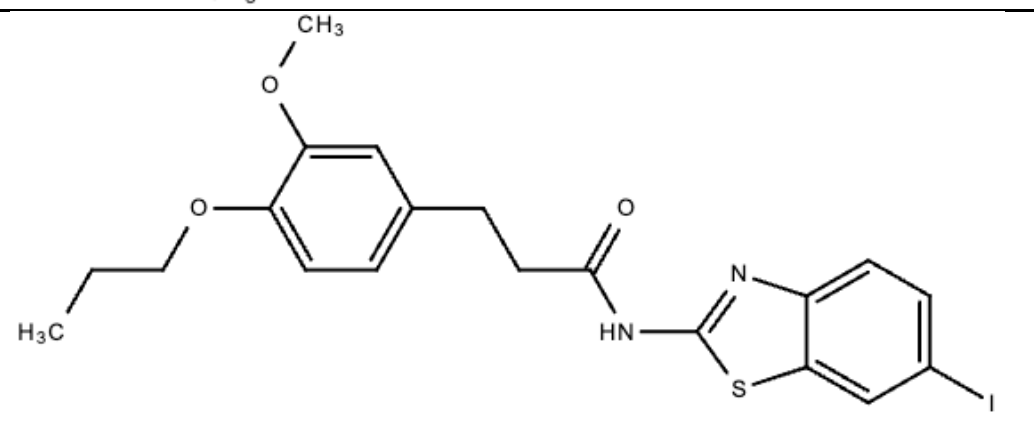 | Strong | 1 |
| 66 | 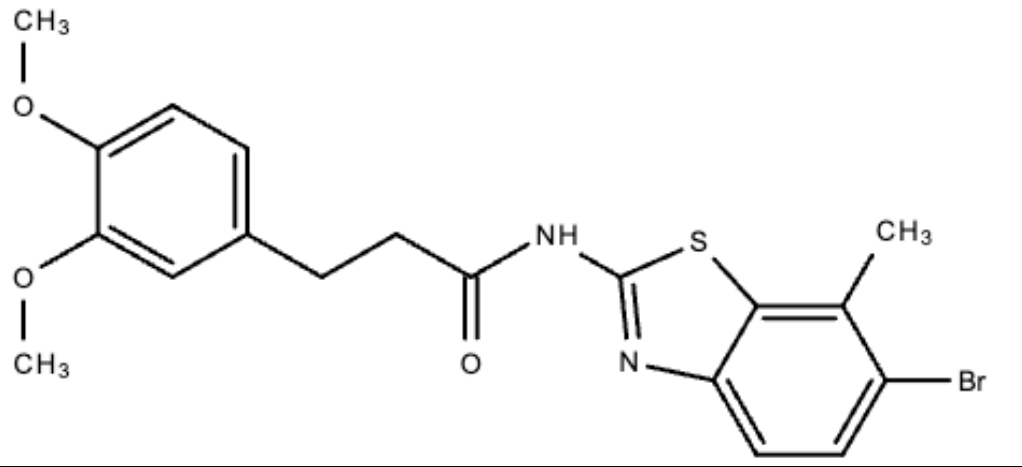 | Medium | 1 |
| 67 | 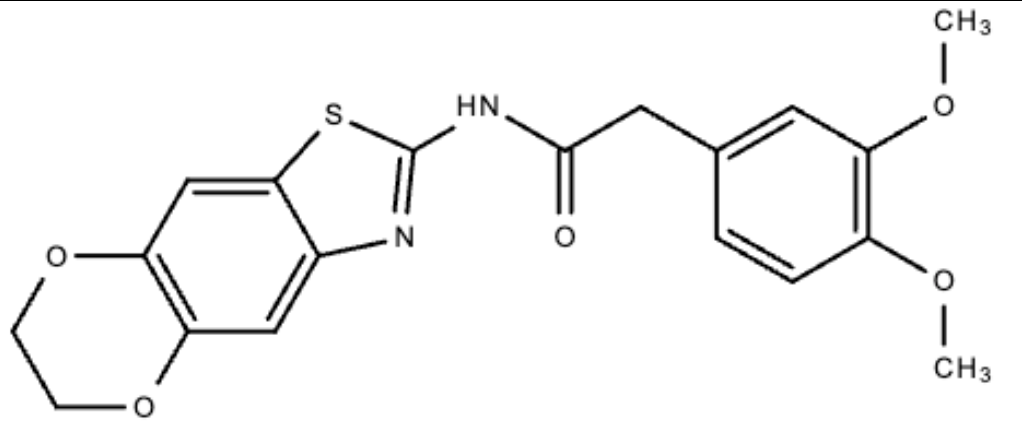 | Weak | 0 |
| 68 | 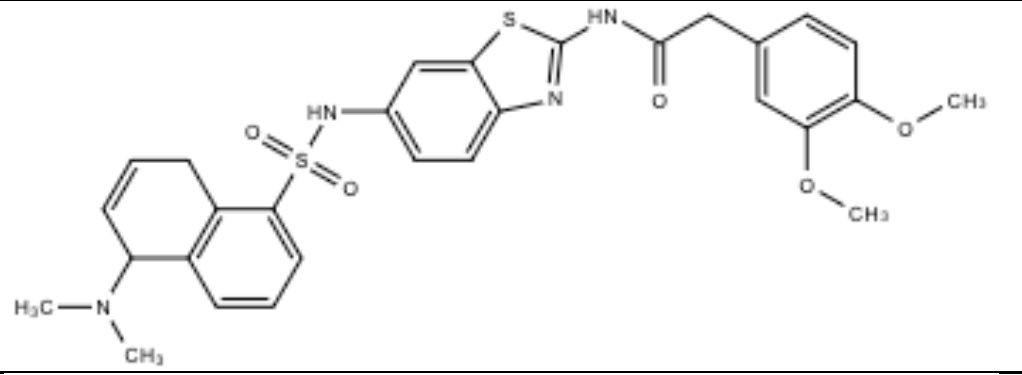 | Very weak | 0 |
| 69 | 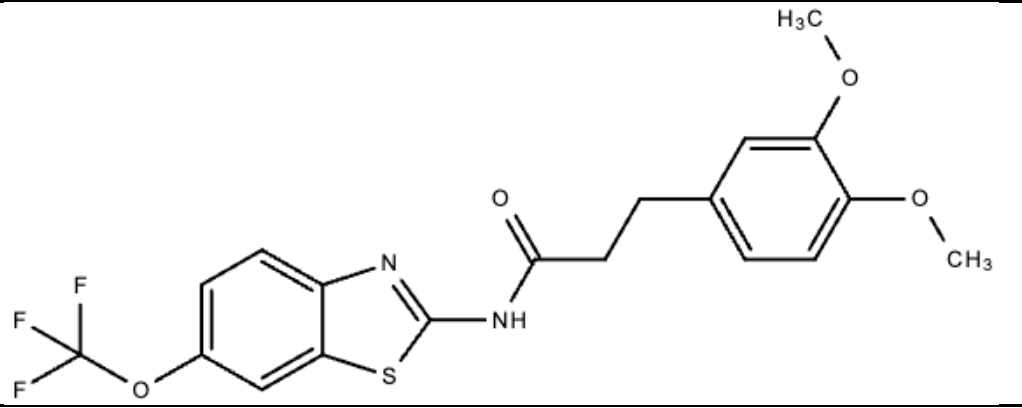 | Weak | 0 |
| 70 | 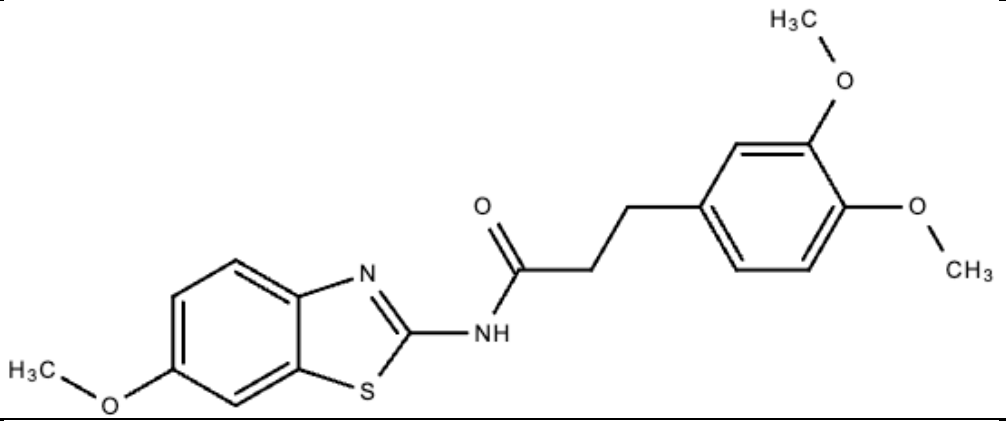 | Weak | 0 |
| 71 | 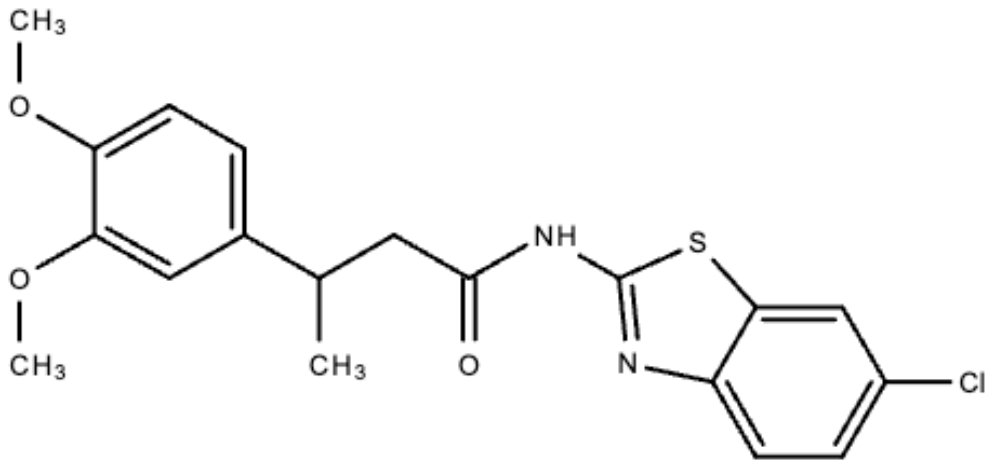 | Very weak | 0 |
| 72 | 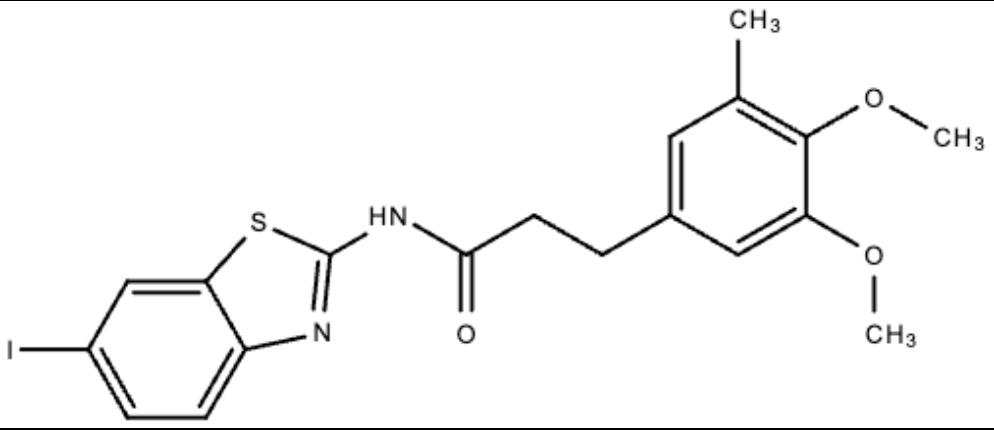 | Medium | 1 |
| 73 | 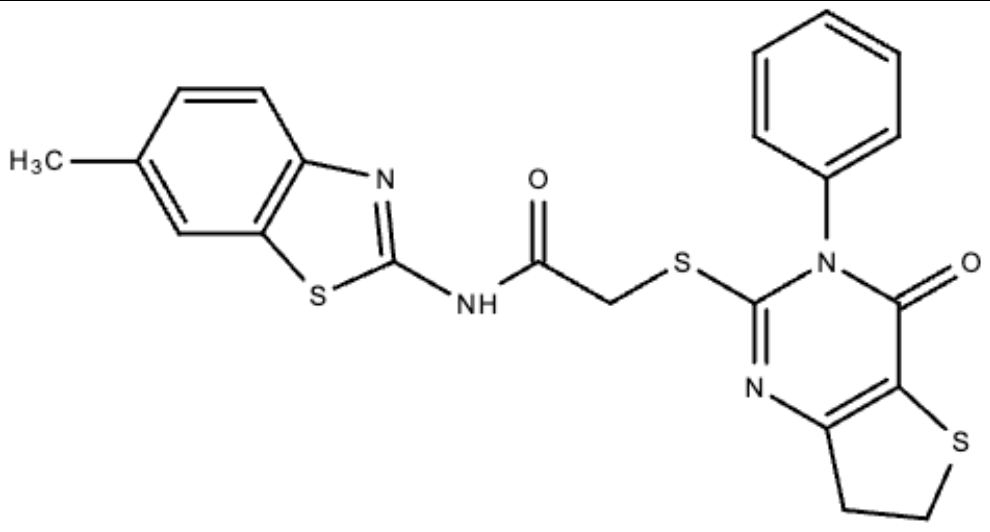 | Strong | 1 |
| 74 | 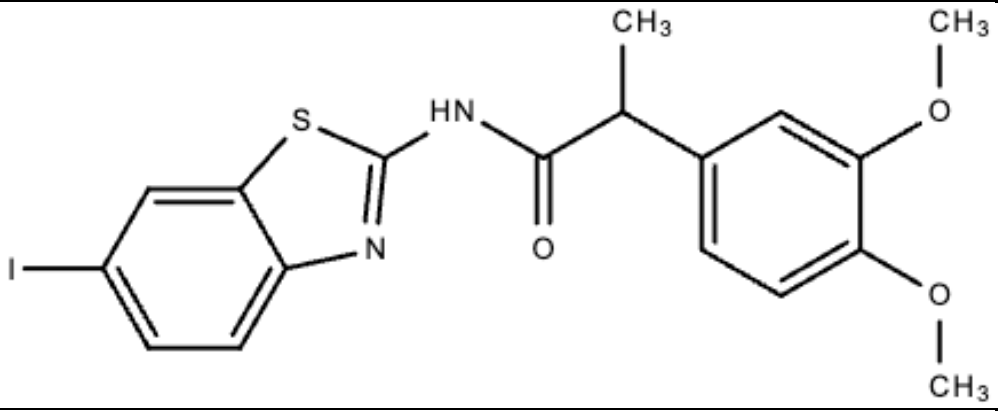 | Super strong | 1 |
| 75 | 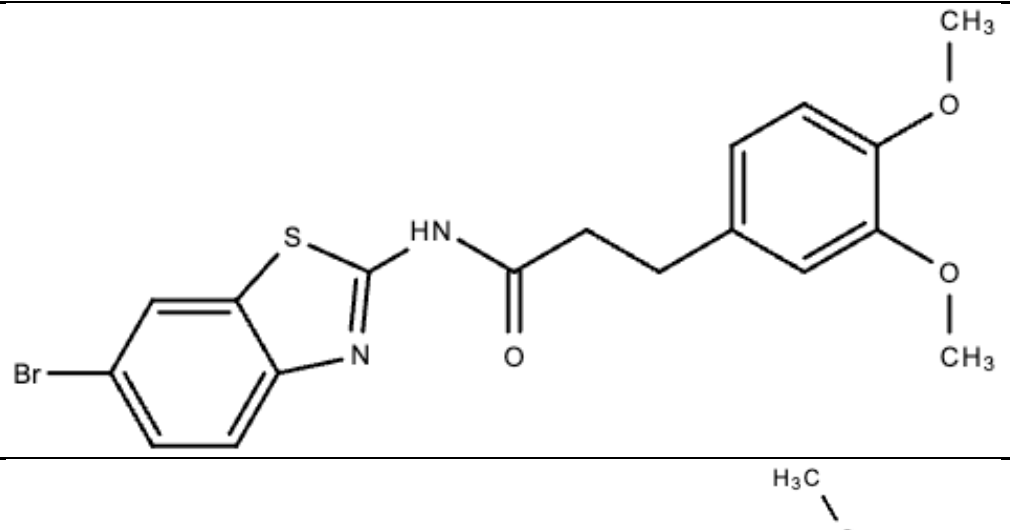 | Super strong | 1 |
| 76 | 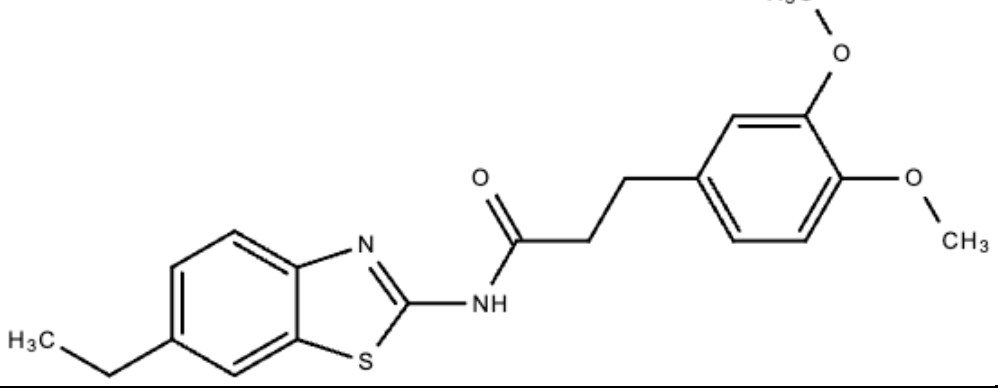 | Medium | 1 |
| 77 | 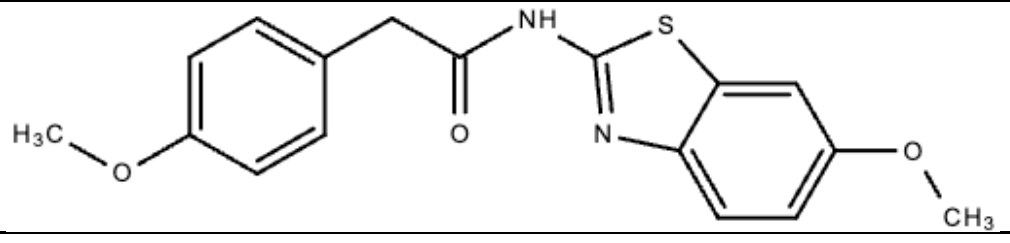 | Very weak | 0 |
| 78 | 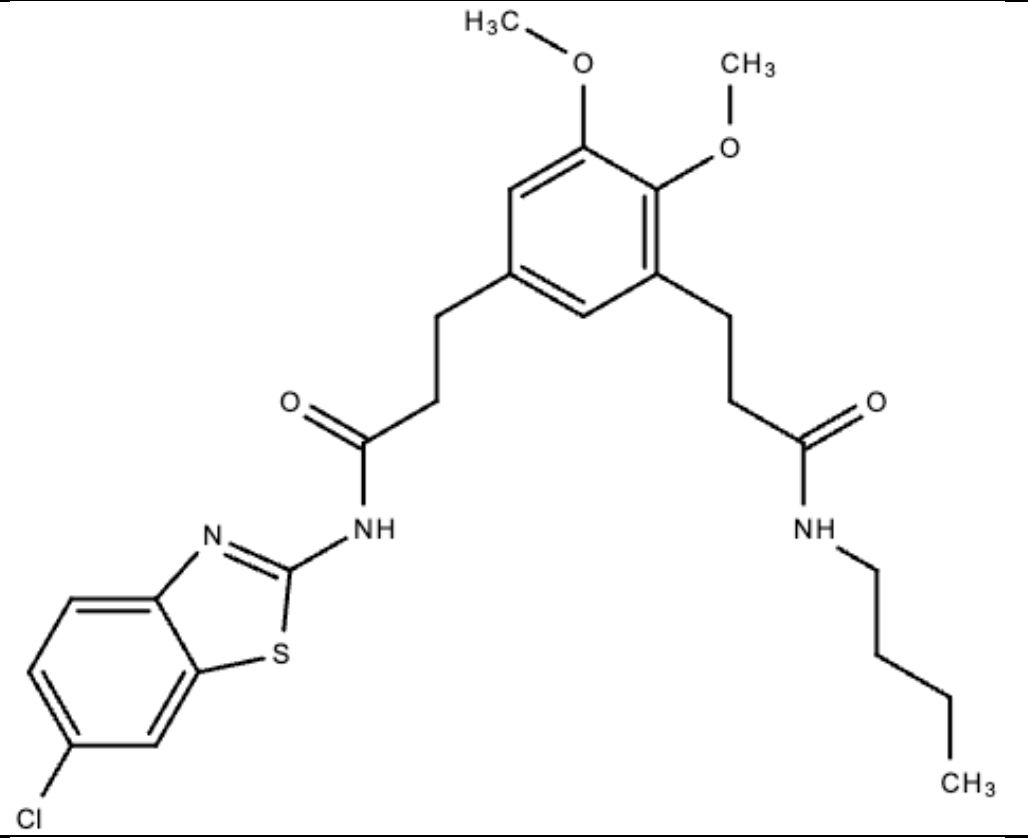 | Very weak | 0 |
| 79 | 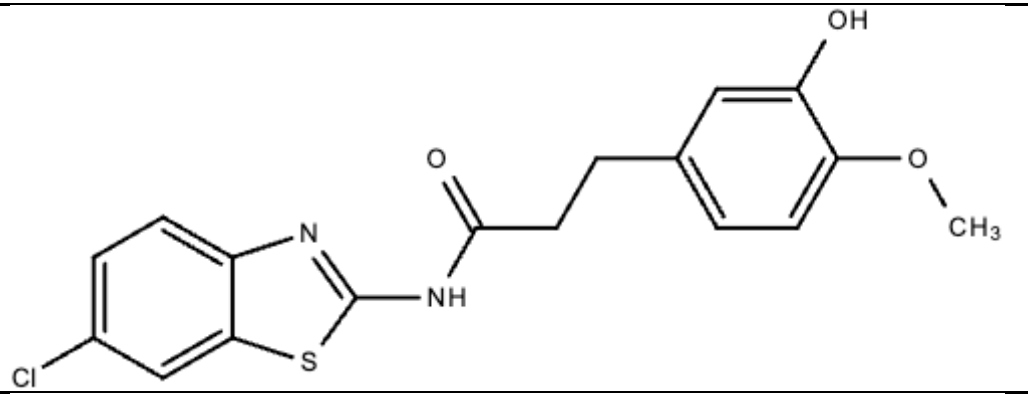 | Very weak | 0 |
| 80 | 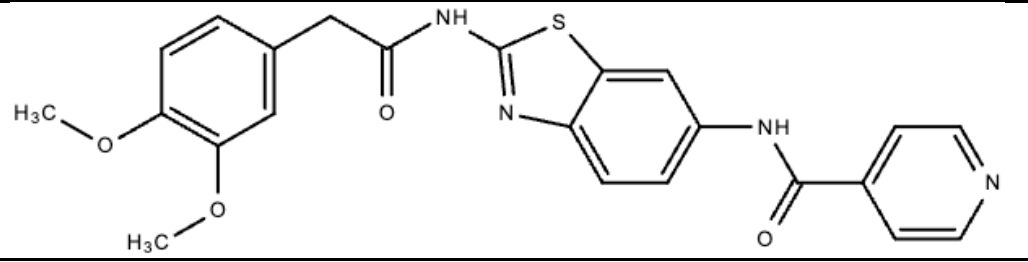 | Very weak | 0 |
| 81 | 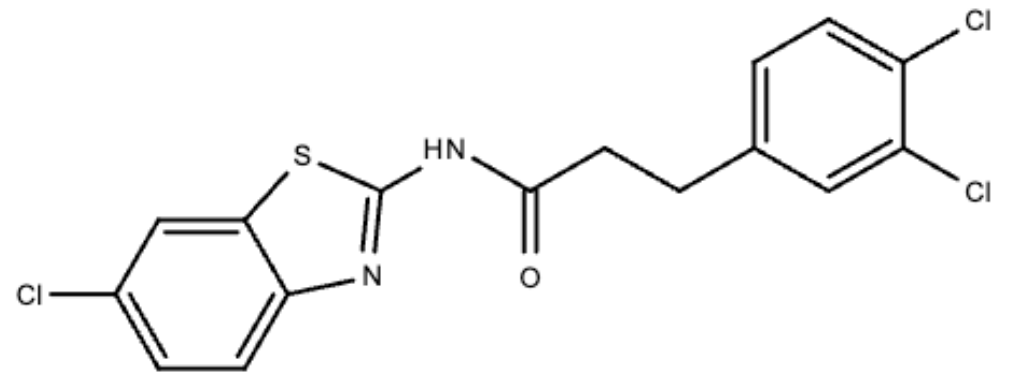 | Very weak | 0 |
| 82 | 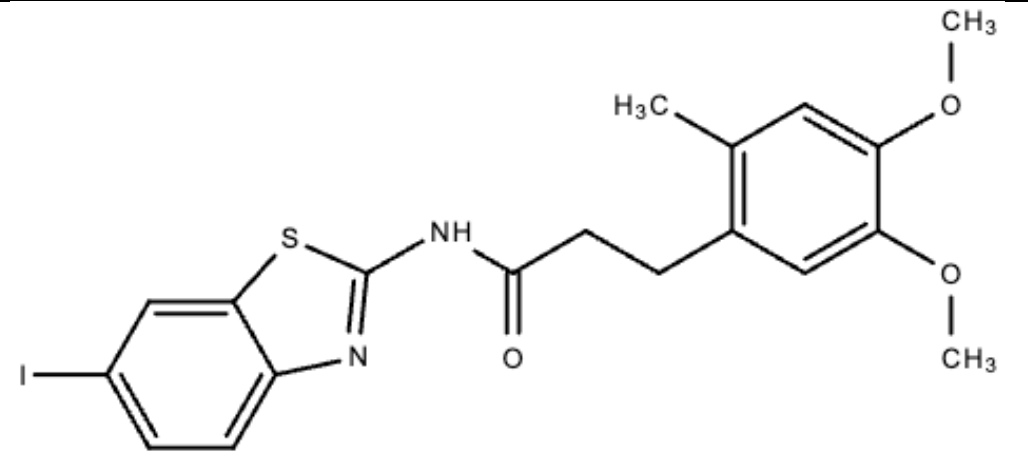 | Weak | 0 |

* Molecule 37 is a duplicate of molecule 13 and was excluded when training the regression models.

**Supporting Information 2**

Initial fragment structures were placed into a 15 x 15 x 15 Å box along with 500 water molecules. Simulations were ran for $10^6$ time steps of size 1 fs, yielding a 1 ns-long trajectory. At every 100th time step, the interaction energy between the fragment and the solvent was calculated (using a 10-way moving average to minimize fluctuations). These interaction energies were then averaged to yield a hydrophilicity value for the fragment. To enable comparison between fragments of different sizes, the hydrophilicity value of each fragment was normalized by the number of non-hydrogen atoms in the fragment.

**Supporting Information 3**

Table of molecular fragments and hydrophilicity values. ‘Type’ refers to the position of the fragment in molecule (X-benzothiozole-linker-benzene-Y). Hydrophilicity values are normalized by the number of non-hydrogen atoms in the fragment.

| Index | Chemical structure | Type | Hydrophilicity (kcal/mol) |
|---|---|---|---|
| 1 | 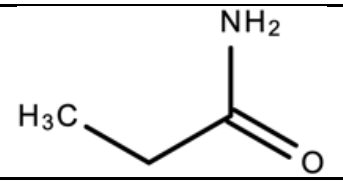 | Linker | -0.38 |
| 2 | $H_3C—OH$ | Y | -1.03 |
| 3 | 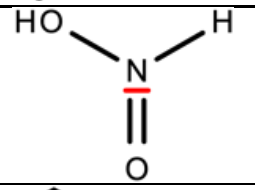 | X | 1.87 |
| 4 | 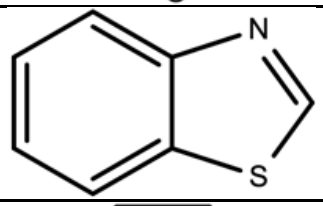 | Benzothiozole | -0.65 |
| 5 | 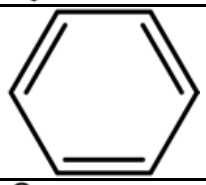 | Benzene | -0.64 |
| 6 | 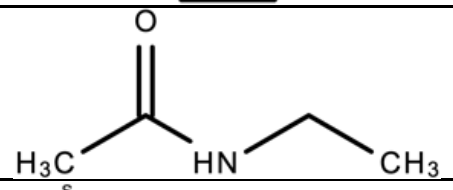 | Linker | -0.70 |
| 7 | 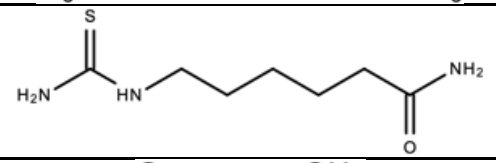 | X | -0.51 |
| 8 | 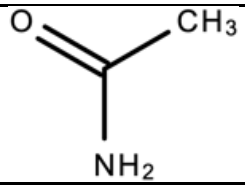 | Linker | -0.38 |
| 9 | $OH_2$ | X | 0.26 |
| 10 | 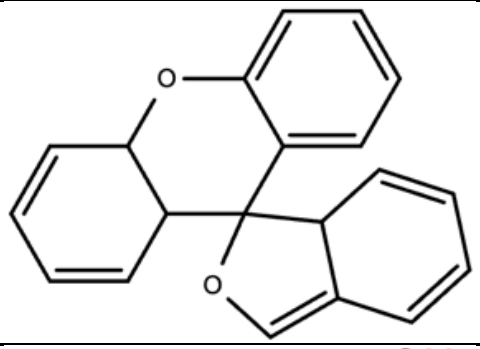 | X | 0.61 |
| 11 | 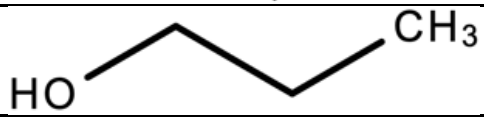 | Y | -0.89 |
| 12 | IH | X | -1.79 |
| 13 | ClH | X | -1.11 |
| 14 | 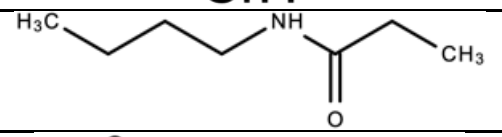 | X | -0.66 |
| 15 | 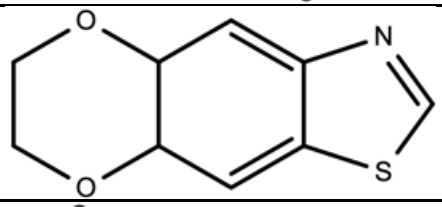 | Benzothiozole | -0.61 |
| 16 | 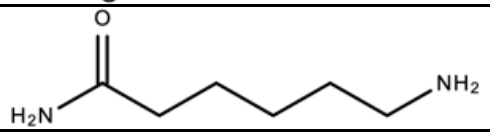 | X | -0.37 |
| 17 | BrH | X | -1.37 |
| 18 | 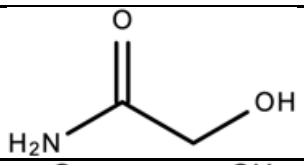 | Linker | -0.35 |
| 19 | 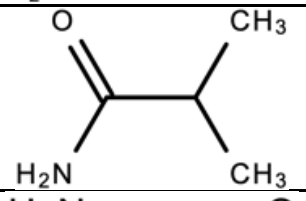 | Linker | -0.42 |
| 20 | 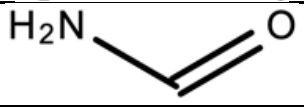 | Linker | -0.18 |
| 21 | $CH_4$ | X | -1.12 |
| 22 | $H_2C=CH_2$ | X | -0.87 |
| 23 | 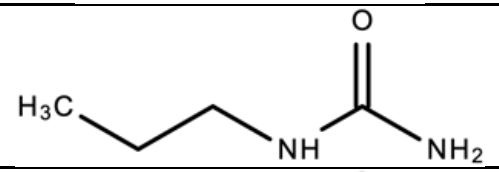 | Linker | -0.49 |
| 24 | 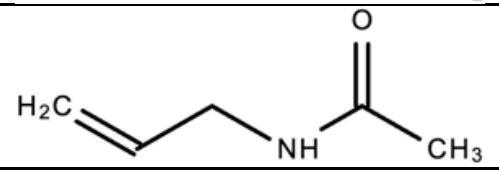 | Linker | -0.66 |
| 25 | $NH_3$ | X | 0.77 |
| 26 | 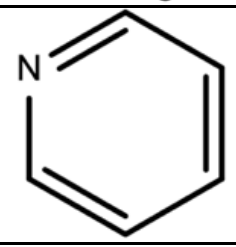 | X | -0.61 |
| 27 | 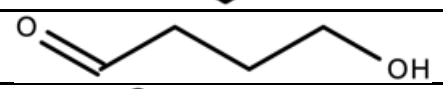 | Linker | -0.64 |
| 28 | 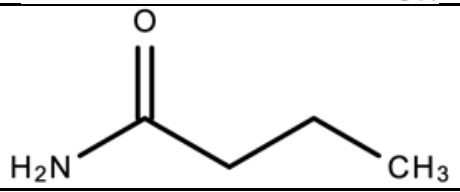 | Linker | -0.41 |
| 29 | FH | X | -0.23 |
| 30 | 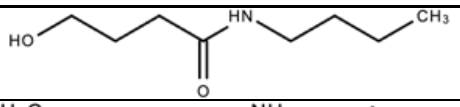 | Linker | -0.71 |
| 31 | 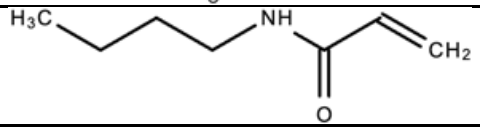 | X | -0.74 |
| 32 | 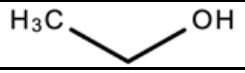 | Y | -0.92 |
| 33 | 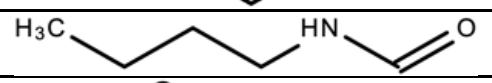 | Linker | -0.72 |
| 34 | 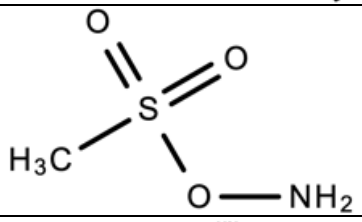 | X | -0.05 |
| 35 | 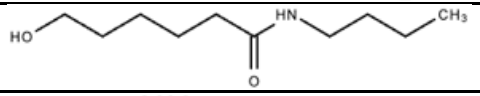 | Linker | -0.76 |
| 36 | 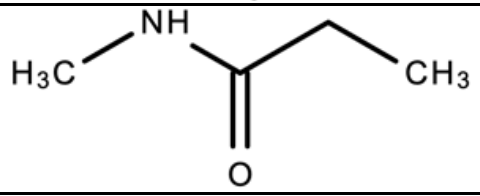 | Linker | -0.73 |
| 38 | 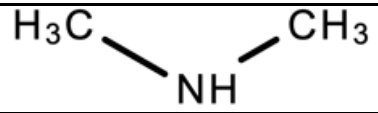 | Linker | -1.09 |
| 39 | 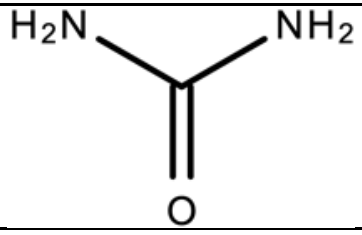 | Linker | -0.29 |
| 40 | $H_2N-CH_3$ | Linker | -0.50 |
| 42 | 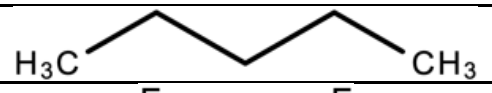 | Linker | -0.67 |
| 43 | 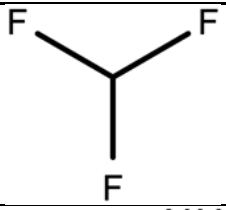 | X | -0.03 |
| 44 | 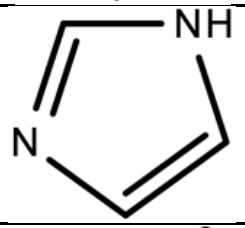 | X | -0.82 |
| 45 | 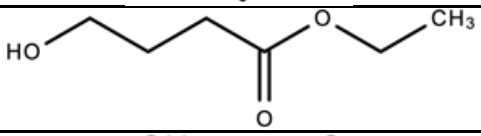 | Linker | -0.58 |
| 46 | 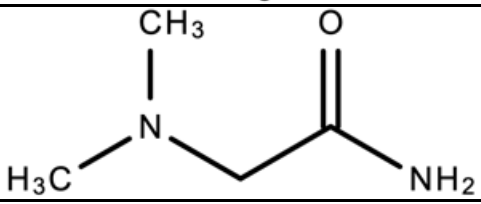 | Linker | -0.40 |
| 48 | 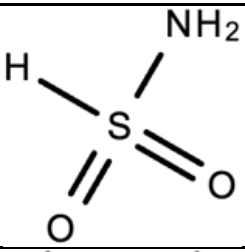 | X | -0.23 |
| 49 | 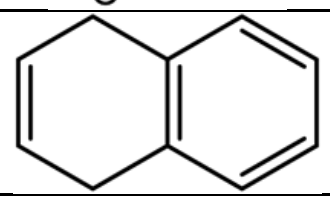 | X | -0.63 |
| 50 | $H_3C-CH_3$ | Linker | -0.92 |
| 51 | 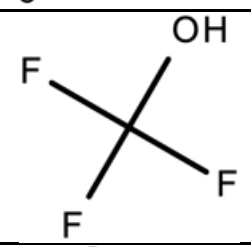 | X | 0.01 |
| 52 | 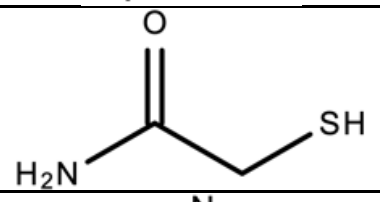 | Linker | -0.48 |
| 53 | 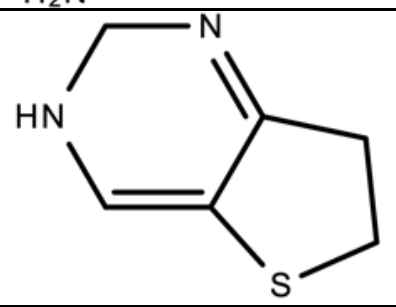 | Benzothiozole | -0.80 |
| 54 | 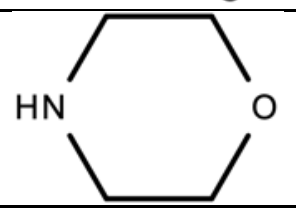 | X | -0.82 |

**Supporting Information 4**

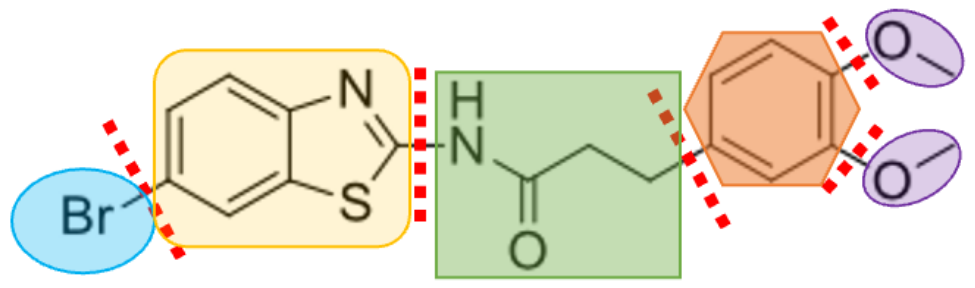

Illustration of the molecular design scheme. First, each molecule in the training data is broken into fragments by cutting all bonds connected to rings (indicated by red lines above). This results in a modifier fragment (blue), a benzothiazole fragment (yellow), an amide linker (green), an aryl ring (orange), and one or more other modifier fragments (purple).

The resulting fragment (build from all molecules in the training set) contains:

- 19 distinct modifier fragments (blue)
- 3 distinct benzothiazole fragments (yellow)
- 12 distinct amide linker fragments (green)
- 3 distinct aryl ring fragments (orange)
- 10 distinct other modifier fragments (purple)

To design new molecules, we combined the fragments in the sequence described above by connecting the sites of the broken covalent bonds. All molecules underwent a classical geometry optimization before computing feature vectors according to Figure 2.

**Supporting Information 5**

To a solution of 2-(3,4-dimethoxyphenyl)-*N*-(6-iodobenzo[*d*]thiazol-2-yl)acetamide (0.44 mmol, 200 mg) and $K_2CO_3$ (2.0 equiv, 0.88 mmol, 121.6 mg) in DMF (4.0 mL), ethyl iodide (1.2 equiv, 0.53 mmol, 42 µL) was added at room temperature. The reaction mixture was stirred for 24h. The reaction mixture was quenched with $H_2O$ and extracted with EtOAc (3 × 30 mL). The organic layer was dried with anhydrous $Na_2SO_4$ and then concentrated under reduced pressure. The residue was purified by flash column chromatography on silica gel to give the desired product (82.4 mg, 39%).

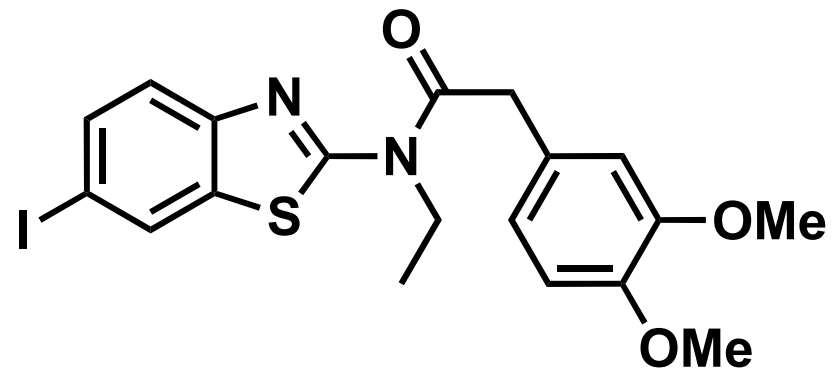

$^1$H-NMR (400 MHz, $CDCl_3$) δ 7.94 (d, *J* = 1.7 Hz, 1H), 7.71 (dd, *J* = 8.6, 1.7 Hz, 1H), 7.06 (d, *J* = 8.6 Hz, 1H), 6.94-6.90 (m, 2H), 6.83 (d, *J* = 8.1 Hz, 1H), 4.38 (q, *J* = 7.2 Hz, 2H), 3.89 (s, 3H), 3.86 (s, 3H), 3.84 (s, 2H), 1.36 (t, *J* = 7.2 Hz, 3H). $^{13}$C-NMR (101 MHz, $CDCl_3$) δ 182.1, 165.9, 148.7, 147.7, 135.8, 135.6, 131.1, 129.3, 128.9, 121.6, 112.8, 112.7, 111.1, 86.1, 55.9, 55.8, 46.7, 40.7, 12.5. HRMS (ESI) calcd for $(M+H)^+$ $C_{19}H_{20}IN_2O_3S^+$ , m/z: 482.3365, found: 482.3367.

**Supporting Information 6**

qPCR primer sequences for human gene markers: forward (F) and reverse (R).

| GENE | Accession Number | Primer sequence (both 5'-3') | Annealing temperature (°C) | Product size (bp) |
|---|---|---|---|---|
| NKX2-5-F | NM_001166176 | ACCCTGAGTCCCCTGGATTT | 60.18 | 125 |
| NKX2-5-R | | TCACTCATTGCACGCTGCAT | 60.96 | |
| TNNT2-F | NM_001001430 | TTCACCAAAGATCTGCTCCTCGCT | 64.18 | 166 |
| TNNT2-R | | TTATTACTGGTGTGGAGTGGGTGTGG | 64.34 | |
| MYH7-F | NM_000257 | ACCAACCTGTCCAAGTTCCG | 55.00 | 128 |
| MYH7-R | | TTCAAGCCCTTCGTGCCAAT | 50.00 | |
| GAPDH-F | NM_001357943 | AATCCCATCACCATCTTCCAG | 57.41 | 122 |
| GAPDH-R | | AAATGAGCCCCAGCCTTC | 57.23 | |

List of primary and secondary antibodies.

| **Cell** | **Protein** | **Primary Antibody** | **Secondary Antibody** | **conditions** |
|---|---|---|---|---|
| Cardiomyocyte | cTnT | Mouse monoclonal IgG2a, MAB 1874, R&D. | Goat anti-mouse IgG2a, Alexa Fluor 594, A-21135, Invitrogen | Primary: room temperature, one hour Secondary: darkness, 30 min |
| | MHY7 | Rabbit Monoclonal Antibody IgG, MAB 90961, R&D. | Goat anti-rabbit IgG, Alexa Fluor 488, A-11008, Invitrogen. | |

**Supporting Information 7**

Examples of molecules contain in each cluster. For each molecule, the chemical structure and a representative 3D structure are shown.

| **Cluster a** | | |
|---|---|---|
| Mol13 | Mol33 | Mol74 |
| 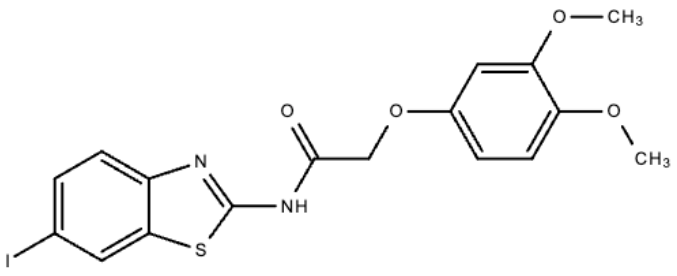 | 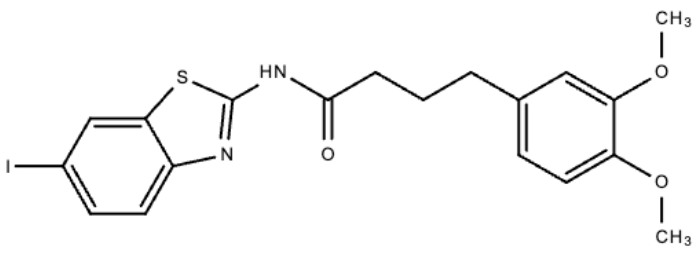 | 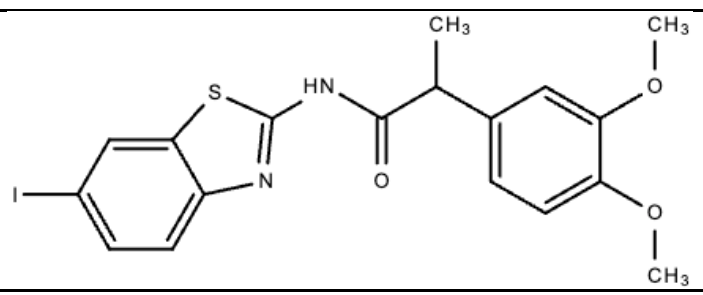 |
| 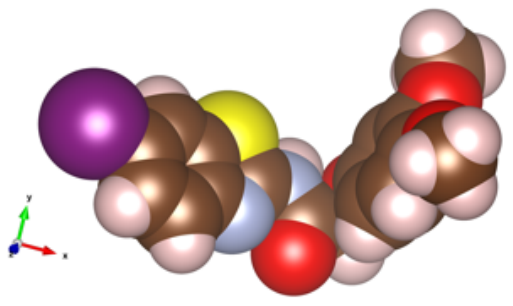 | 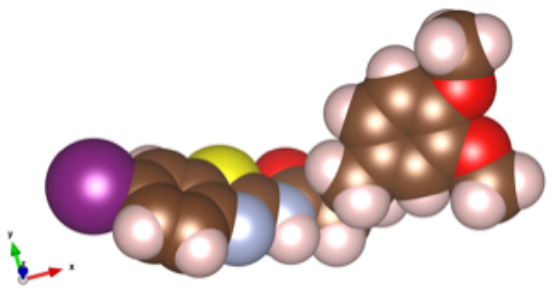 | 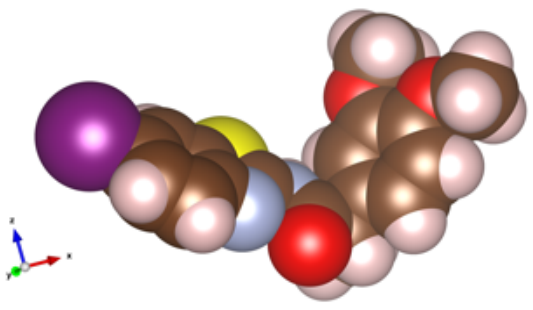 |
| Common features: Iodine atom, open v-shaped conformation. | | |

| **Cluster b** | | |
|---|---|---|
| Mol135 | Mol39 | Mol43 |
| 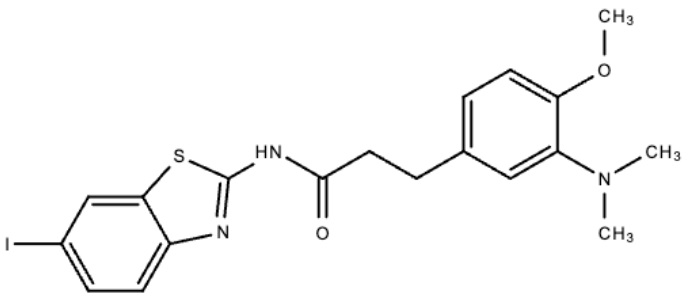 | 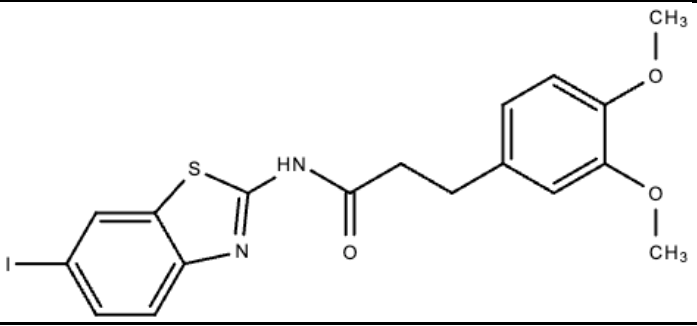 | 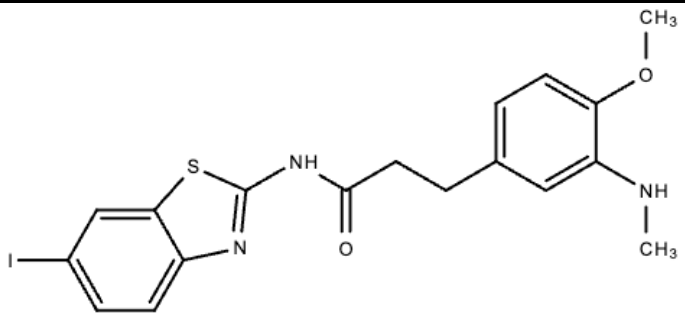 |
| 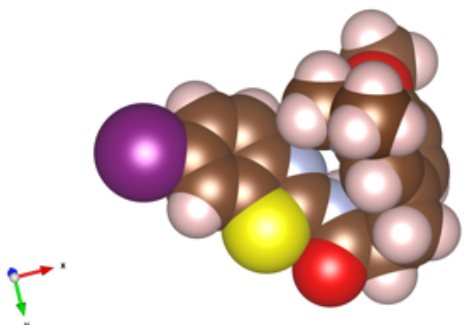 | 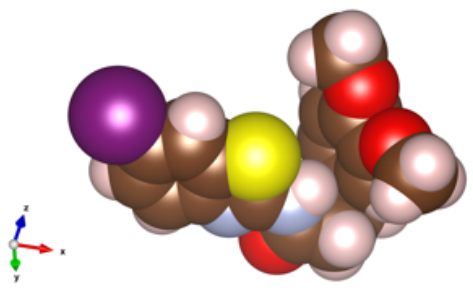 | 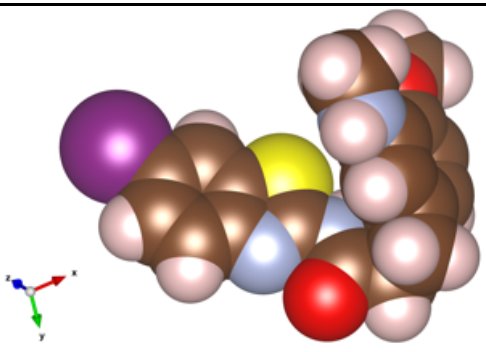 |
| Common features: Iodine atom, closed U-shaped conformation. | | |

| **Cluster c** | | |
|---|---|---|
| Mol152 | Mol73 | Mol76 |
| 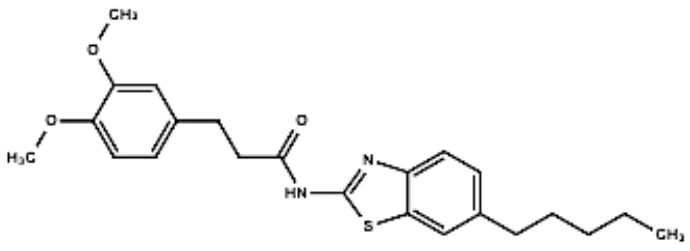 | 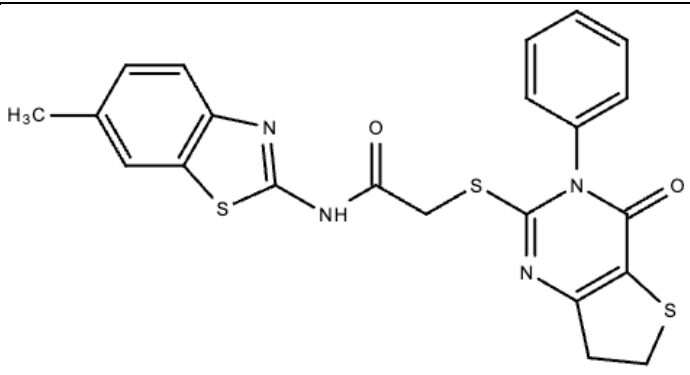 | 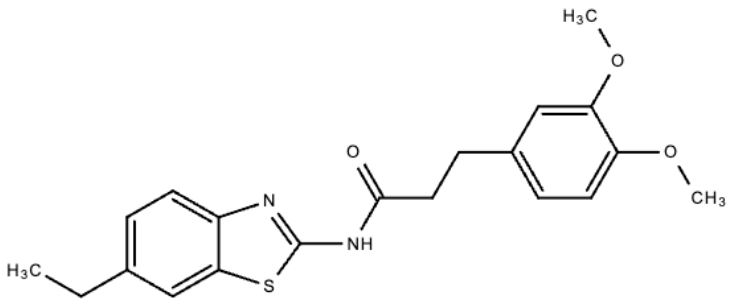 |
| 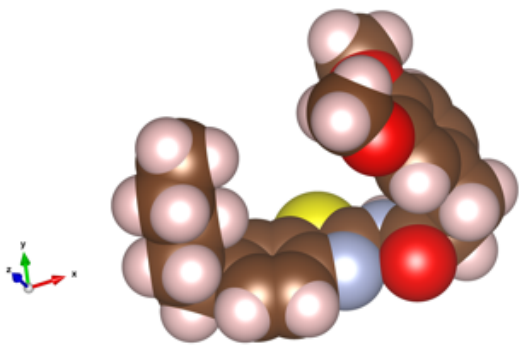 | 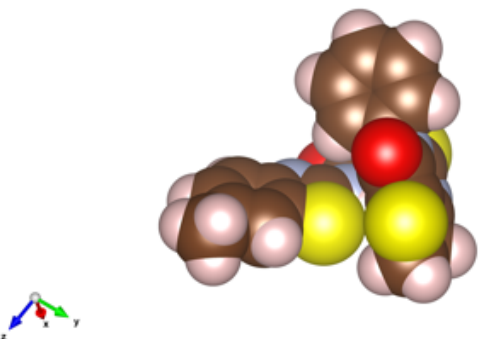 | 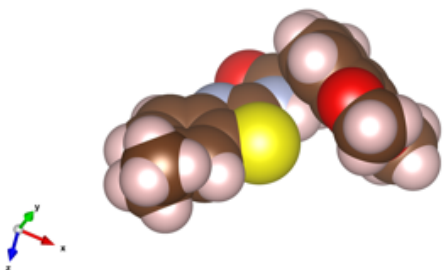 |
| Common features: alkyl group modifier, twisted comformation | | |

| **Cluster d** | | |
|---|---|---|
| Mol1 | Mol14 | Mol31 |
| 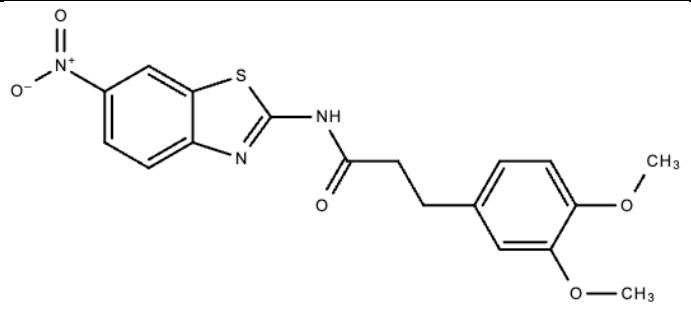 | 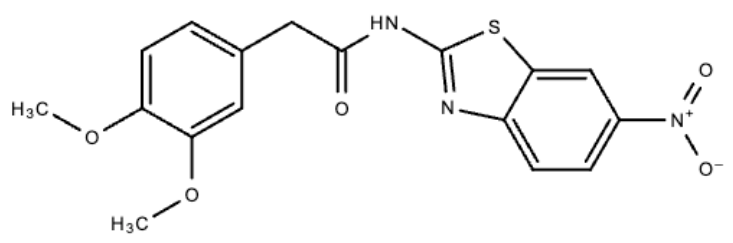 | 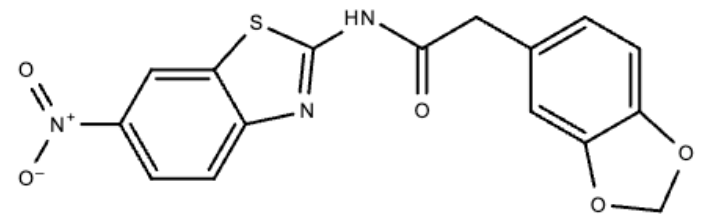 |
| 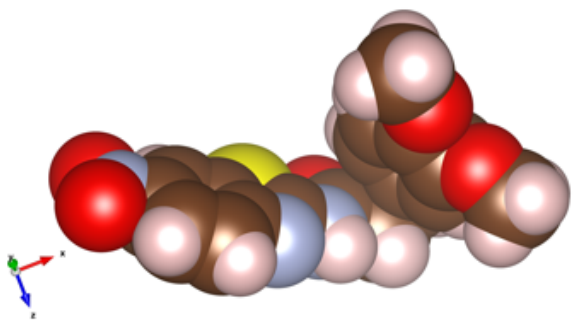 | 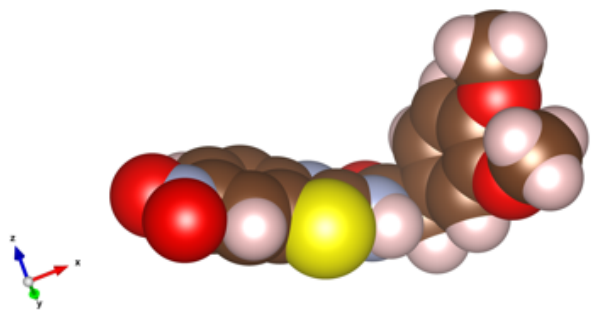 | 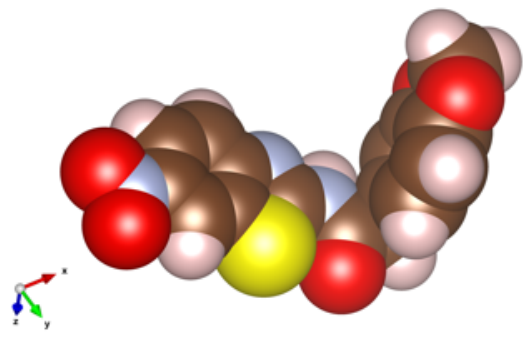 |
| Common features: $NO_2$ modifier, twisted conformation | | |

| **Cluster e** | | |
|---|---|---|
| Mol23 | Mol26 | Mol60 |
| 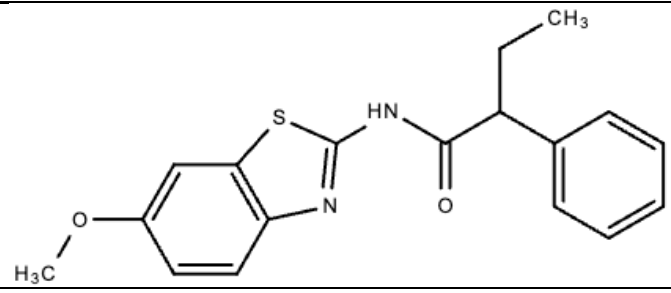 | 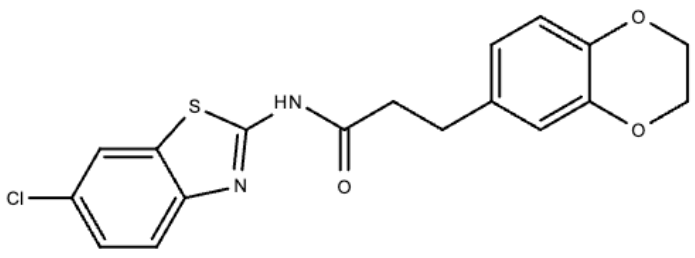 | 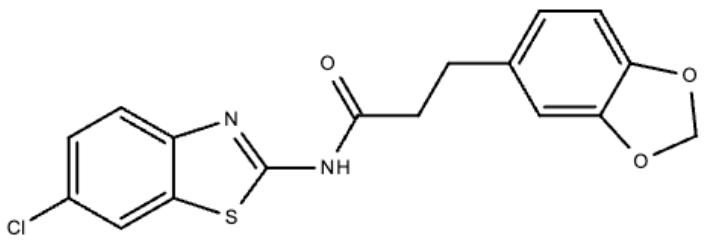 |
| 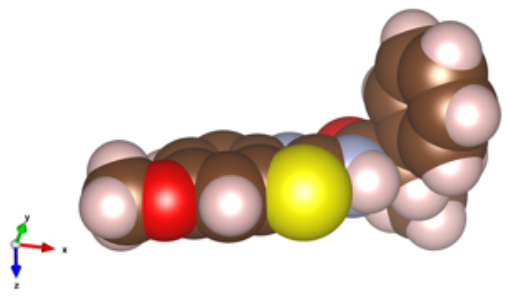 | 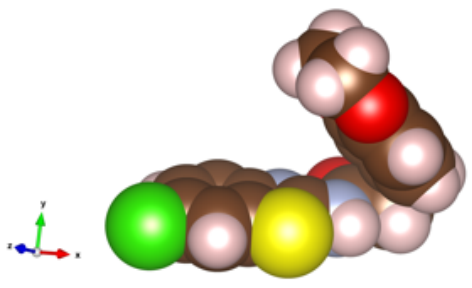 | 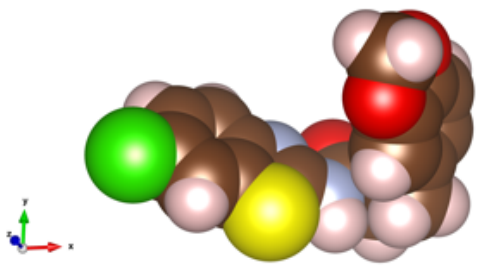 |
| Common features: Bent conformation | | |

| **Cluster f** | | |
|---|---|---|
| Mol20 | Mol62 | Mol80 |
| 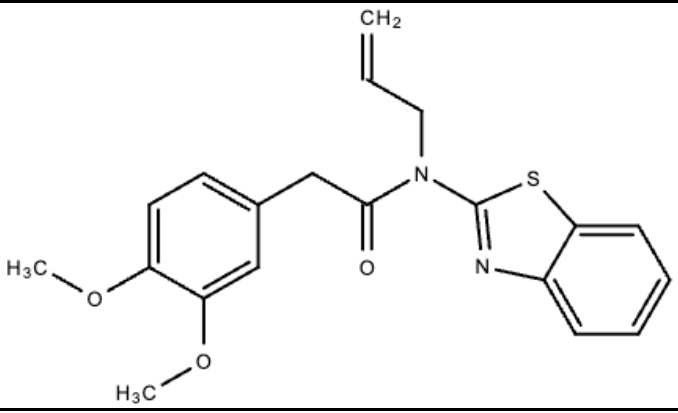 | 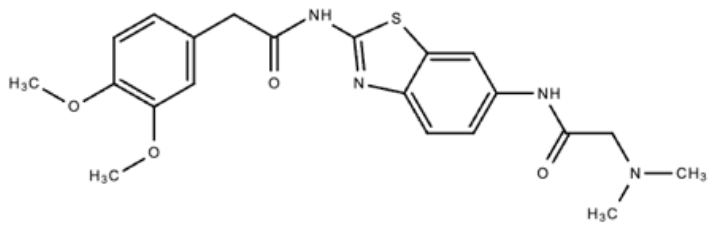 | 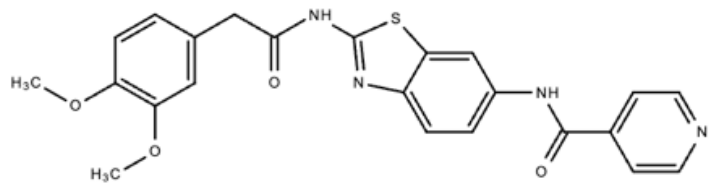 |
| 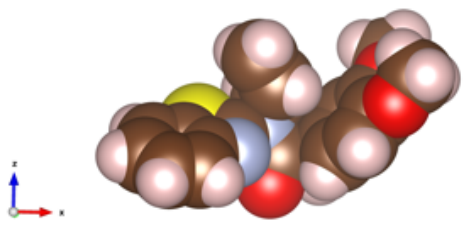 | 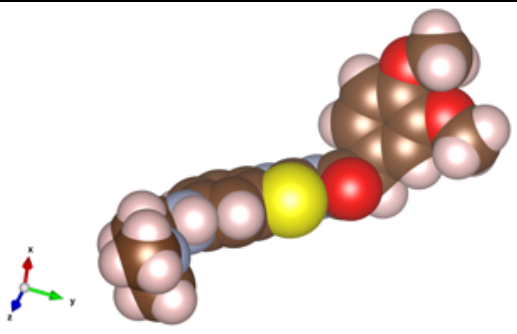 | 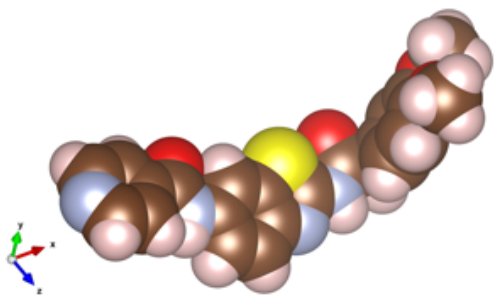 |
| Common features: Open and extended conformations. Less obvious common features than other clusters. | | |

**Supporting Information 8**

(i) Wald and chi-square test results for logistic model trained with all training data and decorated shape features (R output):

```
> summary(model)

Call:
glm(formula = cdif ~ PC1 + PC2 + PC3 + PC4 + PC5 + PC6 + PC7,
    family = binomial(link = "logit"), data = cdat)

Coefficients:
            Estimate Std. Error z value Pr(>|z|)
(Intercept)   -5.999      2.151  -2.789  0.00529 **
PC1          -33.050     13.087  -2.525  0.01156 *
PC2          -45.915     16.564  -2.772  0.00557 **
PC3            2.508     19.922   0.126  0.89982
PC4           -6.767     21.384  -0.316  0.75167
PC5          -38.888     21.020  -1.850  0.06430 .
PC6          -13.756     27.003  -0.509  0.61045
PC7           90.537     58.733   1.542  0.12319
---
Signif. codes:  0 '***' 0.001 '**' 0.01 '*' 0.05 '.' 0.1 ' ' 1

(Dispersion parameter for binomial family taken to be 1)

    Null deviance: 99.374  on 79  degrees of freedom
Residual deviance: 75.489  on 72  degrees of freedom
AIC: 91.489

Number of Fisher Scoring iterations: 5

> anova(model, test="Chisq")
Analysis of Deviance Table

Model: binomial, link: logit

Response: cdif

Terms added sequentially (first to last)


     Df Deviance Resid. Df Resid. Dev Pr(>Chi)
NULL                    79     99.374
PC1   1   7.0763        78     92.297 0.007811 **
PC2   1  10.6862        77     81.611 0.001079 **
PC3   1   0.0166        76     81.595 0.897378
PC4   1   0.0448        75     81.550 0.832305
PC5   1   2.9669        74     78.583 0.084982 .
PC6   1   0.2335        73     78.349 0.628921
PC7   1   2.8607        72     75.489 0.090767 .
---
Signif. codes:  0 '***' 0.001 '**' 0.01 '*' 0.05 '.' 0.1 ' ' 1
```

(2) Test results for logistic model trained with all training data and ordinary shape features:

```
> summary(model)

Call:
glm(formula = cdif ~ PC1 + PC2 + PC3 + PC4 + PC5 + PC6 + PC7,
    family = binomial(link = "logit"), data = cdat)

Coefficients:
            Estimate Std. Error z value Pr(>|z|)
(Intercept)   -2.798      3.009  -0.930   0.3525
PC1          -12.193      9.968  -1.223   0.2212
PC2          -13.015     16.780  -0.776   0.4380
PC3           37.893     25.823   1.467   0.1423
PC4           58.929     28.934   2.037   0.0417 *
PC5           14.577     35.199   0.414   0.6788
PC6           11.112     35.444   0.313   0.7539
PC7         -112.207     53.201  -2.109   0.0349 *
---
Signif. codes:  0 ‘***’ 0.001 ‘**’ 0.01 ‘*’ 0.05 ‘.’ 0.1 ‘ ’ 1

(Dispersion parameter for binomial family taken to be 1)

    Null deviance: 99.374  on 79  degrees of freedom
Residual deviance: 84.661  on 72  degrees of freedom
AIC: 100.66

Number of Fisher Scoring iterations: 6

> anova(model, test="Chisq")
Analysis of Deviance Table

Model: binomial, link: logit

Response: cdif

Terms added sequentially (first to last)


     Df Deviance Resid. Df Resid. Dev Pr(>Chi)
NULL                   79     99.374
PC1   1   1.6326       78     97.741  0.20135
PC2   1   1.4463       77     96.295  0.22913
PC3   1   1.5482       76     94.747  0.21340
PC4   1   3.4508       75     91.296  0.06322 .
PC5   1   0.7119       74     90.584  0.39883
PC6   1   0.0859       73     90.498  0.76947
PC7   1   5.8367       72     84.661  0.01569 *
---
Signif. codes:  0 ‘***’ 0.001 ‘**’ 0.01 ‘*’ 0.05 ‘.’ 0.1 ‘ ’ 1
```

**Supporting Information 9**

(i) Multiple linear regression model for predicting tSNE dimension 1 (R output):

```
Call:
lm(formula = xt ~ PC1 + PC2 + PC3 + PC4 + PC5 + PC6 + PC7, data =
data.frame(Xdat))

Residuals:
     Min       1Q   Median       3Q      Max
-22.1692  -4.4428  -0.0194   4.4867  19.1118

Coefficients:
            Estimate Std. Error t value Pr(>|t|)
(Intercept)   35.488      6.119   5.799 1.59e-07 ***
PC1           44.730     33.856   1.321 0.190568
PC2          590.718     50.339  11.735  < 2e-16 ***
PC3         -230.564     60.122  -3.835 0.000264 ***
PC4         -380.182     70.751  -5.374 8.83e-07 ***
PC5         -279.807     73.497  -3.807 0.000290 ***
PC6          -11.302     91.791  -0.123 0.902345
PC7         -249.718    142.113  -1.757 0.083080 .
---
Signif. codes:  0 '***' 0.001 '**' 0.01 '*' 0.05 '.' 0.1 ' ' 1

Residual standard error: 8.153 on 73 degrees of freedom
Multiple R-squared:  0.7333,  Adjusted R-squared:  0.7077
F-statistic: 28.68 on 7 and 73 DF,  p-value: < 2.2e-16
```

(ii) Multiple linear regression model for predicting tSNE dimension 2:

```
Call:
lm(formula = yt ~ PC1 + PC2 + PC3 + PC4 + PC5 + PC6 + PC7, data =
data.frame(Ydat))

Residuals:
    Min      1Q  Median      3Q     Max
-27.288  -9.988  -0.336   5.293  49.367

Coefficients:
            Estimate Std. Error t value Pr(>|t|)
(Intercept)    58.15      10.92   5.322 1.08e-06 ***
PC1          -726.38      60.44 -12.018  < 2e-16 ***
PC2           487.09      89.87   5.420 7.33e-07 ***
PC3           451.13     107.33   4.203 7.36e-05 ***
PC4           175.68     126.31   1.391   0.1685
PC5           260.63     131.21   1.986   0.0507 .
PC6           -50.17     163.87  -0.306   0.7603
PC7          -216.72     253.70  -0.854   0.3958
---
Signif. codes:  0 '***' 0.001 '**' 0.01 '*' 0.05 '.' 0.1 ' ' 1

Residual standard error: 14.56 on 73 degrees of freedom
Multiple R-squared:  0.7346,  Adjusted R-squared:  0.7092
F-statistic: 28.87 on 7 and 73 DF,  p-value: < 2.2e-16
```

## Supporting Information 10

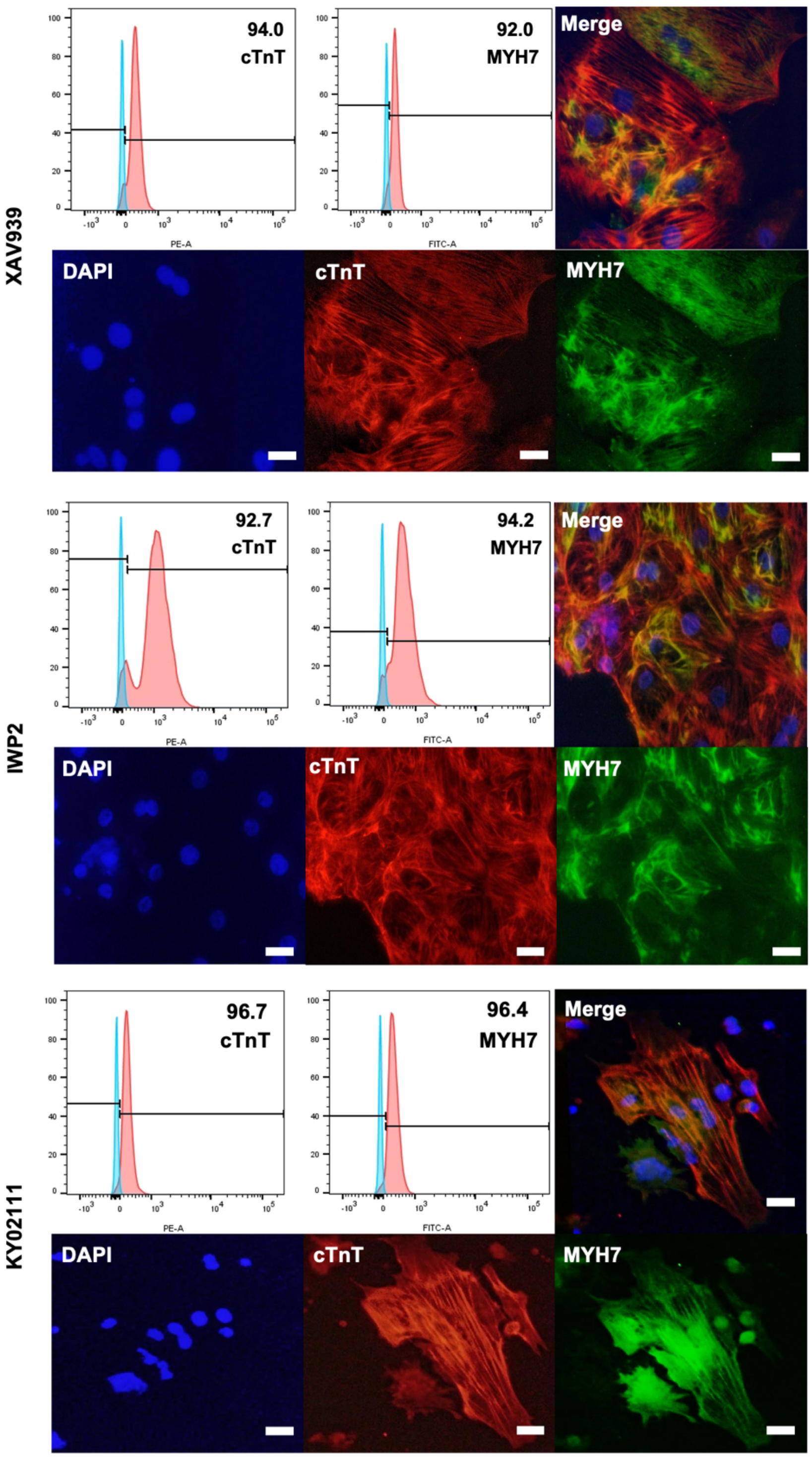

Immunostaining and flow cytometry analysis of human iPS cell differentiation to cardiomyocytes using the compounds XAV939, IWP2, and KY02111. Immunostaining for cTnT and MYH7 expression and flow cytometry analysis of cTnT expression on day 8 of differentiation. (the blue curve represented isotype control, and the red curve represented cTnT-positive cells). Cells are shown at ×40 magnifications.